\documentclass[fleqn,usenatbib]{mnras}

\usepackage{newtxtext,newtxmath}

\usepackage[T1]{fontenc}

\DeclareRobustCommand{\VAN}[3]{#2}
\let\VANthebibliography\thebibliography
\def\thebibliography{\DeclareRobustCommand{\VAN}[3]{##3}\VANthebibliography}

\usepackage{graphicx}	
\usepackage{savesym}
\savesymbol{Bbbk}
\usepackage{amsmath}	
\usepackage{amssymb}	
\restoresymbol{Bbbk1}{Bbbk}
\usepackage[dvipsnames]{xcolor}

\title[SCUBA-2 observations of nearby galaxies]{The JCMT Nearby Galaxies Legacy Survey: SCUBA-2 observations of nearby galaxies}

\author[K. Pattle et al.]{
Kate Pattle,$^{1,2}$\thanks{E-mail: k.pattle@ucl.ac.uk (KP)}
Walter Gear,$^{2}$ and Christine D. Wilson$^{3}$
\\
$^{1}$Department of Physics and Astronomy, University College London, Gower Street, London WC1E 6BT, United Kingdom\\
$^{2}$Centre for Astronomy, School of Natural Sciences, University of Galway, University Road, Galway H91 TK33, Ireland\\
$^{3}$Department of Physics and Astronomy, McMaster University, 1280 Main Street West, Hamilton, Ontario L8S 4M1, Canada 
}

\date{Accepted 2023 February 27. Received 2023 February 15; in original form 2022 December 16}

\pubyear{2023}

\begin{document}
\label{firstpage}
\pagerange{\pageref{firstpage}--\pageref{lastpage}}
\maketitle

\begin{abstract}
We present 850$\mu$m observations of a sample of 8 nearby spiral galaxies, made using the SCUBA-2 camera on the James Clerk Maxwell Telescope (JCMT) as part of the JCMT Nearby Galaxies Legacy Survey (NGLS).  {We corrected our data for the presence of the $^{12}$CO $J=3\to 2$ line in the SCUBA-2 850$\mu$m bandwidth using NGLS HARP data, finding a typical $^{12}$CO contribution of $\sim 20$\%.}  
We measured dust column densities, temperatures and opacity indices by fitting spectral energy distributions constructed from SCUBA-2 and archival \textit{Herschel} observations, {and used archival GALEX and \textit{Spitzer} data to make maps of surface density of star formation ($\Sigma_{\textsc{sfr}}$)}.  {Typically,} comparing SCUBA-2-derived H$_2$ surface densities ($\Sigma_{\rm H_2}$) to $\Sigma_{\textsc{sfr}}$ gives shallow star formation law indices within galaxies, with SCUBA-2-derived values typically being sublinear and \textit{Herschel}-derived values typically being broadly linear.  {This difference is likely due to the effects of atmospheric filtering on the SCUBA-2 data.}  Comparing the mean values of $\Sigma_{\rm H_2}$ and $\Sigma_{\textsc{sfr}}$ of the galaxies in our sample returns a steeper star formation law index, broadly consistent with both the Kennicutt-Schmidt value of 1.4 {and linearity}.
Our results show that a SCUBA-2 detection is a good predictor of star formation.  We suggest that Herschel emission traces {gas in regions which will form stars} on timescales $\sim 5-100$\,Myr, comparable to the star formation timescale traced by GALEX and \textit{Spitzer} data, while SCUBA-2 preferentially traces {the densest} gas {within these regions}, {which likely forms stars on shorter timescales.}
\end{abstract}




\section{Introduction}

The evolution of a galaxy is intrinsically linked to the star formation that takes place within it.  Stars form from the densest phase of the interstellar medium of galaxies, from gravitationally unstable structures within clouds composed primarily of dense molecular hydrogen \citep{bergin2007}.  Understanding the timescale on which, and efficiency with which, molecular gas is converted into stars is crucial to understanding the star-forming histories of galaxies \citep[e.g.][]{kennicutt2012}.  

One of the key metrics by which the link between the gas properties of galaxies and the star formation within them is parametrized is the Kennicutt-Schmidt (KS) star formation law \citep{schmidt1959,kennicutt1998}, a scaling relation between surface gas density ($\Sigma_{gas}$) and surface density of star formation rate ($\Sigma_{\textsc{sfr}}$).  The star formation law can be measured either between a sample of galaxies \citep[e.g.][]{kennicutt1998} or within individual galaxies \citep[e.g.][]{leroy2008}.  $\Sigma_{\textsc{sfr}}$ is typically well-correlated with $\Sigma_{gas}$, and the relationship is parametrized as
\begin{equation}
    \Sigma_{\textsc{sfr}}\propto \Sigma_{gas}^{N}
\end{equation}
\citep{schmidt1959,kennicutt1998}.  \citet{kennicutt1998} found $N=1.4\pm 0.15$, measuring disc-averaged values of both quantities over an ensemble of galaxies.  However, the molecular gas surface density ($\Sigma_{{\rm H}_2}$) is typically much better-correlated with $\Sigma_{\textsc{sfr}}$ than is the total gas surface density \citep{wong2002}, {as might be expected given that star formation occurs within clouds of cold molecular gas \citep[e.g.][]{kennicutt2012}.}  

The relationship between $\Sigma_{SFR}$ and $\Sigma_{gas}$ {or $\Sigma_{{\rm H}_2}$} within individual galaxies (the resolved KS law) has also been extensively investigated \citep[e.g.][]{bigiel2008,bolatto2017,zabel2020,ellison2021}.  {On scales $\gtrsim 1\,$kpc, a correlation is seen between $\Sigma_{SFR}$ and $\Sigma_{{\rm H}_2}$;} \citet{bigiel2008} found an average index $N = 1.0 \pm 0.2$ between $\Sigma_{{\rm H}_2}$ and $\Sigma_{SFR}$ in a sample of spiral galaxies.  This linear relationship, {also found by \citet{bolatto2017},} suggests that {stars form from molecular gas with constant efficiency}  within these galaxies.  {The offset of the resolved KS law varies significantly between galaxies, with galaxies with higher stellar masses, larger Sersic indices, and lower specific star formation rates typically having a lower resolved KS law \citep{ellison2021}}.  {Moreover, a} range of values {of $N$} have been found in nearby galaxies: for example, \citet{ford2013}, observing M31, found a super-KS index of $2.03\pm0.04$ for $\Sigma_{gas}$ (H\textsc{i}, and H$_{2}$ from CO), but sublinear indices for molecular gas only: $0.60\pm0.01$ for $\Sigma_{{\rm H}_2}$ from CO, and $0.55\pm0.01$ for  $\Sigma_{{\rm H}_2}$ from \textit{Herschel} dust emission, assuming a radial gas-to-dust ratio gradient.  A sub-linear star formation law suggests that star formation becomes less efficient at high gas densities, which is difficult to physically motivate.  However, \citet{williams2018}, observing M33, found on kpc scales an index $1.30\pm 0.11$ for molecular gas from CO, but $5.53 \pm 0.75$ for total gas and $5.85 \pm 2.37$ for gas from dust, with all three quantities varying significantly with the spatial scale over which they were measured.  {These differences are likely due to the disc of M33 being H\textsc{i}-dominated \citep{williams2018}; surface density of atomic gas $\Sigma_{{\rm H}\textsc{i}}$ is a poor tracer of $\Sigma_{SFR}$ \citep[e.g.][]{gao2004}.}  The index measured also depends on the amount of diffuse background subtracted in both the gas and the star formation rate tracers \citep{kumari2020}.

Molecular clouds contain an interstellar dust component, consisting principally of silicates, carbonaceous grains, and polyaromatic hydrocarbons (PAHs), which typically makes up $\sim 1$\% of molecular clouds by mass \citep[e.g.][]{draine2007}.  Continuum emission from interstellar dust is a widely-used tracer of molecular gas \citep[e.g.][]{hildebrand1983}.  The James Clerk Maxwell Telescope (JCMT) Nearby Galaxies Legacy Survey \citep[NGLS,][]{wilson2009,wilson2012} is a large programme which mapped the molecular gas and dust in a sample of galaxies within a distance of 25\,Mpc.  In this paper, we present Submillimetre Common-User Bolometer Array 2 (SCUBA-2) 850$\mu$m dust emission observations of 8 galaxies from the NGLS sample.

We present the observations in Section~\ref{sec:obs}, and briefly review the galaxies which we consider in Section~\ref{sec:gals}.  In Section~\ref{sec:g2d} we compare the dust and atomic and molecular gas distributions of the galaxies which we consider.  In Section~\ref{sec:seds} we describe the process of fitting modified black-body functions to the spectral energy distributions of the galaxies which we consider.  In Section~\ref{sec:sf_law} we construct resolved and unresolved star formation laws for the galaxies which we consider.  In Section~\ref{sec:discuss} we discuss our results.  Section~\ref{sec:summary} summarises this work.

\section{Observations}
\label{sec:obs}

\begin{table*}
    \centering
    \begin{tabular}{c cc c c c c}
    \hline
     & R.A. & Dec. & Hubble & & Distance \\
    Galaxy & (J2000) & (J2000) & classification & Inclination & (Mpc) & $12+\log_{10}({\rm O/H})$ \\
    \hline
        NGC 3034 & $09^{h}55^{m}52\farcs43$ & $+69^{\circ}40^{\prime}46\farcs9$ & Scd & 76.9 & 3.61 & -- \\
        NGC 3351 & $10^{h}43^{m}57\fs73$ & $+11^{\circ}42^{\prime}13\farcs0$ & Sb & 54.6 & 9.91 & $8.654^{+0.018}_{-0.020}$ \\ 
        NGC 3521 & $11^{h}05^{m}48\fs57$ & $-00^{\circ}02^{\prime}09\farcs2$ & SABb & 60.0 & 12.42 & $8.604^{+0.026}_{-0.027}$ \\ 
        NGC 4254 & $12^{h}18^{m}49\fs63$ & $+14^{\circ}24^{\prime}59\farcs4$ & Sc & 20.1 & 12.88 & $8.554^{+0.020}_{-0.021}$ \\
        NGC 4569 & $12^{h}36^{m}49\fs82$ & $+13^{\circ}09^{\prime}46\farcs3$ & SABa & 70.8 & 11.86 & -- \\
        NGC 4736 & $12^{h}50^{m}53\fs15$ & $+41^{\circ}07^{\prime}12\farcs6$ & Sab & 31.7 & 4.39 & $8.623^{+0.046}_{-0.046}$ \\ 
        NGC 5055 & $13^{h}15^{m}49\fs27$ & $+42^{\circ}01^{\prime}45\farcs7$ & Sbc & 54.9 & 9.04 & $8.581^{+0.054}_{-0.053}$ \\
        NGC 5194 & $13^{h}29^{m}52\fs70$ & $+47^{\circ}11^{\prime}42\farcs9$ & Sbc & 32.6 & 8.59 & $8.638^{+0.012}_{-0.006}$
\\ 
        \hline
    \end{tabular}
    \caption{The coordinates, Hubble classifications, inclinations, distances and metallicities of our set of galaxies.  Classifications, inclinations, distances and metallicities (where present) are taken from the Dustpedia database \citep{clark2018}.  Dustpedia metallicities are presented by \citet{devis2019};  we use their preferred `PG16S' calibration.  We compare these to a solar metallicity of $12+\log_{10}({\rm O/H})=8.69\pm0.05$ \citep{asplund2009}.} 
    \label{tab:source_list}
\end{table*}

The SCUBA-2 850$\mu$m data presented in this paper was taken under project code MJLSN07.  All data were taken using the SCUBA-2 CV-DAISY mapping mode, with the exception of NGC 5194, which was mapped using the PONG-900 mode \citep{holland2013}.  The CV-DAISY mode was used for these extended sources despite being optimised for compact sources because it produces a high exposure time in the map centre, which is necessary to observe relatively small low-surface-brightness sources in a reasonable amount of time \citep{holland2013}.

We selected 8 bright galaxies (listed in Table~\ref{tab:source_list}) with ancillary observations of the $^{12}$CO $J=1\to 0$ line made using the Nobeyama 45m telescope as part of the COMING \citep{sorai2019} and CO-ATLAS \citep{kuno2007} surveys.  All of the galaxies in our sample were observed in the $^{12}$CO $J=3\to2$ line using HARP \citep{buckle2009} as part of the NGLS.  These galaxies also have ancillary observations with \textit{Herschel}, GALEX and \textit{Spitzer} \citep{clark2018}.  {These galaxies were selected for submillimetre brightness and completeness of ancillary data sets, and are not intended to be a representative sample of nearby galaxies.}
The UT start and end dates of the observations, the number of repeats and integration time per source, and the weather conditions under which they were observed are listed in Table~\ref{tab:obsdetails}.  These data were largely taken in JCMT weather bands 2 and 3.  JCMT weather bands are defined by atmospheric opacity at 225\,GHz ($\tau_{225}$); Band 2 is defined by $0.05<\tau_{225}<0.08$, and Band 3 by $0.08 < \tau_{225} < 0.12$ \citep{dempsey2013}.

\subsection{SCUBA-2 data reduction}
\label{sec:s2dr}

We reduced the data using the \textit{skyloop}\footnote{\url{http://starlink.eao.hawaii.edu/docs/sun258.htx/sun258ss72.html}} implementation of the \textit{makemap} algorithm in \textsc{Smurf} \citep{chapin2013}, in which one iteration of \textit{makemap} is performed on each of the observations in the set in turn, with the set being averaged together at the end of each iteration, rather than each observation being reduced consecutively.  Although SCUBA-2 observes 850$\mu$m and 450$\mu$m simultaneously, we consider only 850$\mu$m data in this paper.

In order to ensure good image fidelity, we ran \textit{makemap} to a tolerance of 1\%.  We also defined a maximum observable size scale of 180$^{\prime\prime}$, in order to prevent the growth of large-scale, low-level emission structures in the output map.  This maximum size scale {is chosen to match the size of the central region of a CV-DAISY map over which exposure times are relatively uniform \citep{holland2013}, and} is more stringent than is usual for SCUBA-2 observations. {The default maximum observable size scale is 300$^{\prime\prime}$, set by the size of a SCUBA-2 subarray}  (see, e.g., \citealt{holland2013}; \citealt{kirk2018}).  {By choosing a smaller maximum size scale, we place stronger constraints on the iterative \textit{makemap} algorithm, which} is necessary in order to accurately recover faint extended sources, particularly when using the CV-DAISY mapping mode, which has a small map size and a non-uniform exposure time.

We used the Nobeyama 45m $^{12}$CO $J=1\to 0$ maps \citep{kuno2007,sorai2019} to define a fixed `mask' for each observation, defining areas of astrophysical emission.  {These observations were chosen because their resolution ($\sim 17^{\prime\prime}$; \citealt{sorai2019}) is comparable to that of the JCMT, while their low-$J$ transition and more transmissive atmospheric window give better sensitivity to extended emission over a larger area than do the NGLS HARP $J=3\to 2$ data.  Our aim in choosing a mask is to define the maximum area over which dust emission is likely to be detected; we want to provide sufficient constraint on the mapmaker to exclude regions without real signal whilst including an area sufficiently large to allow real extended structure to grow.}  The masked area was defined by an SNR $>4$ {in integrated intensity} in the Nobeyama 45m data, {as provided by the CO-ATLAS\footnote{\url{https://www.nro.nao.ac.jp/~nro45mrt/html/COatlas/}} and COMING\footnote{\url{https://astro3.sci.hokudai.ac.jp/~radio/coming/}} archives}.  Areas outside this masked region were set to zero until the final iteration of \textit{makemap} (see \citealt{mairs2015} for a detailed discussion of the role of masking in SCUBA-2 data reduction).  In each case, the SCUBA-2 flux recovered in the output map did not fill the area defined by the mask, indicating that our choice of mask did not encourage the growth of spurious structures in the SCUBA-2 maps.

By choosing these map-making parameters, it is likely that we have sacrificed some potentially-recoverable large-scale structure in order to achieve good image fidelity.  We discuss the implications of this choice at various points throughout this work.

The output maps were gridded to 4$^{\prime\prime}$ pixels and calibrated in mJy\,beam$^{-1}$ and mJy\,arcsec$^{-2}$ using the standard SCUBA-2 850$\mu$m flux conversion factors (FCFs) of 537 Jy\,beam$^{-1}$\,pW$^{-1}$ and 2340 mJy\,arcsec$^{-2}$\,pW$^{-1}$ \citep{dempsey2013}.

The effective resolution of SCUBA-2 at 850$\mu$m is $14\farcs1$.  The RMS noise in our maps is in the range 2.6--4.0 mJy\,beam$^{-1}$, with variation between maps being due to differences in exposure time, elevation, and amounts of large-scale structure present.  RMS noise values and linear resolutions for each individual field are listed in Table~\ref{tab:source_properties}.

\begin{table*}
    \centering
    \begin{tabular}{c cc cc cc}
    \hline
     & \multicolumn{2}{c}{UT date} & & Integration & \multicolumn{2}{c}{$\tau_{225}$} \\ \cline{2-3} \cline{6-7}
    Galaxy  & Start & End & Repeats & time (mins) & Start & End  \\
    \hline
    NGC 3034 & 2012 Apr 01 & 2014 Feb 20 & 8 & $\sim 21$ & 0.055 & 0.099 \\
    NGC 3351 & 2014 Feb 12 & 2014 May 30 & 8 & $\sim 21$ & 0.098 & 0.071 \\
    NGC 3521 & 2013 Dec 05 & 2014 May 30 & 8 & $\sim 21$ & 0.126 & 0.074 \\
    NGC 4254 & 2012 Mar 31 & 2014 Jun 05 & 11 & $\sim 21$ & 0.054 & 0.121 \\
    NGC 4569 & 2012 Mar 31 & 2014 Jun 08 & 11 & $\sim 21$ & 0.037 & 0.119 \\
    NGC 4736 & 2012 Feb 04 & 2014 May 31 & 11 & $\sim 21$ & 0.086 & 0.083 \\
    NGC 5055 & 2013 Dec 11 & 2014 May 29 & 16 & $\sim 21$ & 0.138 & 0.111 \\
    NGC 5194 & 2014 Jan 15 & 2014 May 22 & 16 & $\sim 42$ & 0.092 & 0.117 \\
    \hline
    \end{tabular}
    \caption{Details of the observing dates and conditions for the galaxies in our sample.  Note that integration time is given per repeat.}
    \label{tab:obsdetails}
\end{table*}

\subsection{CO subtraction}

\begin{table*}
    \centering
    \begin{tabular}{@{\extracolsep{3pt}}c c c c c c c c c@{}}
    \hline
     & Linear & \multicolumn{2}{c}{RMS noise} & \multicolumn{2}{c}{Peak 850$\mu$m F.D.} & \multicolumn{2}{c}{Total 850$\mu$m F.D.} & Net CO \\ \cline{5-6} \cline{7-8}
     & resolution & \multicolumn{2}{c}{CO-sub} & With CO & CO-sub & With CO & CO-sub & fraction \\ \cline{2-2} \cline{3-4} \cline{5-6} \cline{7-8}
     & kpc & mJy\,beam$^{-1}$ & mJy\,arcsec$^{-2}$ & \multicolumn{2}{c}{mJy\,arcsec$^{-2}$} & \multicolumn{2}{c}{Jy} & \\
    \hline
        NGC 3034 & 0.25 & 3.7 & 0.016 & 6.53 & 4.45 & 8.72 & 5.43 & 0.38 \\
        NGC 3351 & 0.68 & 3.4 & 0.015 & 0.38 & 0.27 & 0.19 & 0.15 & 0.19 \\
        NGC 3521 & 0.85 & 4.0 & 0.017 & 0.20 & 0.18 & 1.48 & 1.20 & 0.18 \\
        NGC 4254 & 0.88 & 3.8 & 0.017 & 0.22 & 0.16 & 1.02 & 0.80 & 0.21 \\
        NGC 4569 & 0.81 & 3.8 & 0.017 & 0.31 & 0.22 & 0.38 & 0.31 & 0.17 \\
        NGC 4736 & 0.30 & 3.0 & 0.013 & 0.21 & 0.14 & 0.76 & 0.58 & 0.24 \\
        NGC 5055 & 0.62 & 2.6 & 0.011 & 0.36 & 0.28 & 1.92 & 1.65 & 0.14 \\
        NGC 5194 & 0.59 & 3.3 & 0.014 & 0.30 & 0.21 & 0.50* & 0.24* & 0.52 \\
        \hline
    \end{tabular}
    \caption{The key properties of the SCUBA-2 850$\mu$m measurements of the sources in our sample.  The RMS noises listed in this table are measured in the CO-subtracted maps used for analysis. {*Note that these values include summation over the significant negative bowls in the interarm regions of NGC 5194.}} 
    \label{tab:source_properties}
\end{table*}

The SCUBA-2 850$\mu$m filter has a half-power bandwidth of 85$\mu$m \citep{holland2013}, and so SCUBA-2 850$\mu$m data can include a significant contribution from the $^{12}$CO $J=3\to 2$ transition, the rest wavelength of which is 867.6$\mu$m (345.8\,GHz).

We accounted for the contribution of CO to our SCUBA-2 flux densities using the technique described by \citet{drabek2012}.  Each of the 850$\mu$m observations was re-reduced with the integrated HARP \citep{buckle2009} $^{12}$CO data added to the SCUBA-2 bolometer time series as a negative signal.  {By repeating the data reduction process described in Section~\ref{sec:s2dr}, the same spatial filtering is applied to the subtracted HARP $^{12}$CO signal as to the SCUBA-2 data, and so the spatial scales in the two data sets are matched. The SCUBA-2 850$\mu$m and HARP $^{12}$CO data have very similar angular resolutions, agreeing to within 2\%, and so no correction for this small difference is necessary.}

We used the integrated NGLS HARP $^{12}$CO maps\footnote{\url{https://www.physics.mcmaster.ca/~wilson/www_xfer/NGLS/Data_and_Plots/v2-0/SINGS2/}} \citep{wilson2012}, {in which data are supplied for pixels with a SNR $>2$ in total integrated intensity.} The conversion from K\,km\,s$^{-1}$ to pW is dependent on atmospheric opacity; we calculated a conversion factor for each observation from the mean of its start- and end-time 225\,GHz opacities using the relations given by \citet{parsons2018}.

For the galaxies in question, the net contribution of $^{12}$CO to the SCUBA-2 850$\mu$m flux is typically $\sim 20$\%, but reaches 38\% in NGC 3034 and $\sim 50$\% in NGC 5194, {although our data reduction strategy in NGC 5194 may be suboptimal, as discussed in Section~\ref{sec:ngc5194}, below}.  The peak and total flux densities before and after CO subtraction are listed in Table~\ref{tab:source_properties}.  CO fractions of order tens of percent in bright regions indicate that this correction cannot be neglected when calculating dust masses from SCUBA-2 observations.

{As can be seen in Figures~\ref{fig:ngc3034}--\ref{fig:ngc5194}, the CO contamination fraction drops to near zero at the peripheries of the $^{12}$CO SNR $>2$ regions, indicating that little or no $^{12}$CO emission has been retained in the 850$\mu$m images by excluding the lower-SNR data.  With the exception of NGC 5194, the area over which SCUBA-2 signal is detected is well-covered by the footprint of the HARP observations.  However, in the NGC 5194 field, the companion galaxy, NGC 5195, and a small part of the southern spiral arm of NGC 5194, are not mapped by HARP.  The relatively few pixels which are thus bright in 850$\mu$m emission but not corrected for $^{12}$CO contamination transpire not to be well-characterised and are excluded by data quality cuts discussed later in this work, and so do not affect our conclusions.} 

We note that {much lower} CO contamination fractions have been found in SCUBA-2 images of M31 \citep{smith2021}. 
The differences in CO fraction {between M31 and the galaxies in our sample} {are likely} due to CO emission from M31 having low surface brightness, with line fluxes $< 10$\,K\,km\,s$^{-1}$ everywhere \citep{smith2021}.  The galaxies in our sample have significantly higher CO brightnesses (cf. Figures~\ref{fig:ngc3034}--\ref{fig:ngc5194}).  {As a potential green valley galaxy \citep{mutch2011}, M31 is likely to have both less gas and less dust than actively star-forming galaxies.}
There may also be differences in metallicity between the galaxies {in our sample and M31}.
{The contribution of $^{12}$CO $J=3\to 2$ to SCUBA-2 850$\mu$m emission in local Milky Way star-forming regions is typically $ \lesssim 20$\%, but may be higher in the presence of outflows \citep{drabek2012,pattle2015,coude2016}.}

\section{Galaxies under consideration}
\label{sec:gals}

We here briefly introduce the 8 galaxies which we consider in this paper.  The coordinates, classifications, inclinations, distances and metallicities of the galaxies are listed in Table~\ref{tab:source_list}.  The listed metallicities are taken from \citet{devis2019}.  All of the galaxies in our sample for which \citet{devis2019} list metallicities have slightly sub-solar values of $12+\log({\rm O/H})$, with the largest deficit being $-0.136^{+0.070}_{-0.071}$ in NGC 4254 (compared to the solar value of $12+\log({\rm O/H})=8.69\pm 0.05$; \citealt{asplund2009}), equivalent to $Z/Z_{\odot} = 0.72^{+0.14}_{-0.10}$.  {The SCUBA-2 850$\,\mu$m observations of the galaxies in our sample are shown in Figures~\ref{fig:ngc3034}--\ref{fig:ngc5194}, while observations made in the mid-infrared regime with \textit{Spitzer} and in the ultraviolet regime with GALEX are shown in Figures~\ref{fig:mwl_ngc3034}--\ref{fig:mwl_ngc5194} in Appendix~\ref{sec:appendix_mwl}.}

\subsection{NGC 3034}

\begin{figure*}
    \centering
    \includegraphics[width=\textwidth]{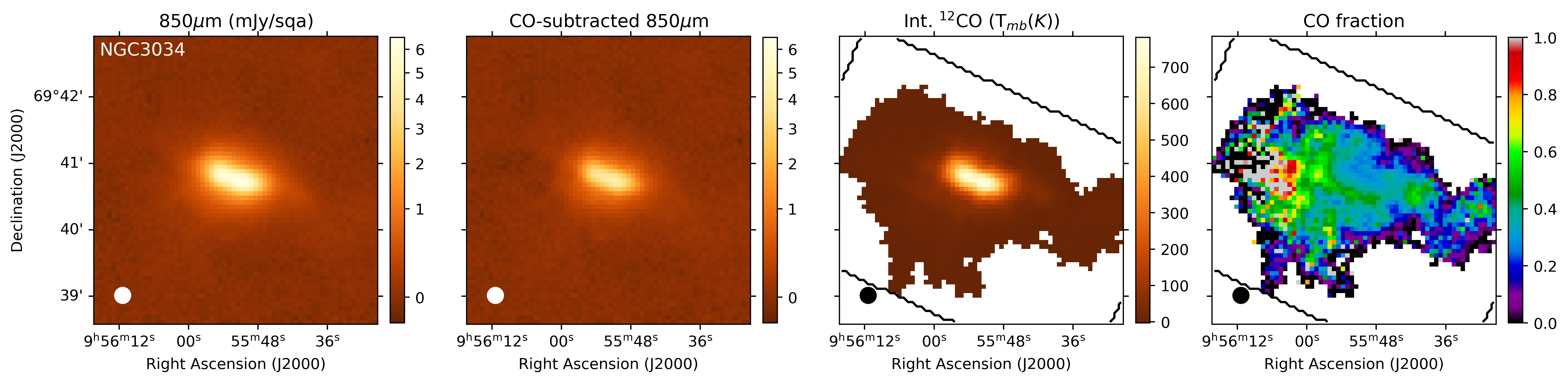}
    \caption{JCMT NGLS observations of NGC 3034.  Far left: SCUBA-2 850\,$\mu$m data.  Centre left: SCUBA-2 850\,$\mu$m data, with $^{12}$CO $J=3\to 2$ contribution subtracted.  Both SCUBA-2 images are shown with square-root scaling, in units of mJy\,arcsec$^{-2}$.  Centre right: integrated HARP $^{12}$CO $J=3\to 2$ emission, in main-beam temperature units.  Far right: fraction of SCUBA-2 emission in far-left panel that arises from $^{12}$CO $J=3\to 2$ emission.  {In the right-hand panels, the footprint of the HARP $^{12}$CO observation is outlined in black.}}
    \label{fig:ngc3034}
\end{figure*}

NGC 3034 (Messier 82), shown in Figure~\ref{fig:ngc3034}, is a highly-inclined galaxy which is interacting with the neighbouring more massive spiral galaxy M81. 
At a distance of 3.61\,Mpc \citep{jacobs2009}, M82 is the nearest starburst galaxy to the Milky Way, and has a well-studied `superwind' emanating from the central starburst \citep[e.g.][]{devine1999}.  M82, although sometimes classed as irregular, has weak bar and spiral arm features \citep{mayya2005}, the magnetic field in which has recently been mapped using the POL-2 polarimeter on SCUBA-2 \citep{pattle2021}, {and was previously mapped with SCUPOL \citep{greaves2000}.}  {HARP $^{12}$CO $J=3\to 2$ observations of NGC 3034 were presented by \citet{wilson2012}.}

{Unlike the other galaxies in our sample, M82 is saturated in the \textit{Spitzer} 24$\mu$m band, and so a star formation rate surface density map cannot be made for it.  We nonetheless include it in our sample as it provides a useful point of comparison against previous observations.}  {M82 was observed with the SCUBA camera, the predecessor to SCUBA-2, by \citet{leeuw2009}.  They measured a peak flux density of 1.4 Jy\,beam$^{-1}$, consistent with the peak flux density of 1.5 Jy/beam which we measure before CO subtraction with SCUBA-2.  The SCUBA camera's 850\,$\mu$m filter was subject to the same CO contamination effects as is that of SCUBA-2 \citep[e.g.][]{meijerink2005}.}

\subsection{NGC 3351}

\begin{figure*}
    \centering
    \includegraphics[width=\textwidth]{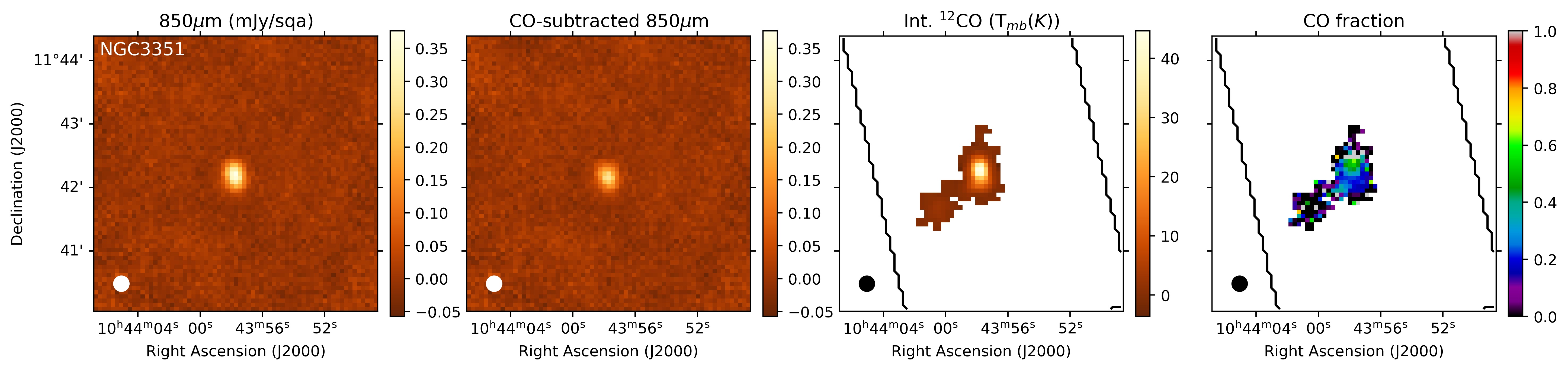}
    \caption{JCMT NGLS observations of NGC 3351.  Far left: SCUBA-2 850\,$\mu$m data.  Centre left: SCUBA-2 850\,$\mu$m data, with $^{12}$CO $J=3\to 2$ contribution subtracted.  Both SCUBA-2 images are shown with linear scaling, in units of mJy\,arcsec$^{-2}$.  Right-hand panels as in Figure~\ref{fig:ngc3034}.}
    \label{fig:ngc3351}
\end{figure*}

NGC3351, shown in Figure~\ref{fig:ngc3351}, is a barred spiral galaxy that displays a very young starburst population within a $15\farcs3\times 11\farcs2$ circumnuclear ring \citep{alloin1982}.  We detect but do not resolve the circumnuclear ring in our observations, but do not recover the extended structure of the galaxy.  HARP $^{12}$CO $J=3\to 2$ observations of NGC 3351 were presented by {\citep{wilson2012} and} \citet{tan2013}.

\subsection{NGC 3521}

\begin{figure*}
    \centering
    \includegraphics[width=\textwidth]{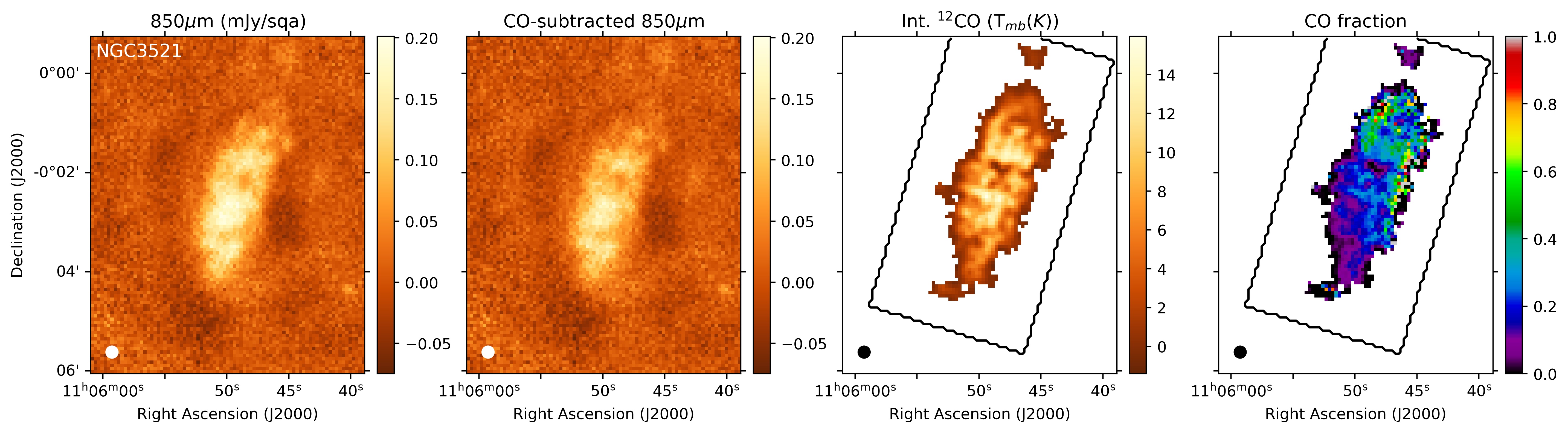}
    \caption{JCMT NGLS observations of NGC 3521.  Panels as in Figure~\ref{fig:ngc3351}.}
    \label{fig:ngc3521}
\end{figure*}

NGC 3521, shown in Figure~\ref{fig:ngc3521} is a flocculent, weakly-barred spiral galaxy with a tightly-wound two-arm pattern \citep{liu2011}.  NGC 3521 is considered comparable to NGC 5194 (discussed below), as both are metal-rich and quiescently star-forming \citep{liu2011}.  We recover a significant amount of the extended structure of NGC 3521, although the spiral arms are not clearly visible in Figure~\ref{fig:ngc3521}.  {HARP $^{12}$CO $J=3\to 2$ observations of NGC 3521 were presented by \citep{wilson2012}.}

\subsection{NGC 4254}

\begin{figure*}
    \centering
    \includegraphics[width=\textwidth]{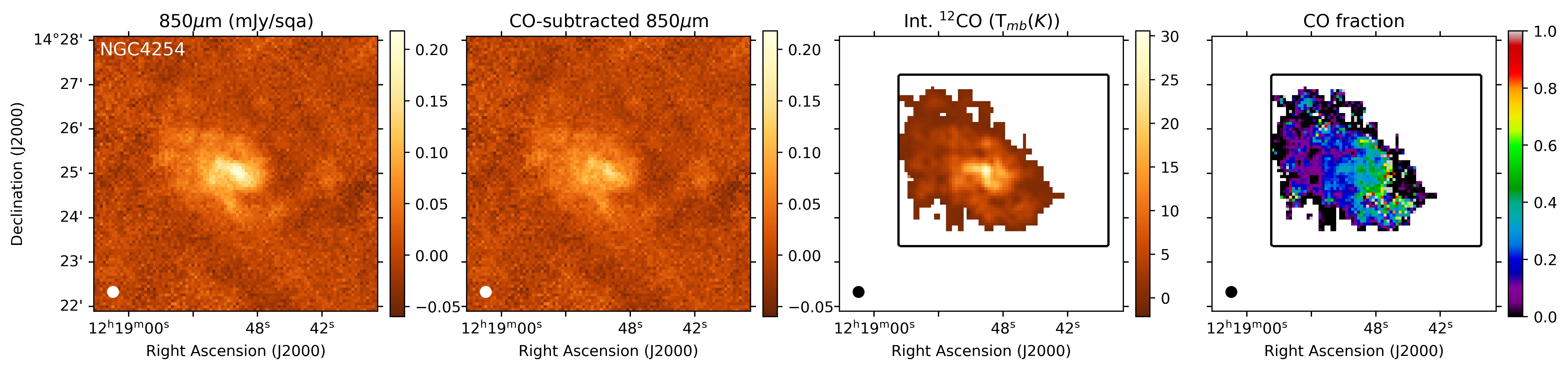}
    \caption{JCMT NGLS observations of NGC 4254.  Panels as in Figure~\ref{fig:ngc3351}.}
    \label{fig:ngc4254}
\end{figure*}

NGC 4254 (Messier 99), shown in Figure~\ref{fig:ngc4254}, is an unbarred Virgo Cluster galaxy with a strong asymmetric spiral pattern which is clearly visible in our SCUBA-2 observations.  NGC 4254 has a high star formation rate, particularly in its southern arm, and a tidal tail extending $\sim250$\,kpc northward from the galaxy (not visible in our observations), suggesting that the galaxy is interacting with the centre of the Virgo Cluster \citep{haynes2007}.  HARP $^{12}$CO $J=3\to 2$ observations of NGC 4254 were presented by \citet{wilson2009} {and \citet{wilson2012}}.

\subsection{NGC 4569}

\begin{figure*}
    \centering
    \includegraphics[width=\textwidth]{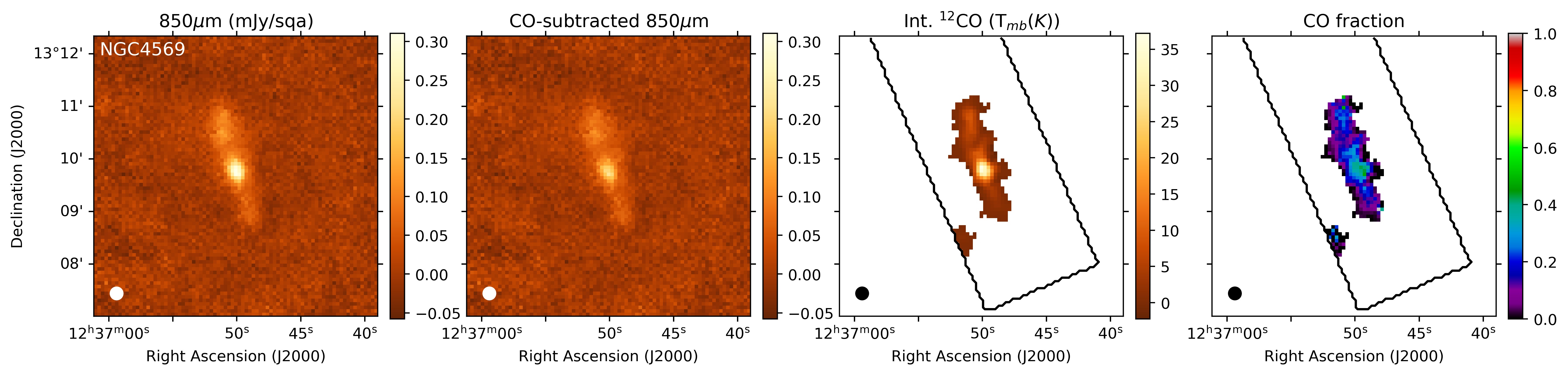}
    \caption{JCMT NGLS observations of NGC 4569.  Panels as in Figure~\ref{fig:ngc3351}.}
    \label{fig:ngc4569}
\end{figure*}

NGC 4569 (Messier 90), shown in Figure~\ref{fig:ngc4569}, is a weakly barred Virgo cluster galaxy which underwent a ram pressure stripping event $\sim 300$\,Myr ago, possibly due to motion relative to the intra-cluster medium \citep{vollmer2004}.  The galaxy has a nuclear outflow and a stripped tail \citep{boselli2016}, which is not visible in the SCUBA-2 data.  HARP $^{12}$CO $J=3\to 2$ observations of NGC 4569 were presented by \citet{wilson2009} {and \citet{wilson2012}}.

\subsection{NGC 4736}

\begin{figure*}
    \centering
    \includegraphics[width=\textwidth]{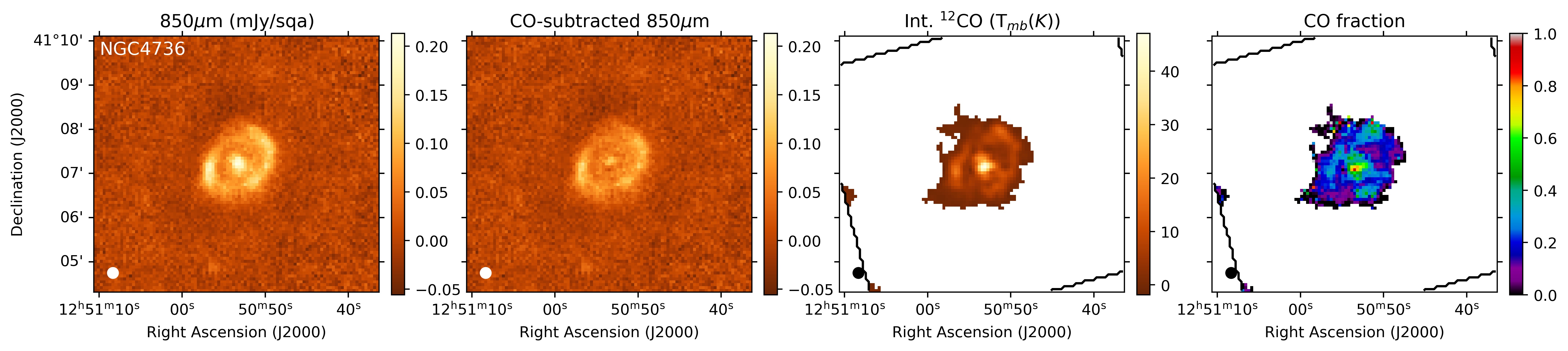}
    \caption{JCMT NGLS observations of NGC 4736.  Panels as in Figure~\ref{fig:ngc3351}.}
    \label{fig:ngc4736}
\end{figure*}

NGC 4736 (Messier 94), shown in Figure~\ref{fig:ngc4736}, is an actively star-forming unbarred ring galaxy.  The galaxy has a weak and diffuse outer ring, which is at best marginally detected in our SCUBA-2 observations, and a bright and actively star-forming inner pseudo-ring \citep{wong2000, waller2001}.  This inner pseudo-ring, with a radius of $\sim 47^{\prime\prime}$ \citep{chyzy2008}, is clearly visible in our SCUBA-2 data.  {HARP $^{12}$CO $J=3\to 2$ observations of NGC 4736 were presented by \citep{wilson2012}.}

\subsection{NGC 5055}

\begin{figure*}
    \centering
    \includegraphics[width=\textwidth]{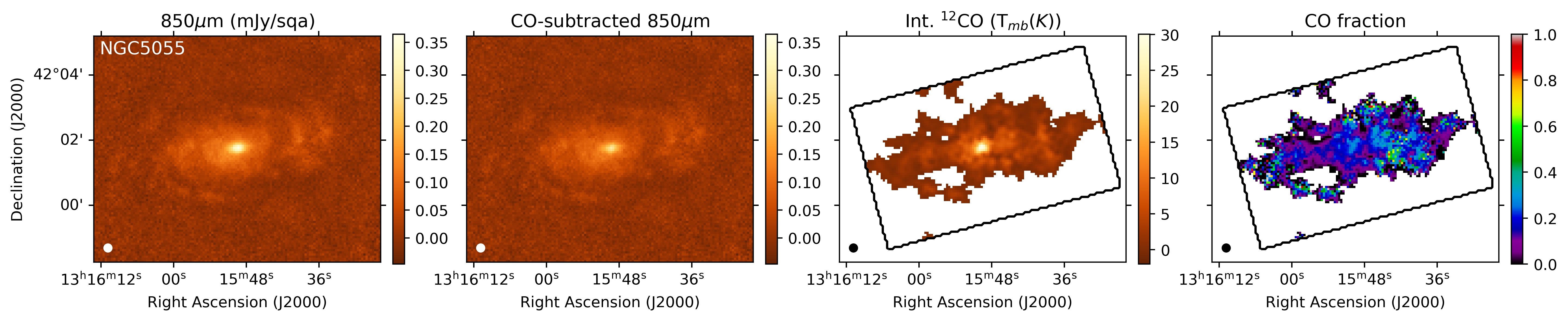}
    \caption{JCMT NGLS observations of NGC 5055.  Panels as in Figure~\ref{fig:ngc3351}.}
    \label{fig:ngc5055}
\end{figure*}

NGC 5055 (Messier 63), shown in Figure~\ref{fig:ngc5055}, is a moderately inclined unbarred spiral galaxy with a very large, warped H\textsc{i} disc \citep{battaglia2006}, in which massive stars have recently formed \citep{thilker2007}.  The spiral structure of the galaxy is clearly visible in our SCUBA-2 observations.  {HARP $^{12}$CO $J=3\to 2$ observations of NGC 5055 were presented by \citep{wilson2012}.}

\subsection{NGC 5194}
\label{sec:ngc5194}

\begin{figure*}
    \centering
    \includegraphics[width=\textwidth]{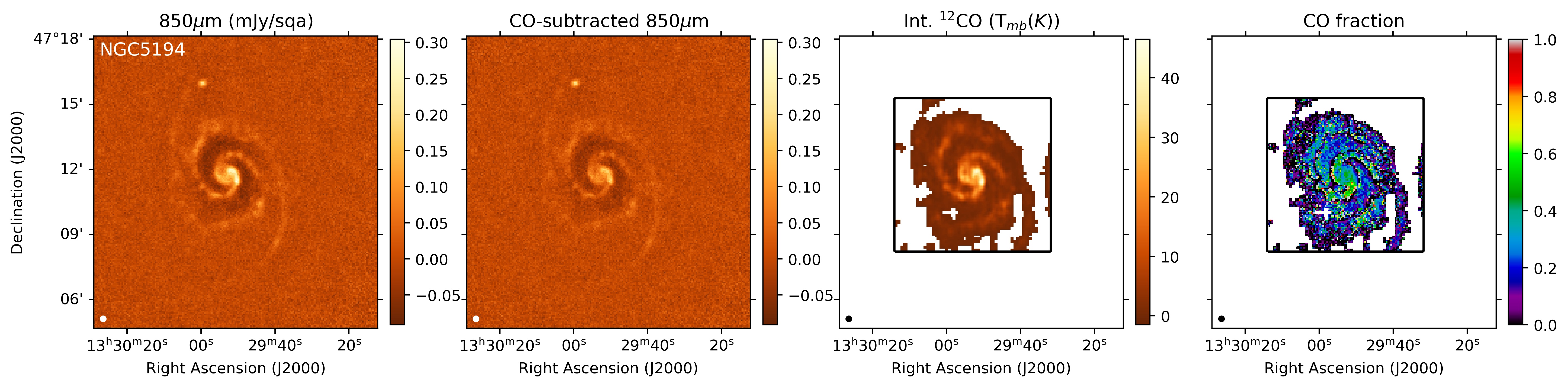}
    \caption{JCMT NGLS observations of NGC 5194.  Panels as in Figure~\ref{fig:ngc3351}.}
    \label{fig:ngc5194}
\end{figure*}

NGC 5194 (Messier 51a) is a face-on two-arm grand-design spiral galaxy, interacting with its neighbour, M51b.  Both are shown in Figure~\ref{fig:ngc5194}.  Star formation in NGC 5194 is mostly taking place in the galactic centre and spiral arms \citep{schinnerer2017}.  CO observations of NGC 5194 show chains of GMCs emerging from the spiral arms into the interarm regions \citep{koda2009, schinnerer2017}, but these features are not resolved in our observations.  {HARP $^{12}$CO $J=3\to 2$ observations of NGC 3521 were presented by \citep{wilson2012} and \citet{vlahakis2013}.}

Unlike the other galaxies in our sample, NGC 5194 was observed using the PONG 900 observing mode \citep{holland2013}.  In order to be as consistent as possible, we have applied the same data reduction parameters to all of the galaxies in our sample.  However, the data reduction process which we describe in Section~\ref{sec:obs} does not appear to be optimal for NGC 5194; the SCUBA-2 images in Figure~\ref{fig:ngc5194} show significant negative bowling in the inter-arm regions.  This suggests that recoverable large-scale emission from NGC 5194 has been lost in the data reduction process, {and affects our estimates of total flux density for the galaxy listed in Table~\ref{tab:source_properties}.}

{NGC 5194 was previously observed with SCUBA \citep{meijerink2005}.  These authors found an exponential disc in their 850$\mu$m observations, which we do not recover here.  They measured a peak flux density $\gtrsim 117$\,mJy\,beam$^{-1}$, including a contribution from CO.  We measure a peak flux density of 69\,mJy\,beam$^{-1}$ before CO subtraction, again suggesting that the data reduction strategy which we employ in this work is not optimal for these observations of NGC 5194.  Interestingly, after subtraction of their modelled exponential disc, \citet{meijerink2005} find a peak flux density of $\gtrsim 72$\,mJy\,beam$^{-1}$, consistent with the peak flux density which we measure.  This suggests that we may be recovering the emission from more compact structures in NGC 5194 quite well.}

\section{Comparison with molecular and atomic gas}
\label{sec:g2d}

We investigated the correlation between SCUBA-2 850$\mu$m flux density and atomic and molecular gas column density.

We measured H\textsc{i} column density ($N({\rm H}\textsc{i})$) using Very Large Array (VLA) H$\textsc{i}$ observations, taken from the THINGS \citep{walter2008} (NGC 3351, 3521, 4736, 5055, 5194) and VIVA \citep{chung2009} (NGC 4254, 4569) surveys, and from \citet{deblok2008} for NGC 3034.  The VLA data were converted from Jy\,beam$^{-1}$ to brightness temperature ($T_{B}$) using the relation 
\begin{equation}
    T_{B} = I_{{\rm H}\textsc{i}}\times \frac{6.07\times 10^{5}}{\theta_{maj}\theta_{min}}\,{\rm K}\,({\rm Jy\,beam}^{-1})^{-1}
\end{equation}
\citep[cf.][]{walter2008}, where $I_{{\rm H}\textsc{i}}$ is H\textsc{i} brightness and $\theta_{maj}$ and $\theta_{min}$ are the major and minor beam widths, respectively. $N({\rm H}\textsc{i})$ was then calculated using the relation
\begin{equation}
    N({\rm H}\textsc{i}) = 1.823\times10^{18}\,{\rm cm}^{-2}\,{\rm K}^{-1}\times T_{B} 
\end{equation}
\citep[cf.][]{walter2008}.
 
We measured H$_{2}$ column density ($N({\rm H}_{2})$) using the Nobeyama $^{12}$CO 1$\to$0 observations from the COMING \citep{sorai2019} and CO ATLAS \citep{kuno2007} surveys which we used to define the data reduction masks as described in Section~\ref{sec:obs}.  $N({\rm H}_{2})$ was calculated using the relation  
\begin{equation}
    N({\rm H}_{2}) = X_{\rm CO}\int T_{mb} {\rm} d \nu,
\end{equation}
taking the integrated main-beam temperature maps from the COMING and CO-ATLAS databases, and using a CO X-factor of $X_{\rm CO} = 2\times 10^{20}$\,cm$^{-2}$ (K\,km\,s$^{-1}$)$^{-1}$ \citep{bolatto2013}. {\citet{bolatto2013} give an uncertainty on $X_{\rm CO}$ of $\pm30$\% in the Milky Way disc, and $X_{\rm CO}$ is expected to vary further with metallicity.  We assume a constant $X_{\rm CO}$ in order to directly compare between $^{12}$CO and SCUBA-2 850$\mu$m emission as molecular gas tracers for the galaxies in our sample, but note that this assumption introduces an additional uncertainty on the values of $N({\rm H}_{2})$ determined from $^{12}$CO observations.}

The Nobeyama 45m $^{12}$CO data have a resolution of 17$^{\prime\prime}$ \citep{sorai2019}.  The VLA data have a variety of resolutions.  Where the VLA resolution was $< 17^{\prime\prime}$ we smoothed both the SCUBA-2 and the $N({\rm H}\textsc{i})$ maps to 17$^{\prime\prime}$.  Where the VLA resolution was $> 17^{\prime\prime}$ (NGC 3034 and NGC 4254), we smoothed the SCUBA-2 and $N({\rm H}_{2})$ maps to the geometric mean of the VLA beam's major and minor axes, 34.3$^{\prime\prime}$ and 29.4$^{\prime\prime}$ for NGC 3034 and NGC 4254 respectively.

We corrected the derived column density for inclination angle using the inclination values given in the Dustpedia database \citep{clark2018} and listed in Table~\ref{tab:source_properties}.  Comparisons between SCUBA-2 850$\mu$m flux density and $N({\rm H\textsc{i}})$, $N({\rm H_{2}})$ and total hydrogen column density ($N({\rm H\textsc{i}})+ 2N({\rm H_{2}})$) are shown in Figure~\ref{fig:co_hi_ngc3034}--\ref{fig:co_hi_ngc5194} in Appendix~\ref{sec:appendix_g2d}. 

$^{12}$CO emission and 850$\mu$m flux density are consistently correlated with one another, {suggesting that both are tracing similar material, i.e. molecular gas (e.g. \citealt{bolatto2013})}.  We typically see no strong correlation between H\textsc{i} and 850$\mu$m, with the possible exceptions of NGC 4569 and 4736.  Generally, the less well-resolved sources show a stronger correlation between H\textsc{i} and 850$\mu$m dust emission.  This {further suggests} that the 850$\mu$m dust emission detected by SCUBA-2 is {preferentially} tracing the molecular gas, {as H\textsc{i} and 850$\mu$m dust emission are better-correlated where H\textsc{i} and $^{12}$CO emission occupy the same beam}.  We note that in NGC 3034 the VLA data are saturated at high column densities, so there is no overlap between the $^{12}$CO and H\textsc{i} measurements.

In the following section, we measure column density of $N({\rm H}_{2})$ from dust emission, taking the dust emission to trace the molecular gas.  However, to do so, we must assume a gas-to-dust ratio, as discussed below.  The gas-to-dust ratio depends on metallicity and varies somewhat between the galaxies in our sample (\citealt{devis2019}; cf. Table~\ref{tab:source_list}).  {As discussed above, we expect $X_{\rm CO}$ to vary systematically within and between galaxies in our sample.}  We are not able to distinguish between variation in gas-to-dust ratio and in $X_{\rm CO}$, and so do not attempt to use the ratio between $^{12}$CO emission and 850$\mu$m flux density to determine gas-to-dust ratios for the galaxies which we consider.  We comment on the consequences of using a constant gas-to-dust ratio further below.

\section{SED fitting}
\label{sec:seds}

We measured dust column densities, temperatures and opacity indices for the galaxies in our sample by fitting their spectral energy distributions (SEDs) using SCUBA-2 850$\mu$m and \textit{Herschel} Space Observatory data.  We took \textit{Herschel} 70$\mu$m, 100$\mu$m (where present), 160$\mu$m, 250$\mu$m and 350$\mu$m data from the Dustpedia database \citep{clark2018}.  We excluded the \textit{Herschel} 500$\mu$m data as being too low-resolution (36.6$^{\prime\prime}$; \citealt{griffin2010}), and, being away from the SED peak but shorter-wavelength than the SCUBA-2 850$\mu$m data, not necessary in order to produce a well-constrained fit.  The lowest-resolution data set which we use is \textit{Herschel} 350$\mu$m, with a resolution of 25.2$^{\prime\prime}$ \citep{griffin2010}.

\subsection{Spatial filtering}

SCUBA-2 is restricted in the spatial scales to which it is sensitive due to the need to distinguish between astrophysical and atmospheric signal \citep[e.g.][]{chapin2013}.  SCUBA-2 is fundamentally insensitive to signal on scales larger than its array size (600$^{\prime\prime}$), but in practice the maximum size scale recovered is set by a combination of the mask used and the maximum size scale set in the reduction process, as discussed in Section~\ref{sec:obs}.  \textit{Herschel} images, having been taken above the atmosphere, are not subject to such constraints.  It is therefore necessary to match the spatial scales in the Herschel and SCUBA-2 observations before comparing the data sets.

We removed the large-scale structure from the \textit{Herschel} observations by passing them through the SCUBA-2 pipeline in the manner described by \citet{sadavoy2013}.  Similarly to the method for CO subtraction, the \textit{Herschel} data are added to the SCUBA-2 bolometer time series, and the reduction process is repeated, including the application of the mask.  In this case the \textit{Herschel} data are scaled to be a small positive perturbation on the SCUBA-2 signal in order to minimise the effect of the Herschel data on map convergence. The SCUBA-2 map is then subtracted from the \textit{Herschel}+SCUBA-2 map and the scaling applied to the \textit{Herschel} data is reversed, leaving the spatially-filtered \textit{Herschel} signal.

\subsection{SED fitting}

We used the $Starlink$ \textsc{Kappa} package \citep{currie2014} to convolve the 850$\mu$m data and all of the filtered \textit{Herschel} maps to 350$\mu$m resolution (25.2$^{\prime\prime}$), and to grid all of the maps to 16$^{\prime\prime}$ pixels.  We then fitted SEDs pixel-by-pixel using the relation
\begin{equation}
    F_{\nu} = \Sigma_{\rm dust}B_{\nu}(T)\kappa_{0}\left(\frac{\nu}{\nu_{0}}\right)^{\beta}
    \label{eq:hildebrand}
\end{equation}
\citep[cf.][]{hildebrand1983}, where $\Sigma_{\rm dust}$ is surface density of dust {in units of g\,cm$^{-2}$}, $B_{\nu}(T)$ is the Planck function at temperature $T$, dust opacity $\kappa_{0} = 1.92$\,cm$^{2}$\,g$^{-1}$ at a reference frequency $\nu_{0} = 0.857\,$THz (350$\mu$m) \citep{draine2003}, and $\beta$ is dust opacity index.

We restricted our fitting to pixels where all the flux density values have an SNR greater than 3.  We first fitted with $\Sigma_{\rm dust}$, $T$ and $\beta$ as free parameters.  We then repeated the fitting process, with $\beta$ fixed at its median value from the previous fit.  In both cases, we rejected any pixels where the fitted values of any of $\Sigma_{\rm dust}$, $T$ or $\beta$ have uncertainties of $> 50\%$.  This criterion only excluded a significant number of pixels in the $\beta$-free fit for NGC 3034.

We converted $\Sigma_{\rm dust}$ to molecular hydrogen column density ($N({\rm H}_{2})$) using the relation
\begin{equation}
    N({\rm H}_{2}) = 100\times\frac{\Sigma_{\rm dust}}{\mu m_{\textsc{h}}},
\end{equation}
where mean molecular mass is taken to be $\mu = 2.8$, $m_{\textsc{h}}$ is the mass of hydrogen and the gas-to-dust ratio is taken to be 100.  We note that the galaxies in our sample have metallicities similar to, but slightly less than, the solar value (cf. Table~\ref{tab:source_list}), and so these values of $N({\rm H}_{2})$ may be slightly underestimated.  {Moreover, the value of the dust-to-gas ratio may vary significantly within galaxies (e.g. \citealt{williams2018}), creating a further systematic uncertainty on $N({\rm H}_{2})$.}  The values of $N({\rm H}_{2})$ determined using this equation are not {yet} corrected for the effect of inclination angle.

We show maps of best-fit $N({\rm H}_2)$, $T$, $\beta$, their uncertainties and reduced $\chi^{2}$ values for each galaxy in Appendix~\ref{sec:appendix_seds}.  The mean, median maximum and minimum values for each of the fitted parameters are listed in Table~\ref{tab:sed_fits_free} for the $\beta$-free case, and in Table~\ref{tab:sed_fits_median} for the median-$\beta$ case.  {$N({\rm H}_2)$ values are at this stage shown without correction for inclination angle in order to show the fitted values.}  The median value of $\beta$ varies in the range $1.63-2.26$.  The mean and median values of the fitted parameters are similar, except in the case of $\Sigma_{dust}$ for NGC 3034 and, to a lesser extent, NGC 5194, where the mean value is significantly higher than the median.  This is due to the large dynamic range in the NGC 3034 and NGC 5194 observations.  The values of $\Sigma_{dust}$ and $T$ are similar in the $\beta$-free and median-$\beta$ cases, suggesting that the variation in $\beta$ within the galaxies is not very large.

\subsection{SCUBA-2 flux loss}

We compared the total flux density in the Herschel 250$\mu$m maps before and after they were passed through the SCUBA-2 pipeline in order to assess the likely amount of large-scale structure lost in the SCUBA-2 data reduction process.  The 250$\mu$m data were chosen as having the resolution most similar to the SCUBA-2 850$\mu$m data: 18.1$^{\prime\prime}$ \citep{griffin2010} compared to 14.1$^{\prime\prime}$ for SCUBA-2.  The total flux densities before and after filtering are listed in Table~\ref{tab:filtering}.  The mean fraction of flux lost in the filtering process is $0.58\pm 0.15$, and the median is 0.59.  However, this flux is not lost evenly across the map, with small-scale structures being recovered well, and extended, diffuse emission being lost.  The best-recovered galaxy is the bright and relatively compact NGC 3034, where only 32\% of the Herschel 250$\mu$m flux is lost.  The worst-recovered is NGC 3351, only the central region of which is visible in the SCUBA-2 image (see Figure~\ref{fig:ngc3351}).  In this case, 85\% of the Herschel flux density -- mostly associated with the galactic disc -- is lost.  Despite this, the compact and strongly-peaked central region is recovered quite well.

\begin{table}
    \centering
    \begin{tabular}{c c c c}
    \hline
    Galaxy & \multicolumn{2}{c}{250$\mu$m Flux Density} & Fraction lost \\ \cline{2-3}
     & Original & Filtered & \\ \cline{2-3}
     & \multicolumn{2}{c}{(Jy)} & \\
    \hline
    NGC 3034 & 451.0 & 305.8 & 0.32 \\  
    NGC 3351 & 27.0 & 5.5 & 0.85 \\
    NGC 3521 & 104.1 & 58.3 & 0.56 \\
    NGC 4254 & 69.1 & 27.3 & 0.56 \\
    NGC 4569 & 41.0 & 13.6 & 0.47 \\
    NGC 4736 & 59.1 & 22.5 & 0.65 \\
    NGC 5055 & 134.4 & 52.3 & 0.63 \\
    NGC 5194 & 183.5 & 82.2 & 0.61 \\
    \hline
    \end{tabular}
    \caption{Comparison of Herschel 250$\mu$m flux densities before and after filtering.}
    \label{tab:filtering}
\end{table}

\begin{table*}
    \centering
    \begin{tabular}{c c c c c}
    \hline
         & SED,$\beta$-free & SED,median-$\beta$ & 250,$\beta$-free & 250,median-$\beta$ \\
    \hline
    $N({\rm H}_{2})$ & Free parameter in SED fit & Free parameter in SED fit & From SPIRE $F_{\nu}(250\mu{\rm m})$ & From SPIRE $F_{\nu}(250\mu{\rm m})$ \\
    $T$ & Free parameter in SED fit & Free parameter in SED fit & From $\beta$-free SED fit & From $\beta$-median SED fit \\
    $\beta$ & Free parameter in SED fit & Fixed at median value from $\beta$-free fit & From $\beta$-free SED fit & Fixed at median value from $\beta$-free fit \\
    Filtered? & Yes & Yes & No & No \\
    \hline
    \end{tabular}
    \caption{{A summary of our four measures of $N({\rm H}_{2})$.}}
    \label{tab:models}
\end{table*}

We note again that we have optimised the SCUBA-2 DR process for image fidelity, rather than for large-scale structure recovery, as described in Section~\ref{sec:obs}.  An alternative data reduction scheme might be able to recover a {larger} fraction of the large-scale emission.  However, these results show that SCUBA-2 will inherently be insensitive to a significant fraction of the submillimetre flux density from well-resolved galaxies such as those in our sample. 

In subsequent analysis, in order to evaluate the effects of the SCUBA-2 flux loss, we {use measures of $\Sigma_{dust}$ and $N({\rm H}_{2})$ determined in two ways: (1) from SED fitting, and therefore subject to flux loss due to atmospheric filtering; and (2) from \textit{Herschel} SPIRE 250$\mu$m flux density using equation~\ref{eq:hildebrand}, using the $T$ and $\beta$ values from SED fitting, and therefore calculated from the full column of emission.  These measures of $\Sigma_{dust}$ and $N({\rm H}_{2})$ are summarised in Table~\ref{tab:models} and discussed further below.}

\begin{table*}
    \centering
    \begin{tabular}{@{\extracolsep{2pt}}c cccccccccccc@{}}
\hline
 & \multicolumn{12}{c}{$\beta$ free} \\ \cline{2-13}
 & \multicolumn{4}{c}{$\Sigma_{dust}$ ($\times 10^{-5}$ g\,cm$^{-2}$)} & \multicolumn{4}{c}{Temperature (K)} & \multicolumn{4}{c}{Dust Opacity Index $\beta$} \\ \cline{2-5} \cline{6-9} \cline{10-13}
Galaxy & Mean & Median & Min & Max & Mean & Median & Min & Max & Mean & Median & Min & Max \\ 
\hline
NGC 3034 & $38.0\pm 72.9$ & $10.4\pm5.4$ & 1.5 & 304 & $30.5\pm 4.4$ & $30.7\pm 2.8$ & 21.6 & 38.9 & $2.05\pm 0.45$ & $2.03\pm 0.30$ & 1.24 & 3.13 \\
NGC 3351 & $12.7\pm 3.4$ & $12.4\pm 2.7$ & 8.4 & 17.6 & $27.8\pm 0.9$ & $27.6 \pm 0.6$ & 26.6 & 29.1 & $2.01 \pm 0.17$ & $2.03\pm 0.15$ & 1.78 & 2.19 \\
NGC 3521 & $10.9\pm 4.9$ & $10.4\pm 3.6$ & 2.3 & 21.2 & $25.5\pm 1.9$ & $25.2 \pm 1.3$ & 22.0 & 29.3 & $1.81\pm 0.39$ & $1.86\pm 0.32$ & 1.07 & 2.43 \\
NGC 4254 & $6.2\pm 4.9$ & $4.2\pm 2.4$ & 1.1 & 20.9 & $25.1\pm 2.5$ & $25.1\pm 1.9$ & 19.7 & 32.0 & $2.02\pm 0.40$ & $1.98 \pm 0.30$ & 1.17 & 3.21 \\
NGC 4569 & $7.8\pm 3.3$ & $8.0\pm 2.5$ & 2.4 & 13.6 & $24.2\pm 2.5$ & $24.0\pm 1.8$ & 19.4 & 28.8 & $1.84\pm 0.30$ & $1.75\pm 0.12$ & 1.46 & 2.79 \\
NGC 4736 & $4.7\pm 1.8$ & $4.8\pm 1.3$ & 1.1 & 9.1 & $32.2\pm 2.6$ & $32.8\pm 1.5$ & 23.8 & 36.3 & $1.69\pm 0.23$ & $1.63\pm 0.09$ & 1.37 & 2.69 \\
NGC 5055 & $6.5\pm 5.6$ & $4.7\pm 2.2$ & 1.1 & 33.8 & $23.4\pm 2.4$ & $23.7\pm 1.3$ & 14.9 & 27.4 & $1.98\pm 0.56$ & $1.86\pm 0.28$ & 1.09 & 4.69 \\
NGC 5194 & $9.1\pm 10.5$ & $4.2\pm 2.3$ & 0.78 & 43.3 & $23.1\pm 2.4$ & $23.3\pm 1.7$ & 17.5 & 29.4 & $2.32\pm 0.56$ & $2.26\pm 0.32$ & 1.20 & 3.74 \\
\hline
    \end{tabular}
    \caption{A summary of the SED fitting results for the galaxies in our sample, with $\beta$ as a free parameter.}
    \label{tab:sed_fits_free}
\end{table*}

\begin{table*}
    \centering
    \begin{tabular}{@{\extracolsep{4pt}}c cccccccc c@{}}
\hline
 & \multicolumn{8}{c}{Median-$\beta$} \\ \cline{2-10}
 & \multicolumn{4}{c}{$\Sigma_{dust}$ ($\times 10^{-5}$ g\,cm$^{-2}$)} & \multicolumn{4}{c}{Temperature (K)} & $\beta$ \\ \cline{2-5} \cline{6-9} \cline{10-10}
Galaxy & Mean & Median & Min & Max & Mean & Median & Min & Max & \\ 
\hline
NGC 3034 & $35.7\pm 68.4$ & $10.6\pm 6.9$ & 1.14 & 293 & $30.7\pm3.3$ & $31.2\pm 2.5$ & 22.8 & 38.5 & 2.03\\
NGC 3351 & $12.6\pm 2.4$ & $12.7\pm 1.9$ & 9.1 & 15.9 & $27.6\pm 0.9$ & $27.5\pm 0.8$ & 26.7 & 28.8 & 2.03 \\
NGC 3521 & $10.5 \pm 4.6$ & $10.1\pm 3.1$ & 2.0 & 20.5 & $25.1 \pm 1.1$ & $25.2 \pm 0.9$ & 22.9 & 26.9 & 1.86 \\
NGC 4254 & $6.4\pm 4.7$ & $4.9\pm 2.8$ & 1.0 & 20.4 & $25.0 \pm 0.7$ & $25.0\pm 0.4$ & 23.8 & 27.2 & 1.98 \\
NGC 4569 & $7.7\pm 3.4$ & $8.5\pm 2.0$ & 1.7 & 14.8 & $24.6\pm 1.3$ & $24.5\pm 0.6$ & 22.2 & 27.1 & 1.75 \\
NGC 4736 & $4.5\pm 1.6$ & $4.6\pm 1.2$ & 1.1 & 7.5 & $32.9\pm 3.1$ & $32.3\pm 1.1$ & 27.9 & 41.4 & 1.63 \\
NGC 5055 & $6.7\pm 5.0$ & $5.6\pm 2.3$ & 1.2 & 27.9 & $23.8\pm 1.3$ & $23.9\pm 0.8$ & 20.3 & 26.4 & 1.86 \\
NGC 5194 & $6.5\pm 6.7$ & $3.9\pm 1.8$ & 0.83 & 33.3 & $23.5 \pm 2.3$ & $24.2\pm 1.0$ & 16.6 & 28.9 & 2.26 \\
\hline
    \end{tabular}
    \caption{A summary of the SED fitting results for the galaxies in our sample, with $\beta$ fixed at its median value from the free-parameter fit.}
    \label{tab:sed_fits_median}
\end{table*}

\section{Star Formation Law}
\label{sec:sf_law}

We measured surface density of star formation ($\Sigma_{\textsc{sfr}}$) for the galaxies in our sample, in order to investigate the relationship between $\Sigma_{\textsc{sfr}}$ and $\Sigma_{gas}$ both within and between the galaxies.

\subsection{Creating star formation surface density maps}
\label{sec:ssfr}

We measured the star formation rate in the galaxies in our sample using the method described by \citet{leroy2008}.  We measured unobscured star formation using GALEX far-ultraviolet (FUV) emission ($I_{\textsc{fuv}}$) and obscured star formation using \textit{Spitzer} Space Telescope 24$\mu$m emission ($I_{24}$).  The older stellar population was accounted for using \textit{Spitzer} 3.6$\mu$m emission ($I_{3.6}$).  All of these maps were taken from the Dustpedia database \citep{clark2018}.  NGC 3034 is saturated in the \textit{Spitzer} 24$\mu$m band, and so is excluded from further analysis.  {The \textit{Spitzer} and GALEX observations are shown in Figures~\ref{fig:mwl_ngc3034}--\ref{fig:mwl_ngc5194} in Appendix~\ref{sec:appendix_mwl}.}
We smoothed all of the data sets to $25\farcs2$ to match the resolution of the SED-fitted maps, and again gridded the maps to 16$^{\prime\prime}$ pixels.

Any pixels with significant GALEX near-ultraviolet (NUV) emission, i.e. where $I_{\textsc{nuv}}/I_{\textsc{fuv}} > 15$, were excluded from the analysis, in order to avoid contamination by foreground stars \citep{leroy2008}.
We then estimated the FUV flux density associated with star formation ($I_{\textsc{fuv,sf}}$) using the relation
\begin{equation}
    I_{\textsc{fuv,sf}} = I_{\textsc{fuv}} - \alpha_{\textsc{fuv}}I_{3.6},
\end{equation}
taking $\alpha_{\textsf{fuv}}=3\times10^{-3}$, and the 24$\mu$m flux density associated with star formation ($I_{24,\textsc{sf}}$) using the relation
\begin{equation}
    I_{24,\textsc{sf}} = I_{24} - \alpha_{24}I_{3.6},
\end{equation}
taking $\alpha_{24}=0.1$ \citep{leroy2008}.
We then estimated the surface density of star formation to be
\begin{equation}
    \Sigma_{\textsc{sfr}} = 8.1\times10^{-2}I_{\textsc{fuv,sf}} + 3.2\times10^{-3}I_{24,\textsc{sf}}.
    \label{eq:ssfr}
\end{equation}
By using these values, we have implicitly assumed a \citet{chabrier2003} Initial Mass Function (IMF) \citep{leroy2008}.

We assumed calibration uncertainties of 4.5\% on GALEX FUV, 5\% on \textit{Spitzer} 24$\mu$m and 3\% on \textit{Spitzer} 3.6$\mu$m measurements, as per \citet{clark2018}.  We assumed a 33\% uncertainty on the $I_{24,\textsc{sf}}$ term in equation~\ref{eq:ssfr}, following \citet{leroy2008} and \citet{ford2013}.

The final $\Sigma_{\textsc{sfr}}$ maps are shown in Appendix~\ref{sec:appendix_sfr}.  {These maps are expected to trace star formation on timescales $<100$\,Myr \citep{leroy2008}.  The mean age of the stellar population contributing to the unobscured star formation rate is $\sim 10$\,Myr, while the stellar population contributing to the obscured star formation rate has a mean age of $\sim 5$\,Myr \citep{kennicutt2012}. 90\% of the stellar population contributing to each measure is expected to have an age $<100$\,Myr \citep{kennicutt2012}.}

\begin{figure}
    \centering
    \includegraphics[width=0.47\textwidth]{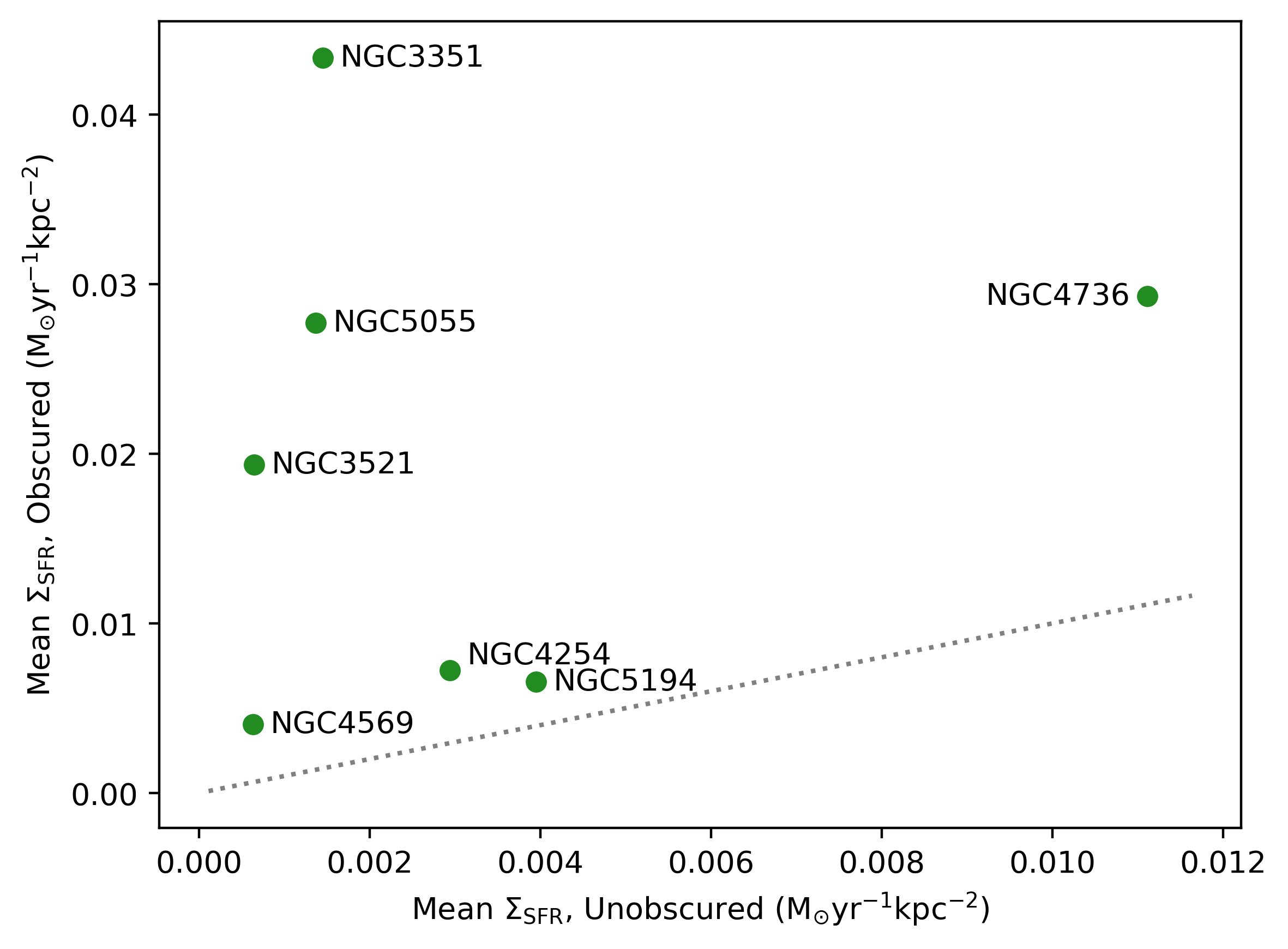}
    \caption{Mean contributions to $\Sigma_{\textsc{sfr}}$ from obscured (24$\mu$m-traced) and unobscured (FUV-traced) star formation for the galaxies in our sample.  Dotted grey line marks the 1:1 relationship.}
    \label{fig:ssfr_means}
\end{figure}

{We note that in every case, the surface density of obscured (24$\mu$m-traced) star formation is greater than that of unobscured (FUV-traced) star formation. The mean contributions from obscured and unobscured star formation to $\Sigma_{\textsc{sfr}}$ are shown in Figure~\ref{fig:ssfr_means}.  This suggests that the star formation being traced may be preferentially occurring on timescales as short as $\sim 5$\,Myr \citep{rieke2009,kennicutt2012}.}

\subsection{Resolved Kennicutt-Schmidt Law}

\setlength{\tabcolsep}{3pt}
\begin{table*}
    \centering
    \begin{tabular}{@{\extracolsep{3pt}}c ccc ccc ccc ccc@{}}
    \hline
       & \multicolumn{6}{c}{SED-fit} & \multicolumn{6}{c}{H250} \\ \cline{2-7} \cline{8-13}
       & \multicolumn{3}{c}{$\beta$ free} & \multicolumn{3}{c}{median $\beta$} & \multicolumn{3}{c}{$\beta$ free} & \multicolumn{3}{c}{median $\beta$} \\ \cline{2-4} \cline{5-7} \cline{8-10} \cline{11-13}
    Galaxy & $N$ & $C$ & $r^{2}$ & $N$ & $C$ & $r^{2}$ & $N$ & $C$ & $r^{2}$ & $N$ & $C$ & $r^{2}$ \\
    \hline
    NGC 3351 & 0.60$\pm$0.04 & $-$1.84$\pm$0.06 & 0.991 & 0.77$\pm$0.19 & $-$2.10$\pm$0.30 & 0.887 & 0.97$\pm$0.15 & $-$2.49$\pm$0.25 & 0.952 & 1.49$\pm$0.28 & $-$3.35$\pm$0.45 & 0.935 \\
    NGC 3521 & 0.75$\pm$0.02 & $-$2.59$\pm$0.03 & 0.958 & 0.71$\pm$0.05 & $-$2.54$\pm$0.07 & 0.794 & 1.06$\pm$0.04 & $-$3.22$\pm$0.06 & 0.947 & 1.10$\pm$0.07 & $-$3.31$\pm$0.12 & 0.810 \\
    NGC 4254 & 0.58$\pm$0.04 & $-$2.58$\pm$0.05 & 0.836 & 0.57$\pm$0.03 & $-$2.57$\pm$0.04 & 0.869 & 0.68$\pm$0.08 & $-$2.87$\pm$0.14 & 0.709 & 0.89$\pm$0.05 & $-$3.24$\pm$0.08 & 0.850 \\
    NGC 4569 & 1.20$\pm$0.26 & $-$3.76$\pm$0.27 & 0.508 & 1.18$\pm$0.21 & $-$3.72$\pm$0.22 & 0.605 & 1.51$\pm$0.47 & $-$4.26$\pm$0.56 & 0.358 & 1.45$\pm$0.29 & $-$4.16$\pm$0.34 & 0.542 \\
    NGC 4736 & 0.69$\pm$0.08 & $-$1.95$\pm$0.10 & 0.674 & 0.63$\pm$0.09 & $-$1.87$\pm$0.11 & 0.567 & 1.08$\pm$0.15 & $-$2.63$\pm$0.21 & 0.605 & 1.03$\pm$0.16 & $-$2.55$\pm$0.22 & 0.521 \\
    NGC 5055 & 0.38$\pm$0.01 & $-$2.07$\pm$0.02 & 0.881 & 0.41$\pm$0.02 & $-$2.10$\pm$0.02 & 0.855 & 0.62$\pm$0.02 & $-$2.53$\pm$0.03 & 0.892 & 0.63$\pm$0.02 & $-$2.54$\pm$0.03 & 0.861 \\
    NGC 5194 & 0.66$\pm$0.06 & $-$2.73$\pm$0.08 & 0.648 & 0.84$\pm$0.08 & $-$2.94$\pm$0.11 & 0.626 & 0.91$\pm$0.09 & $-$3.29$\pm$0.14 & 0.609 & 1.01$\pm$0.12 & $-$3.46$\pm$0.19 & 0.527 \\
    \hline
    \end{tabular}
    \caption{Resolved KS law fitting results for the galaxies in our sample.  The function $\log_{10}\Sigma_{\textsc{sfr}} = N\log_{10}\Sigma_{\rm gas}+C$ was fitted to the data shown in Figure~\ref{fig:SK}.  Best-fit values of $N$ and $C$ are listed, along with the $r^{2}$ correlation coefficient.  The best-fit models are shown in Figure~\ref{fig:SK}.}
    \label{tab:SK_results}
\end{table*}

For each of the galaxies in our sample, we converted $\Sigma_{\rm dust}$ to $\Sigma_{\rm H_{2}}$, again assuming a gas-to-dust ratio of 100.  We considered values of $\Sigma_{\rm dust}$ derived (1) from SED fitting of the SCUBA-2 and filtered Herschel data, (a) with $\beta$ as a free parameter {(suffixed `SED,$\beta$-free')} and (b) with $\beta$ fixed at the median value from the free-parameter fit {(suffixed `SED,median-$\beta$')}, and (2) from \textit{Herschel} 250$\mu$m flux density using equation~\ref{eq:hildebrand}, taking $T$ and $\beta$ from the SED-fitted values (a) with $\beta$ as a free parameter {(suffixed `250,$\beta$-free')} and (b) with $\beta$ fixed at its median value {(suffixed `250,median-$\beta$')}.  {These permutations are summarised in Table~\ref{tab:models}.}  We note that the applicability of these $T$ and $\beta$ values {derived from SED fitting to the filtered data} to the extended emission which is not seen by SCUBA-2 is not certain, particularly as \textit{Herschel} is likely to be tracing a slightly warmer dust population.  The values of $\Sigma_{\rm H_{2}}$ determined from the \textit{Herschel} 250$\mu$m data may broadly represent upper limits on the true values, while the SED-fitted values are likely to be underestimates.  Uncertainties on values of $\Sigma_{\rm H_{2}}$ determined using method (1) were derived from the uncertainties returned by the SED fitting process.
{For method (2), uncertainties on values of $\Sigma_{\rm H_{2}}$ were determined assuming that the uncertainty on the Herschel 250$\mu$m flux density is given by the 10\% \textit{Herschel} SPIRE calibration uncertainty \citep{griffin2010}, and propagating the uncertainties on $T$ and $\beta$ returned by the SED fitting process.  For consistency with method (1), any pixels with an uncertainty on $\Sigma_{\rm H_{2}}$ of $> 50\%$ were rejected.  Our assumption of a constant dust-to-gas ratio adds a further source of uncertainty which is difficult to quantify, but we note that within any given galaxy, this systematic uncertainty ought to affect the values of $\Sigma_{\rm H_{2}}$ returned by methods (1) and (2) consistently.}

\begin{figure*}
    \centering
    \includegraphics[width=0.85\textwidth]{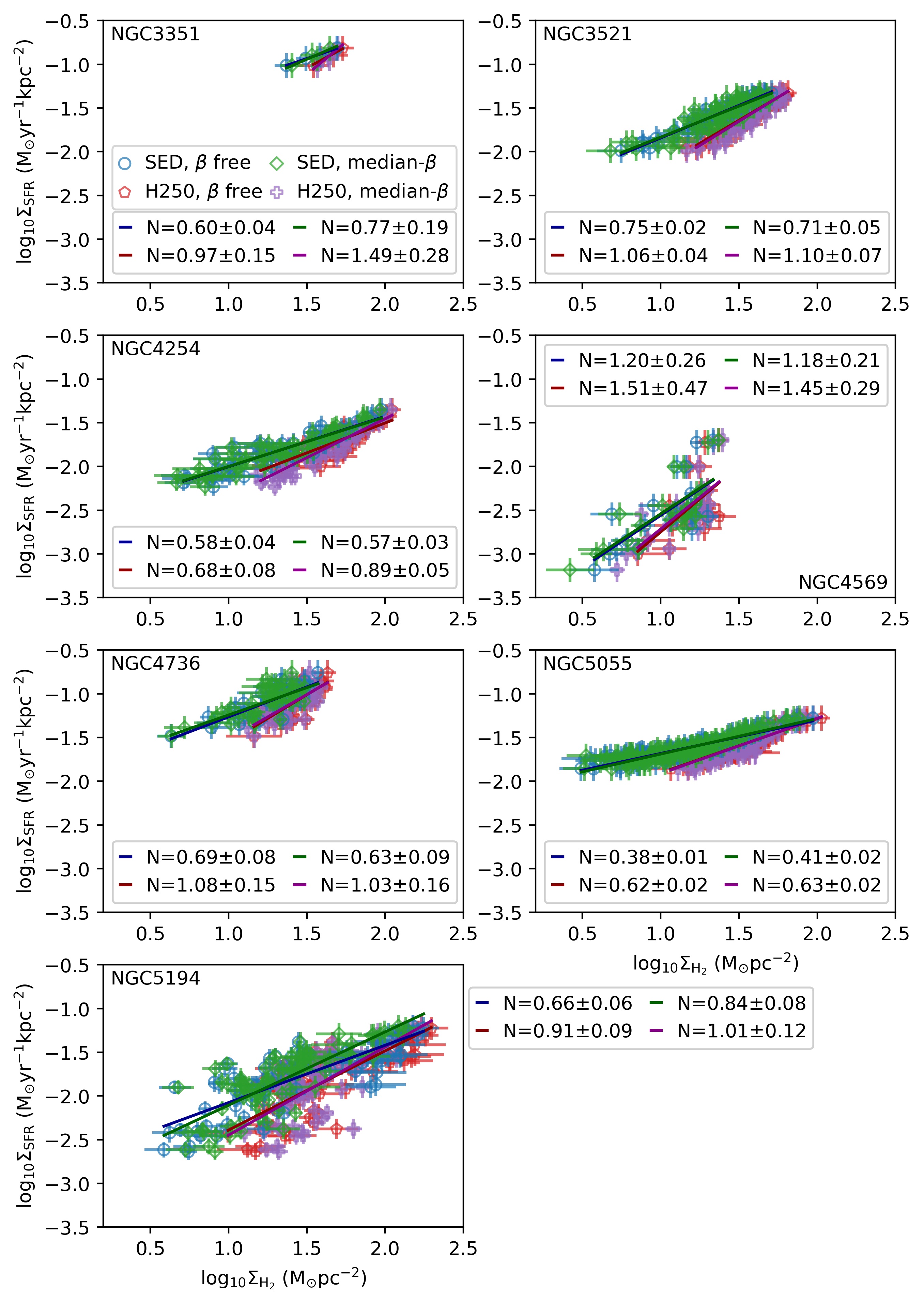}
    \caption{Resolved star formation law plots for the galaxies in our sample.  Blue circles mark values of $\Sigma_{\rm H_{2}}$ derived from SED-fitted values of $\Sigma_{dust}$ with $\beta$ as a free parameter.  Green diamonds; as blue circles but with $\beta$ fixed at its median value.  Red pentagons mark values of $\Sigma_{\rm H_{2}}$ derived from 250$\mu$m dust emission, using values of $T$ and $\beta$ from SED fitting with $\beta$ as a free parameter.  Purple pluses; as red pentagons but with $\beta$ fixed at its median value.  In each case the data points are fitted with the function $\log_{10}\Sigma_{\textsc{sfr}} = N\log_{10}\Sigma_{\rm gas}+C$; best-fit lines are plotted in each panel.  Best-fit relationships and correlation coefficients are listed in Table~\ref{tab:SK_results}.  All values are corrected for inclination.}
    \label{fig:SK}
\end{figure*}

In order to directly compare between the galaxies in our sample, we corrected both $\Sigma_{\rm H_{2}}$ and $\Sigma_{\textsc{sfr}}$ for inclination angle.  Each quantity was multiplied by a factor $\cos i$, where $i$ is the inclination of the galaxy, as listed in Table~\ref{tab:source_list}.  Plots of $\Sigma_{\textsc{sfr}}$ against $\Sigma_{\rm H_{2}}$ for the galaxies in our sample are shown in Figure~\ref{fig:SK}.

We fitted the function $\log_{10}\Sigma_{\textsc{sfr}} = N\log_{10}\Sigma_{\rm H_{2}}+C$ to the data shown in Figure~\ref{fig:SK}.  The results of this fitting are shown in Table~\ref{tab:SK_results}, and the best-fit lines are plotted on Figure~\ref{fig:SK}.
 
Figure~\ref{fig:SK} shows that there is a strong correlation between dust-derived surface mass of H$_{2}$ and surface density of star formation for all of the galaxies in our sample, for both the SED-fit and the 250$\mu$m-derived values of $\Sigma_{\rm H_{2}}$.  Allowing $\beta$ to vary or fixing it at its median value {typically} results in little difference in the best-fit indices.  All of the best-fit indices are smaller than the standard KS value of $N=1.4$, with the exceptions of the 250$\mu$m median-$\beta$ fit for NGC 3351 (in which only 4 data points are fitted), and {both 250$\mu$m fits for} NGC 4569, in which $\Sigma_{\rm H_{2}}$ and $\Sigma_{\textsc{sfr}}$ are less well-correlated than is the case in the other galaxies in our sample.  {This is not a surprising result, as our submillimetre dust emission observations are tracing molecular gas surface density, rather than total gas surface density, and so a linear relationship, implying that molecular gas is converted into stars with a constant efficiency, is typically expected \citep[e.g.][]{bigiel2008,bolatto2017}.}

{In every case}, the SED-fit values of $\Sigma_{\rm H_{2}}$ produce significantly shallower values of $N$ than do the 250$\mu$m-derived values.  For most of the galaxies, the SED-fit values of $N$ are somewhat sublinear, while the 250$\mu$m-derived values are broadly consistent with linearity.  The exceptions are NGC 4569, in which the fitted indices are {super-linear}, but in which the correlation coefficients are relatively poor, and the well-correlated NGC 5055, in which the fitted indices are significantly sub-linear in both cases.  There are significant offsets in $\Sigma_{\textsc{sfr}}$ between the galaxies in our sample.  {This is in keeping with the $\pm 0.30$ dex rms dispersion which is seen around the KS law \citep{kennicutt2012}, which {they note} is greater than can be attributed to measurement uncertainties, and so which likely indicates physical differences in star formation rate between galaxies of comparable surface density \citep[e.g.][]{ellison2021}.  {As shown in Figure~\ref{fig:SK}, the uncertainties on our values of $\Sigma_{\textsc{sfr}}$ are typically smaller than those on $\Sigma_{\rm H_{2}}$, and not subject to the uncertainty on gas-to-dust ratio which affect the $\Sigma_{\rm H_{2}}$ measurements.}

\subsection{Unresolved Kennicutt-Schmidt Law}

\begin{figure}
    \centering
    \includegraphics[width=0.47\textwidth]{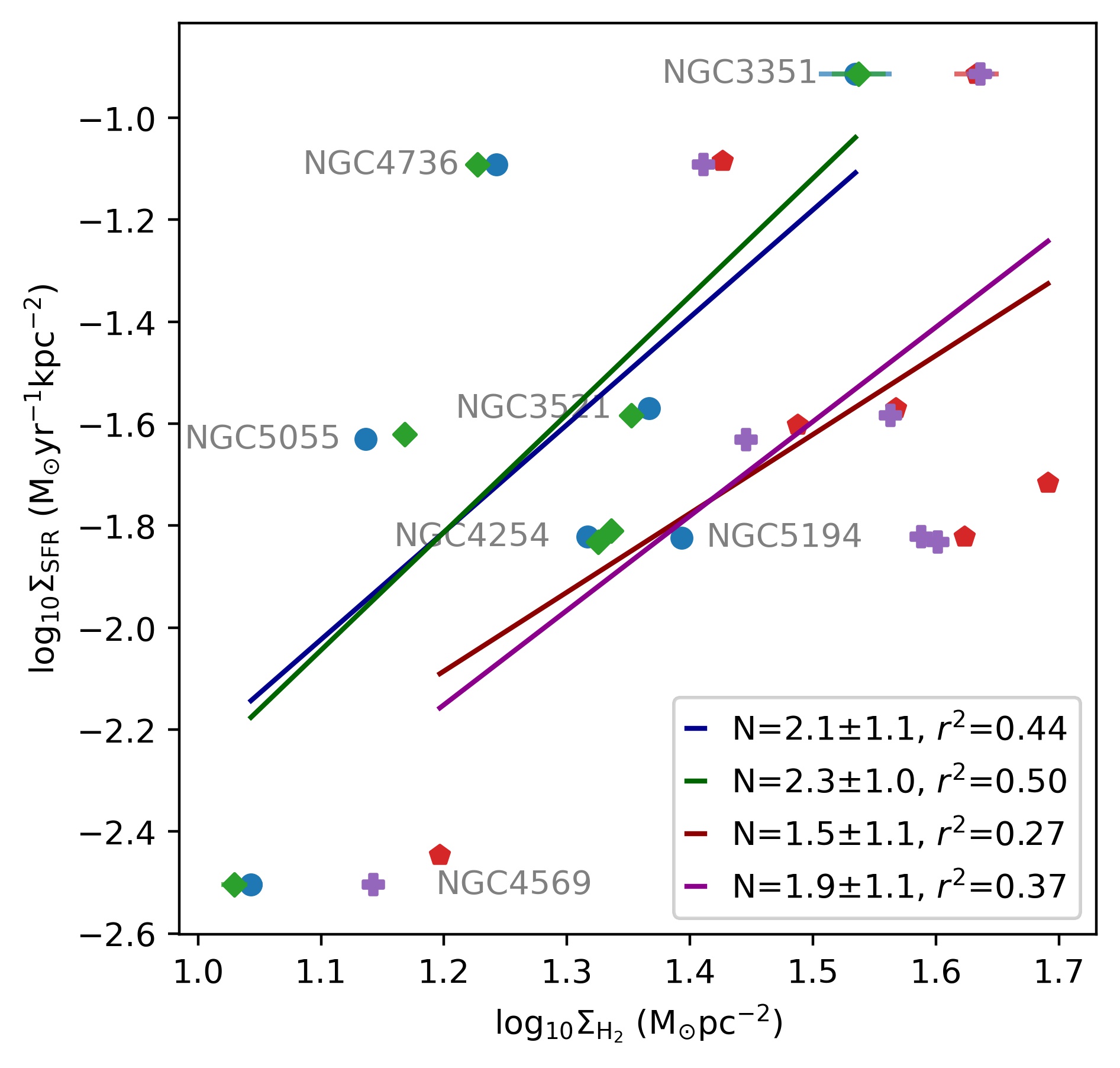}
    \caption{Mean values of $\Sigma_{\rm H_{2}}$ and $\Sigma_{\textsc{sfr}}$ for the galaxies in our sample.  Lines of best-fit are plotted.  Symbols and colour coding as in Figure~\ref{fig:SK}.}
    \label{fig:SK_ures}
\end{figure}

We measured the relationship between the global values of $\Sigma_{\rm H_{2}}$ and $\Sigma_{\textsc{sfr}}$ for the galaxies in our sample.  To do so, we took the mean value of $\Sigma_{\rm H_{2}}$ and $\Sigma_{\textsc{sfr}}$ for each galaxy.  {We  are thus to some extent measuring luminosity-weighted values, although our use of SED-fitted values of $T$ and $\beta$ in our measurements of $\Sigma_{\rm H_{2}}$ should mitigate against this effect.  This is intended to be analogous to observing the galaxies as unresolved sources.  However, the SED-fit values of $\Sigma_{\rm H_{2}}$ remain subject to loss of extended emission: if we were truly observing unresolved sources with SCUBA-2, the full column of submillimetre emission would be accounted for.}  The mean values {that we measure} are shown in Figure~\ref{fig:SK_ures}.  We see a positively correlated and broadly linear relationship between $\log_{10}(\Sigma_{\rm H_{2}})$ and $\log_{10}(\Sigma_{\textsc{sfr}})$, albeit with considerable scatter, for all four of the methods of measuring $\Sigma_{\rm H_{2}}$ which we consider.  The mean values of $\Sigma_{\rm H_{2}}$ are consistently lower for the SED-fit cases than for the 250$\mu$m-derived cases, as would be expected due to the loss of large-scale emission in the SCUBA-2 data.  Allowing $\beta$ to vary or fixing it at its median value produces little difference in the mean values of $\Sigma_{\rm H_{2}}$.  The best-fit indices are $N_{\rm SED,\beta-free}=2.1\pm 1.1$, $N_{\rm SED,median-\beta}=2.3\pm 1.0$, $N_{\rm 250\mu m,\beta-free}=1.5\pm 1.1$, and $N_{\rm 250\mu m,median-\beta}=1.9\pm 1.1$.  These values are steeper than those seen within the galaxies, and are consistent both with one another and with the KS value of $N=1.4$ within their large uncertainties.  The best-fit relationships are plotted on Figure~\ref{fig:SK_ures}.

\section{Discussion}
\label{sec:discuss}

\begin{figure}
    \centering
    \includegraphics[width=0.47\textwidth]{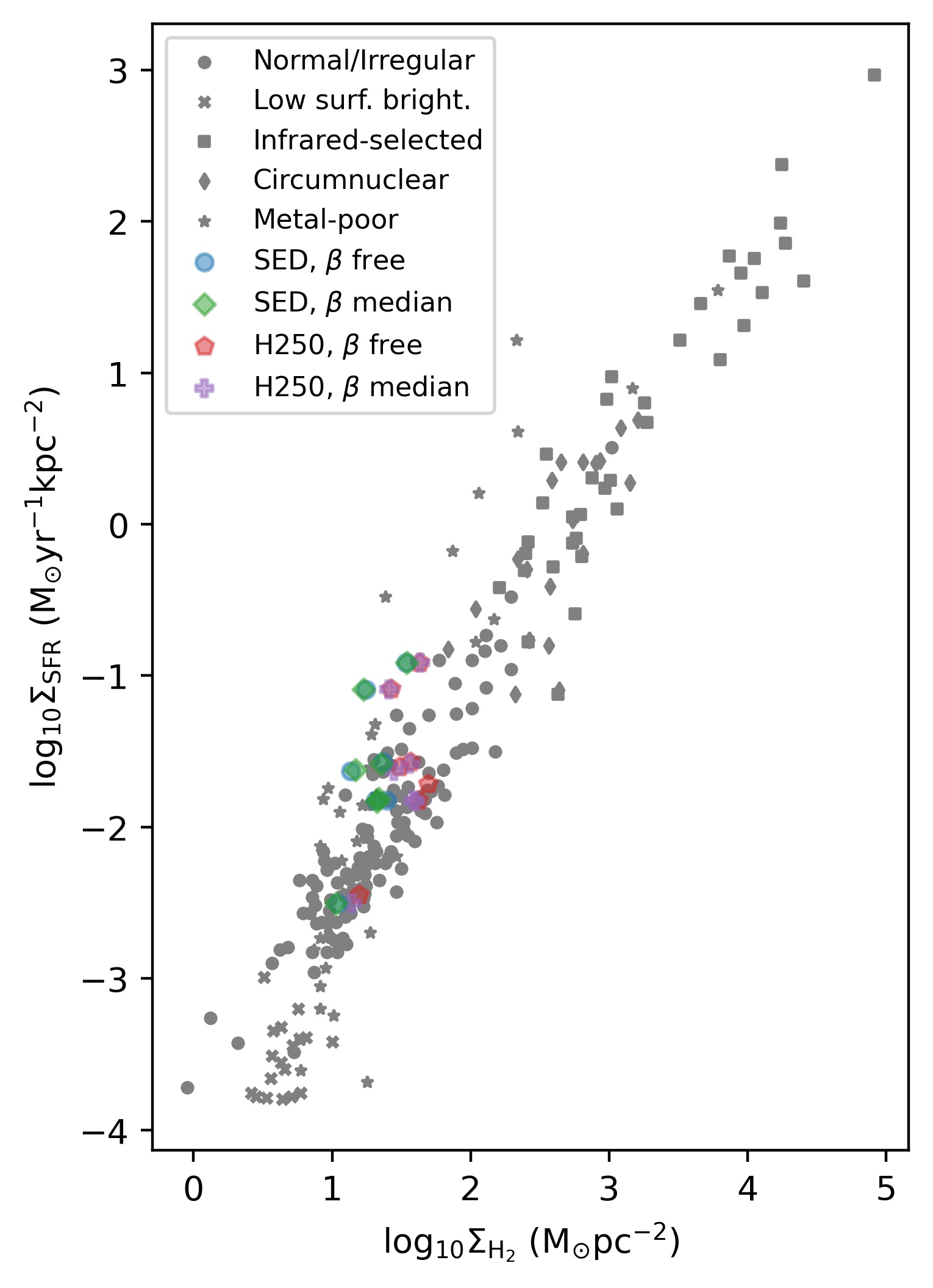}
    \caption{The mean values of $\Sigma_{\rm H_{2}}$ and $\Sigma_{\textsc{sfr}}$ for the galaxies in our sample, compared to archival measurements.  The grey circles mark normal or irregular galaxies presented by \citet{gavazzi2003}, \citet{james2004}, \citet{hameed2005} and \citet{kennicutt2008}.  Grey crosses mark the low-surface-brightness subset of these galaxies.  Grey squares mark infrared-selected starburst galaxies presented by \citet{scoville2000} and \citet{dopita2002}. Grey diamonds mark the circumnuclear starbursts presented by \citet{kormendy2004}.  Grey stars mark any low-metallicity galaxies in the above samples, i.e., those where $Z < 0.3\,Z_{\odot}$.  }
    \label{fig:SK_compare}
\end{figure}

\subsection{Resolved KS Law}

The fact that the values of $N$ for the SED-fit cases are consistently shallower than the 250$\mu$m-derived values indicates that the atmospheric filtering to which SCUBA-2 is subject results in the loss of a higher fraction of emission in low-surface-brightness regions.  This is as expected, as it is extended emission that is preferentially lost in the filtering process, as discussed in Section~\ref{sec:obs}.  Figure~\ref{fig:SK} shows good agreement between the SED-fit and 250$\mu$m-derived values of $\Sigma_{\rm H_{2}}$ in the highest-density regions.

If the sub-linear values of $N$ recovered from the SED-fit cases were physical, it would suggest that star formation becomes less efficient at high gas densities \citep[e.g.][]{ford2013}.  It appears likely that the shallow indices seen in the SED-fit data typically result from spatially-filtered dust emission data being compared to $\Sigma_{\textsc{sfr}}$ maps derived from GALEX and \textit{Spitzer} data, the extended components of which have not been correspondingly lost to atmospheric filtering.  {Nonetheless, the} SED-fit data is well-correlated with $\Sigma_{\textsc{sfr}}$, indicating that the SCUBA-2 850$\mu$m emission is a good tracer of star formation despite missing some fraction of the extended dust emission. 

The $\Sigma_{\textsc{sfr}}$ maps constructed from GALEX and \textit{Spitzer} data trace star formation on timescales {$\sim 5-100$\,Myr} \citep[][and refs. therein]{leroy2008,kennicutt2012}.  Therefore the 250$\mu$m-derived relationship between $\Sigma_{\rm H_{2}}$ and $\Sigma_{\textsc{sfr}}$ is more accurate than the SED-fit relationship only if the gas traced by \textit{Herschel} but not by SCUBA-2 will {arises from molecular clouds which will form stars} in the next {$\sim 5-100$\,Myr}.  Observations of local star-forming regions have shown that in resolved observations of molecular clouds SCUBA-2 effectively selects for dense and gravitationally bound material which is likely to form stars in the immediate future \citep{wardthompson2016}.  While each SCUBA-2 beam {in our observations} will integrate over many star-forming molecular cloud complexes, the strong correlation between the SED-fit {$\Sigma_{\rm H_{2}}$ values} and $\Sigma_{\textsc{sfr}}$ {nonetheless} suggests that {a gas density peak sufficiently strong to be traceable by SCUBA-2 may be required for significant amounts of star formation to occur.}

Most of the galaxies in our sample are, in the 250$\mu$m-derived case, broadly consistent with $N\sim 1$, suggesting a linear relationship between molecular gas mass and star formation rate {\citep[e.g.][]{bigiel2008,lada2012}}.  NGC 3351 and NGC 4569 show somewhat steeper relationships, but the former includes only four pixels in the galactic centre, and the latter is relatively weakly correlated.  Uniquely in our sample, NGC 5055 shows an index which is significantly $<1$ in both the SED-fit and 250$\mu$m-derived cases, and has a very high correlation coefficient between $\Sigma_{\rm H_{2}}$ and $\Sigma_{\textsc{sfr}}$.

The indices which we derive from dust emission for the galaxies in our sample are somewhat similar to those derived from dust emission by \citet{ford2013} for M31.  \citet{ford2013} found an index of $0.55\pm0.01$ in M31 from \textit{Herschel} dust emission, quite similar to the values which we typically measure in the SED-fit case, but shallower than the indices which we typically measure using 250$\mu$m-derived values. However, the \citet{ford2013} value for M31 is similar to our 250$\mu$m-derived values for NGC 5055, $0.62\pm0.02$ and $0.63\pm 0.02$, perhaps suggesting a commonality between the two galaxies.  However, none of our galaxies have indices similar to that measured on kpc scales in M33 from dust emission by \citep{williams2018}, of $5.85 \pm 2.37$.

The broadly linear relationship between 250$\mu$m-derived $\Sigma_{\rm H_{2}}$ and $\Sigma_{\textsc{sfr}}$ suggests that the {molecular clouds or cloud complexes traced by \textit{Herschel} will go on to form stars on timescales $\sim 5-100$\,Myr.  The star formation efficiency per freefall time of a molecular cloud is very low ($\sim 1$\%; e.g., \citealt{mckee2007}), and so we expect only a very small fraction of the material traced by these observations to be incorporated into new stars on this timescale.}  However, the {equally} good, {albeit sub-linear,} correlation between the SED-fit $\Sigma_{\rm H_{2}}$ and $\Sigma_{\textsc{sfr}}$ suggests that SCUBA-2 is also a good tracer of star formation.

{The fact that 250$\mu$m-derived $\Sigma_{\rm H_{2}}$ and SED-fit $\Sigma_{\rm H_{2}}$ are equally well-correlated with $\Sigma_{\textsc{sfr}}$ suggests that the relatively diffuse material traced by \textit{Herschel} but not by SCUBA-2 is (a) not forming stars independently of the material traced by SCUBA-2 and (b) associated with the same star-forming peaks in column density which are detected by SCUBA-2.}  We suggest that SCUBA-2 is preferentially detecting {the} denser material {within these molecular clouds}.  We hypothesise that this denser material may be likely to form stars on a timescale shorter than that of the material exclusively traced by \textit{Herschel}.

\subsection{Unresolved KS law}

Figure~\ref{fig:SK_ures} shows that the best-fit index for the star formation law over the mean values $\langle\Sigma_{\rm H_{2}}\rangle$ and $\langle\Sigma_{\textsc{sfr}}\rangle$ of our sample of galaxies is steeper, and more comparable to the standard KS value of $N=1.4$, than are those in the individual galaxies shown in Figure~\ref{fig:SK}, {although the uncertainties on the fits are sufficiently large that the comparison may not be meaningful}.  {The fitted values are also broadly consistent with linearity, as would be expected for tracers of molecular gas \citep{gao2004}.}  The presence or absence of the full column of dust emission changes only the offset of the slope and not the slope itself, again to within the very large uncertainties on the fit.  This again suggests that the presence of a SCUBA-2 detection is a good predictor of star formation.  {We emphasise that the loss of the contribution from extended emission in the SCUBA-2 data affects these results because we are averaging over observations of resolved sources; if we were observing more distant unresolved galaxies, the full column of submillimetre emission would be accounted for.}

We note that in this paper we have assumed a single gas-to-dust ratio and mean molecular mass for all of the galaxies in our sample.  Some amount of the scatter in $\langle \Sigma_{\rm H_{2}}\rangle$ seen in Figure~\ref{fig:SK_ures} could thus be caused by systematic errors in the conversion from dust to gas mass.  However, the variations in metallicity in our sample are not large (see Table~\ref{tab:source_list}), suggesting that not all of the observed scatter is due to errors in in this conversion.

We compared our results to archival measurements of a range of galaxies in Figure~\ref{fig:SK_compare}.  Our galaxies correspond well to the normal/irregular galaxies presented by \citet{gavazzi2003}, \citet{james2004}, \citet{hameed2005} and \citet{kennicutt2008}.  This is as expected, as all of the galaxies for which we have calculated $\Sigma_{\textsc{sfr}}$ are non-starbust spirals.  The SED-fit values -- in which the extended component of the SFR tracers is included but that of {the} dust emission is not -- correspond better to the metal-poor archival galaxies ($Z/Z_{\odot} < 0.3$) than to the `normal' sample.  Although the galaxies which we consider have sub-solar metallicities, the lowest-metallicity galaxy, NGC 4254, has $Z/Z_{\odot}=0.72$ \citep{devis2019}.  Using the 250$\mu$m-derived data puts the galaxies in our sample in the midst of the normal galaxies.  This is as expected because, as discussed above, if dust continuum emission is lost to atmospheric filtering, the dust mass derived for the galaxy will be systematically lowered, and the galaxy will artificially appear metal-poor.  {Feathering of JCMT data with lower-resolution \textit{Planck} or \textit{Herschel} data \citep{smith2021} could aid in direct comparison of the cold dust and star formation tracers.}

\section{Summary}
\label{sec:summary}

In this paper we have presented 850$\mu$m dust continuum observations of 8 nearby galaxies made with the SCUBA-2 camera on the James Clerk Maxwell Telescope (JCMT) as part of the JCMT Nearby Galaxies Legacy Survey (NGLS).  The galaxies which we present are NGC 3034, NGC 3351, NGC 3521, NGC 4254, NGC 4569, NGC 4736, NGC 5055 and NGC 5194.  {These galaxies were selected for their high surface brightness and the presence of ancillary data, and are not a representative sample of local galaxies}.

We find that there is a significant contribution from the $^{12}$CO $J=3\to 2$ line in the SCUBA-2 850$\mu$m observations of all of the galaxies in our sample, typically $\sim 20$\%, but higher for NGC 3034 and NGC 5194.  {We corrected the SCUBA-2 maps for this CO contamination} using NGLS HARP $^{12}$CO $J=3\to 2$ observations.

Comparison of our observations to VLA H\textsc{i} and Nobeyama 45m $^{12}$CO $J=1\to 0$ observations shows that SCUBA-2 {850$\mu$m} emission is correlated with $^{12}$CO emission, suggesting that the dust emission detected by SCUBA-2 is tracing molecular hydrogen gas.  SCUBA-2 {850$\mu$m} emission is not well-correlated with atomic hydrogen {emission}.

We fitted spectral energy distributions to each of the galaxies in our sample in order to measure surface dust mass, dust temperature and dust opacity index, using our SCUBA-2 850$\mu$m data and archival \textit{Herschel} Space Observatory observations.  To do this, we filtered the \textit{Herschel} data to match the spatial scales present in the SCUBA-2 data.  For our chosen SCUBA-2 data reduction scheme, a large fraction of the Herschel emission {(a mean of 58\%)} was lost, principally the extended, low-surface-brightness component.

We constructed resolved and unresolved star formation law plots for 7 of the galaxies in our sample, using archival \textit{Spitzer} and GALEX data to measure surface density of star formation.  

In the resolved case, we found that comparing surface density of star formation rate ($\Sigma_{\textsc{sfr}}$) to SED-fit-derived (i.e. subject to atmospheric filtering) values of H$_{2}$ surface density ($\Sigma_{\rm H_{2}}$) typically produces sublinear star formation law indices, while comparing to \textit{Herschel} 250$\mu$m-derived values typically produces indices which are broadly linear.  The exceptions to this are the poorly-fitted NGC 4569, which is significantly superlinear in both cases, and the well-fitted NGC 5055, which is significantly sublinear in both cases.  The \textit{Herschel} 250$\mu$m-derived star formation law index for NGC 5055 is similar to that found in M31 by \citet{ford2013}, suggesting a commonality between the two galaxies.

In the unresolved case, we found that comparing the mean values of $\Sigma_{\textsc{sfr}}$ and $\Sigma_{\rm H_{2}}$ of the galaxies in our sample returns {star formation law indices which are} broadly consistent with {both} the Kennicutt-Schmidt value of 1.4 {and linearity}, {within the large error bars on {the best-fit indices}. The loss of large-scale emission in the SCUBA-2 data changes the offset, but not the measured index, of the star formation law measured across the galaxies in our sample.  The galaxies which we consider have mean $\Sigma_{\textsc{sfr}}$ and $\Sigma_{\rm H_{2}}$ {values} consistent with their being `normal' spiral galaxies, when compared to archival measurements.

We find that SCUBA-2 emission is very well-correlated with star formation, but that SCUBA-2 cannot capture the extended dust emission component of the galaxies in our sample.  We suggest that \textit{Herschel} emission traces material {in molecular clouds which will form stars} on {timescales comparable} to the star formation timescale traced by GALEX and \textit{Spitzer} data, while SCUBA-2 preferentially traces {the densest gas within these clouds}, {which we hypothesise may} form stars on a shorter timescale.

\section*{Acknowledgements}

K.P. is a Royal Society University Research Fellow, supported by grant number URF\textbackslash R1\textbackslash 211322.  The research of C.D.W. is supported by grants from the Natural Sciences and Engineering Research Council of Canada and the Canada Research Chairs program.  The James Clerk Maxwell Telescope is operated by the East Asian Observatory on behalf of The National Astronomical Observatory of Japan; Academia Sinica Institute of Astronomy and Astrophysics; the Korea Astronomy and Space Science Institute; Center for Astronomical Mega-Science (as well as the National Key R\&D Program of China with No. 2017YFA0402700).  Additional funding support is provided by the Science and Technology Facilities Council (STFC) of the United Kingdom (UK) and participating universities in the UK, Canada and Ireland.  The JCMT has historically been operated by the Joint Astronomy Centre on behalf of the STFC of the UK, the National Research Council of Canada and the Netherlands Organisation for Scientific Research.  Additional funds for the construction of SCUBA-2 were provided by the Canada Foundation for Innovation. 
The authors wish to recognize and acknowledge the very significant cultural role and reverence that the summit of Maunakea has always had within the indigenous Hawaiian community. We are most fortunate to have the opportunity to conduct observations from this mountain.

\section*{Data Availability}

The raw SCUBA-2 data used in this paper are available in the JCMT archive at the Canadian Astronomy Data Centre under project code MJLSN07.  The reduced SCUBA-2 data presented in this paper are available at \url{https://dx.doi.org/10.11570/23.0007}.



\typeout{}
\bibliographystyle{mnras}

\begin{thebibliography}{}
\makeatletter
\relax
\def\mn@urlcharsother{\let\do\@makeother \do\$\do\&\do\#\do\^\do\_\do\%\do\~}
\def\mn@doi{\begingroup\mn@urlcharsother \@ifnextchar [ {\mn@doi@}
  {\mn@doi@[]}}
\def\mn@doi@[#1]#2{\def\@tempa{#1}\ifx\@tempa\@empty \href
  {http://dx.doi.org/#2} {doi:#2}\else \href {http://dx.doi.org/#2} {#1}\fi
  \endgroup}
\def\mn@eprint#1#2{\mn@eprint@#1:#2::\@nil}
\def\mn@eprint@arXiv#1{\href {http://arxiv.org/abs/#1} {{\tt arXiv:#1}}}
\def\mn@eprint@dblp#1{\href {http://dblp.uni-trier.de/rec/bibtex/#1.xml}
  {dblp:#1}}
\def\mn@eprint@#1:#2:#3:#4\@nil{\def\@tempa {#1}\def\@tempb {#2}\def\@tempc
  {#3}\ifx \@tempc \@empty \let \@tempc \@tempb \let \@tempb \@tempa \fi \ifx
  \@tempb \@empty \def\@tempb {arXiv}\fi \@ifundefined
  {mn@eprint@\@tempb}{\@tempb:\@tempc}{\expandafter \expandafter \csname
  mn@eprint@\@tempb\endcsname \expandafter{\@tempc}}}

\bibitem[\protect\citeauthoryear{{Alloin} \& {Nieto}}{{Alloin} \&
  {Nieto}}{1982}]{alloin1982}
{Alloin} D.,  {Nieto} J.~L.,  1982, \aaps, \href
  {https://ui.adsabs.harvard.edu/abs/1982A&AS...50..491A} {50, 491}

\bibitem[\protect\citeauthoryear{{Asplund}, {Grevesse}, {Sauval}  \&
  {Scott}}{{Asplund} et~al.}{2009}]{asplund2009}
{Asplund} M.,  {Grevesse} N.,  {Sauval} A.~J.,   {Scott} P.,  2009, \mn@doi
  [\araa] {10.1146/annurev.astro.46.060407.145222}, \href
  {https://ui.adsabs.harvard.edu/abs/2009ARA&A..47..481A} {47, 481}

\bibitem[\protect\citeauthoryear{{Battaglia}, {Fraternali}, {Oosterloo}  \&
  {Sancisi}}{{Battaglia} et~al.}{2006}]{battaglia2006}
{Battaglia} G.,  {Fraternali} F.,  {Oosterloo} T.,   {Sancisi} R.,  2006,
  \mn@doi [\aap] {10.1051/0004-6361:20053210}, \href
  {https://ui.adsabs.harvard.edu/abs/2006A&A...447...49B} {447, 49}

\bibitem[\protect\citeauthoryear{{Bergin} \& {Tafalla}}{{Bergin} \&
  {Tafalla}}{2007}]{bergin2007}
{Bergin} E.~A.,  {Tafalla} M.,  2007, \mn@doi [\araa]
  {10.1146/annurev.astro.45.071206.100404}, \href
  {https://ui.adsabs.harvard.edu/abs/2007ARA&A..45..339B} {45, 339}

\bibitem[\protect\citeauthoryear{{Bigiel}, {Leroy}, {Walter}, {Brinks}, {de
  Blok}, {Madore}  \& {Thornley}}{{Bigiel} et~al.}{2008}]{bigiel2008}
{Bigiel} F.,  {Leroy} A.,  {Walter} F.,  {Brinks} E.,  {de Blok} W.~J.~G.,
  {Madore} B.,   {Thornley} M.~D.,  2008, \mn@doi [\aj]
  {10.1088/0004-6256/136/6/2846}, \href
  {https://ui.adsabs.harvard.edu/abs/2008AJ....136.2846B} {136, 2846}

\bibitem[\protect\citeauthoryear{{Bolatto}, {Wolfire}  \& {Leroy}}{{Bolatto}
  et~al.}{2013}]{bolatto2013}
{Bolatto} A.~D.,  {Wolfire} M.,   {Leroy} A.~K.,  2013, \mn@doi [\araa]
  {10.1146/annurev-astro-082812-140944}, \href
  {https://ui.adsabs.harvard.edu/abs/2013ARA&A..51..207B} {51, 207}

\bibitem[\protect\citeauthoryear{{Bolatto} et~al.,}{{Bolatto}
  et~al.}{2017}]{bolatto2017}
{Bolatto} A.~D.,  et~al., 2017, \mn@doi [\apj] {10.3847/1538-4357/aa86aa},
  \href {https://ui.adsabs.harvard.edu/abs/2017ApJ...846..159B} {846, 159}

\bibitem[\protect\citeauthoryear{{Boselli} et~al.,}{{Boselli}
  et~al.}{2016}]{boselli2016}
{Boselli} A.,  et~al., 2016, \mn@doi [\aap] {10.1051/0004-6361/201527795},
  \href {https://ui.adsabs.harvard.edu/abs/2016A&A...587A..68B} {587, A68}

\bibitem[\protect\citeauthoryear{{Buckle} et~al.,}{{Buckle}
  et~al.}{2009}]{buckle2009}
{Buckle} J.~V.,  et~al., 2009, \mn@doi [\mnras]
  {10.1111/j.1365-2966.2009.15347.x}, \href
  {https://ui.adsabs.harvard.edu/abs/2009MNRAS.399.1026B} {399, 1026}

\bibitem[\protect\citeauthoryear{{Chabrier}}{{Chabrier}}{2003}]{chabrier2003}
{Chabrier} G.,  2003, \mn@doi [\pasp] {10.1086/376392}, \href
  {https://ui.adsabs.harvard.edu/abs/2003PASP..115..763C} {115, 763}

\bibitem[\protect\citeauthoryear{{Chapin}, {Berry}, {Gibb}, {Jenness}, {Scott},
  {Tilanus}, {Economou}  \& {Holland}}{{Chapin} et~al.}{2013}]{chapin2013}
{Chapin} E.~L.,  {Berry} D.~S.,  {Gibb} A.~G.,  {Jenness} T.,  {Scott} D.,
  {Tilanus} R. P.~J.,  {Economou} F.,   {Holland} W.~S.,  2013, \mn@doi
  [\mnras] {10.1093/mnras/stt052}, \href
  {https://ui.adsabs.harvard.edu/abs/2013MNRAS.430.2545C} {430, 2545}

\bibitem[\protect\citeauthoryear{{Chung}, {van Gorkom}, {Kenney}, {Crowl}  \&
  {Vollmer}}{{Chung} et~al.}{2009}]{chung2009}
{Chung} A.,  {van Gorkom} J.~H.,  {Kenney} J. D.~P.,  {Crowl} H.,   {Vollmer}
  B.,  2009, \mn@doi [\aj] {10.1088/0004-6256/138/6/1741}, \href
  {https://ui.adsabs.harvard.edu/abs/2009AJ....138.1741C} {138, 1741}

\bibitem[\protect\citeauthoryear{{Chy{\.z}y} \& {Buta}}{{Chy{\.z}y} \&
  {Buta}}{2008}]{chyzy2008}
{Chy{\.z}y} K.~T.,  {Buta} R.~J.,  2008, \mn@doi [\apjl] {10.1086/587958},
  \href {https://ui.adsabs.harvard.edu/abs/2008ApJ...677L..17C} {677, L17}

\bibitem[\protect\citeauthoryear{{Clark} et~al.,}{{Clark}
  et~al.}{2018}]{clark2018}
{Clark} C.~J.~R.,  et~al., 2018, \mn@doi [\aap] {10.1051/0004-6361/201731419},
  \href {https://ui.adsabs.harvard.edu/abs/2018A&A...609A..37C} {609, A37}

\bibitem[\protect\citeauthoryear{{Coud{\'e}} et~al.,}{{Coud{\'e}}
  et~al.}{2016}]{coude2016}
{Coud{\'e}} S.,  et~al., 2016, \mn@doi [\mnras] {10.1093/mnras/stv3009}, \href
  {https://ui.adsabs.harvard.edu/abs/2016MNRAS.457.2139C} {457, 2139}

\bibitem[\protect\citeauthoryear{{Currie}, {Berry}, {Jenness}, {Gibb}, {Bell}
  \& {Draper}}{{Currie} et~al.}{2014}]{currie2014}
{Currie} M.~J.,  {Berry} D.~S.,  {Jenness} T.,  {Gibb} A.~G.,  {Bell} G.~S.,
  {Draper} P.~W.,  2014, in {Manset} N.,  {Forshay} P.,  eds,  Astronomical
  Society of the Pacific Conference Series Vol. 485, Astronomical Data Analysis
  Software and Systems XXIII. p.~391

\bibitem[\protect\citeauthoryear{{De Vis} et~al.,}{{De Vis}
  et~al.}{2019}]{devis2019}
{De Vis} P.,  et~al., 2019, \mn@doi [\aap] {10.1051/0004-6361/201834444}, \href
  {https://ui.adsabs.harvard.edu/abs/2019A&A...623A...5D} {623, A5}

\bibitem[\protect\citeauthoryear{{Dempsey} et~al.,}{{Dempsey}
  et~al.}{2013}]{dempsey2013}
{Dempsey} J.~T.,  et~al., 2013, \mn@doi [\mnras] {10.1093/mnras/stt090}, \href
  {https://ui.adsabs.harvard.edu/abs/2013MNRAS.430.2534D} {430, 2534}

\bibitem[\protect\citeauthoryear{{Devine} \& {Bally}}{{Devine} \&
  {Bally}}{1999}]{devine1999}
{Devine} D.,  {Bally} J.,  1999, \mn@doi [\apj] {10.1086/306582}, \href
  {https://ui.adsabs.harvard.edu/abs/1999ApJ...510..197D} {510, 197}

\bibitem[\protect\citeauthoryear{{Dopita}, {Pereira}, {Kewley}  \&
  {Capaccioli}}{{Dopita} et~al.}{2002}]{dopita2002}
{Dopita} M.~A.,  {Pereira} M.,  {Kewley} L.~J.,   {Capaccioli} M.,  2002,
  \mn@doi [\apjs] {10.1086/342624}, \href
  {https://ui.adsabs.harvard.edu/abs/2002ApJS..143...47D} {143, 47}

\bibitem[\protect\citeauthoryear{{Drabek} et~al.,}{{Drabek}
  et~al.}{2012}]{drabek2012}
{Drabek} E.,  et~al., 2012, \mn@doi [\mnras]
  {10.1111/j.1365-2966.2012.21140.x}, \href
  {https://ui.adsabs.harvard.edu/abs/2012MNRAS.426...23D} {426, 23}

\bibitem[\protect\citeauthoryear{{Draine}}{{Draine}}{2003}]{draine2003}
{Draine} B.~T.,  2003, \mn@doi [\araa]
  {10.1146/annurev.astro.41.011802.094840}, \href
  {https://ui.adsabs.harvard.edu/abs/2003ARA&A..41..241D} {41, 241}

\bibitem[\protect\citeauthoryear{{Draine} \& {Li}}{{Draine} \&
  {Li}}{2007}]{draine2007}
{Draine} B.~T.,  {Li} A.,  2007, \mn@doi [\apj] {10.1086/511055}, \href
  {https://ui.adsabs.harvard.edu/abs/2007ApJ...657..810D} {657, 810}

\bibitem[\protect\citeauthoryear{{Ellison}, {Lin}, {Thorp}, {Pan}, {Scudder},
  {S{\'a}nchez}, {Bluck}  \& {Maiolino}}{{Ellison} et~al.}{2021}]{ellison2021}
{Ellison} S.~L.,  {Lin} L.,  {Thorp} M.~D.,  {Pan} H.-A.,  {Scudder} J.~M.,
  {S{\'a}nchez} S.~F.,  {Bluck} A. F.~L.,   {Maiolino} R.,  2021, \mn@doi
  [\mnras] {10.1093/mnras/staa3822}, \href
  {https://ui.adsabs.harvard.edu/abs/2021MNRAS.501.4777E} {501, 4777}

\bibitem[\protect\citeauthoryear{{Ford} et~al.,}{{Ford}
  et~al.}{2013}]{ford2013}
{Ford} G.~P.,  et~al., 2013, \mn@doi [\apj] {10.1088/0004-637X/769/1/55}, \href
  {https://ui.adsabs.harvard.edu/abs/2013ApJ...769...55F} {769, 55}

\bibitem[\protect\citeauthoryear{{Gao} \& {Solomon}}{{Gao} \&
  {Solomon}}{2004}]{gao2004}
{Gao} Y.,  {Solomon} P.~M.,  2004, \mn@doi [\apjs] {10.1086/383003}, \href
  {https://ui.adsabs.harvard.edu/abs/2004ApJS..152...63G} {152, 63}

\bibitem[\protect\citeauthoryear{{Gavazzi}, {Boselli}, {Donati}, {Franzetti}
  \& {Scodeggio}}{{Gavazzi} et~al.}{2003}]{gavazzi2003}
{Gavazzi} G.,  {Boselli} A.,  {Donati} A.,  {Franzetti} P.,   {Scodeggio} M.,
  2003, \mn@doi [\aap] {10.1051/0004-6361:20030026}, \href
  {https://ui.adsabs.harvard.edu/abs/2003A&A...400..451G} {400, 451}

\bibitem[\protect\citeauthoryear{{Greaves}, {Holland}, {Jenness}  \&
  {Hawarden}}{{Greaves} et~al.}{2000}]{greaves2000}
{Greaves} J.~S.,  {Holland} W.~S.,  {Jenness} T.,   {Hawarden} T.~G.,  2000,
  \mn@doi [\nat] {10.1038/35008010}, \href
  {https://ui.adsabs.harvard.edu/abs/2000Natur.404..732G} {404, 732}

\bibitem[\protect\citeauthoryear{{Griffin} et~al.,}{{Griffin}
  et~al.}{2010}]{griffin2010}
{Griffin} M.~J.,  et~al., 2010, \mn@doi [\aap] {10.1051/0004-6361/201014519},
  \href {https://ui.adsabs.harvard.edu/abs/2010A&A...518L...3G} {518, L3}

\bibitem[\protect\citeauthoryear{{Hameed} \& {Devereux}}{{Hameed} \&
  {Devereux}}{2005}]{hameed2005}
{Hameed} S.,  {Devereux} N.,  2005, \mn@doi [\aj] {10.1086/430211}, \href
  {https://ui.adsabs.harvard.edu/abs/2005AJ....129.2597H} {129, 2597}

\bibitem[\protect\citeauthoryear{{Haynes}, {Giovanelli}  \& {Kent}}{{Haynes}
  et~al.}{2007}]{haynes2007}
{Haynes} M.~P.,  {Giovanelli} R.,   {Kent} B.~R.,  2007, \mn@doi [\apjl]
  {10.1086/521188}, \href
  {https://ui.adsabs.harvard.edu/abs/2007ApJ...665L..19H} {665, L19}

\bibitem[\protect\citeauthoryear{{Hildebrand}}{{Hildebrand}}{1983}]{hildebrand1983}
{Hildebrand} R.~H.,  1983, \qjras, \href
  {https://ui.adsabs.harvard.edu/abs/1983QJRAS..24..267H} {24, 267}

\bibitem[\protect\citeauthoryear{{Holland} et~al.,}{{Holland}
  et~al.}{2013}]{holland2013}
{Holland} W.~S.,  et~al., 2013, \mn@doi [\mnras] {10.1093/mnras/sts612}, \href
  {https://ui.adsabs.harvard.edu/abs/2013MNRAS.430.2513H} {430, 2513}

\bibitem[\protect\citeauthoryear{{Jacobs}, {Rizzi}, {Tully}, {Shaya}, {Makarov}
   \& {Makarova}}{{Jacobs} et~al.}{2009}]{jacobs2009}
{Jacobs} B.~A.,  {Rizzi} L.,  {Tully} R.~B.,  {Shaya} E.~J.,  {Makarov} D.~I.,
   {Makarova} L.,  2009, \mn@doi [\aj] {10.1088/0004-6256/138/2/332}, \href
  {https://ui.adsabs.harvard.edu/abs/2009AJ....138..332J} {138, 332}

\bibitem[\protect\citeauthoryear{{James} et~al.,}{{James}
  et~al.}{2004}]{james2004}
{James} P.~A.,  et~al., 2004, \mn@doi [\aap] {10.1051/0004-6361:20031568},
  \href {https://ui.adsabs.harvard.edu/abs/2004A&A...414...23J} {414, 23}

\bibitem[\protect\citeauthoryear{{Kennicutt}}{{Kennicutt}}{1998}]{kennicutt1998}
{Kennicutt} Robert~C. J.,  1998, \mn@doi [\apj] {10.1086/305588}, \href
  {https://ui.adsabs.harvard.edu/abs/1998ApJ...498..541K} {498, 541}

\bibitem[\protect\citeauthoryear{{Kennicutt} \& {Evans}}{{Kennicutt} \&
  {Evans}}{2012}]{kennicutt2012}
{Kennicutt} R.~C.,  {Evans} N.~J.,  2012, \mn@doi [\araa]
  {10.1146/annurev-astro-081811-125610}, \href
  {https://ui.adsabs.harvard.edu/abs/2012ARA&A..50..531K} {50, 531}

\bibitem[\protect\citeauthoryear{{Kennicutt}, {Lee}, {Funes}, {J.}, {Sakai}  \&
  {Akiyama}}{{Kennicutt} et~al.}{2008}]{kennicutt2008}
{Kennicutt} Robert~C. J.,  {Lee} J.~C.,  {Funes} J.~G.,  {J.} S.,  {Sakai} S.,
   {Akiyama} S.,  2008, \mn@doi [\apjs] {10.1086/590058}, \href
  {https://ui.adsabs.harvard.edu/abs/2008ApJS..178..247K} {178, 247}

\bibitem[\protect\citeauthoryear{{Kirk} et~al.,}{{Kirk}
  et~al.}{2018}]{kirk2018}
{Kirk} H.,  et~al., 2018, \mn@doi [\apjs] {10.3847/1538-4365/aada7f}, \href
  {https://ui.adsabs.harvard.edu/abs/2018ApJS..238....8K} {238, 8}

\bibitem[\protect\citeauthoryear{{Koda} et~al.,}{{Koda}
  et~al.}{2009}]{koda2009}
{Koda} J.,  et~al., 2009, \mn@doi [\apjl] {10.1088/0004-637X/700/2/L132}, \href
  {https://ui.adsabs.harvard.edu/abs/2009ApJ...700L.132K} {700, L132}

\bibitem[\protect\citeauthoryear{{Kormendy} \& {Kennicutt}}{{Kormendy} \&
  {Kennicutt}}{2004}]{kormendy2004}
{Kormendy} J.,  {Kennicutt} Robert~C. J.,  2004, \mn@doi [\araa]
  {10.1146/annurev.astro.42.053102.134024}, \href
  {https://ui.adsabs.harvard.edu/abs/2004ARA&A..42..603K} {42, 603}

\bibitem[\protect\citeauthoryear{{Kumari}, {Irwin}  \& {James}}{{Kumari}
  et~al.}{2020}]{kumari2020}
{Kumari} N.,  {Irwin} M.~J.,   {James} B.~L.,  2020, \mn@doi [\aap]
  {10.1051/0004-6361/201732467}, \href
  {https://ui.adsabs.harvard.edu/abs/2020A&A...634A..24K} {634, A24}

\bibitem[\protect\citeauthoryear{{Kuno} et~al.,}{{Kuno}
  et~al.}{2007}]{kuno2007}
{Kuno} N.,  et~al., 2007, \mn@doi [\pasj] {10.1093/pasj/59.1.117}, \href
  {https://ui.adsabs.harvard.edu/abs/2007PASJ...59..117K} {59, 117}

\bibitem[\protect\citeauthoryear{{Lada}, {Forbrich}, {Lombardi}  \&
  {Alves}}{{Lada} et~al.}{2012}]{lada2012}
{Lada} C.~J.,  {Forbrich} J.,  {Lombardi} M.,   {Alves} J.~F.,  2012, \mn@doi
  [\apj] {10.1088/0004-637X/745/2/190}, \href
  {https://ui.adsabs.harvard.edu/abs/2012ApJ...745..190L} {745, 190}

\bibitem[\protect\citeauthoryear{{Leeuw} \& {Robson}}{{Leeuw} \&
  {Robson}}{2009}]{leeuw2009}
{Leeuw} L.~L.,  {Robson} E.~I.,  2009, \mn@doi [\aj]
  {10.1088/0004-6256/137/1/517}, \href
  {https://ui.adsabs.harvard.edu/abs/2009AJ....137..517L} {137, 517}

\bibitem[\protect\citeauthoryear{{Leroy}, {Walter}, {Brinks}, {Bigiel}, {de
  Blok}, {Madore}  \& {Thornley}}{{Leroy} et~al.}{2008}]{leroy2008}
{Leroy} A.~K.,  {Walter} F.,  {Brinks} E.,  {Bigiel} F.,  {de Blok} W.~J.~G.,
  {Madore} B.,   {Thornley} M.~D.,  2008, \mn@doi [\aj]
  {10.1088/0004-6256/136/6/2782}, \href
  {https://ui.adsabs.harvard.edu/abs/2008AJ....136.2782L} {136, 2782}

\bibitem[\protect\citeauthoryear{{Liu}, {Koda}, {Calzetti}, {Fukuhara}  \&
  {Momose}}{{Liu} et~al.}{2011}]{liu2011}
{Liu} G.,  {Koda} J.,  {Calzetti} D.,  {Fukuhara} M.,   {Momose} R.,  2011,
  \mn@doi [\apj] {10.1088/0004-637X/735/1/63}, \href
  {https://ui.adsabs.harvard.edu/abs/2011ApJ...735...63L} {735, 63}

\bibitem[\protect\citeauthoryear{{Mairs} et~al.,}{{Mairs}
  et~al.}{2015}]{mairs2015}
{Mairs} S.,  et~al., 2015, \mn@doi [\mnras] {10.1093/mnras/stv2192}, \href
  {https://ui.adsabs.harvard.edu/abs/2015MNRAS.454.2557M} {454, 2557}

\bibitem[\protect\citeauthoryear{{Mayya}, {Carrasco}  \& {Luna}}{{Mayya}
  et~al.}{2005}]{mayya2005}
{Mayya} Y.~D.,  {Carrasco} L.,   {Luna} A.,  2005, \mn@doi [\apjl]
  {10.1086/432644}, \href
  {https://ui.adsabs.harvard.edu/abs/2005ApJ...628L..33M} {628, L33}

\bibitem[\protect\citeauthoryear{{McKee} \& {Ostriker}}{{McKee} \&
  {Ostriker}}{2007}]{mckee2007}
{McKee} C.~F.,  {Ostriker} E.~C.,  2007, \mn@doi [\araa]
  {10.1146/annurev.astro.45.051806.110602}, \href
  {https://ui.adsabs.harvard.edu/abs/2007ARA&A..45..565M} {45, 565}

\bibitem[\protect\citeauthoryear{{Meijerink}, {Tilanus}, {Dullemond}, {Israel}
  \& {van der Werf}}{{Meijerink} et~al.}{2005}]{meijerink2005}
{Meijerink} R.,  {Tilanus} R.~P.~J.,  {Dullemond} C.~P.,  {Israel} F.~P.,
  {van der Werf} P.~P.,  2005, \mn@doi [\aap] {10.1051/0004-6361:20040469},
  \href {https://ui.adsabs.harvard.edu/abs/2005A&A...430..427M} {430, 427}

\bibitem[\protect\citeauthoryear{{Mutch}, {Croton}  \& {Poole}}{{Mutch}
  et~al.}{2011}]{mutch2011}
{Mutch} S.~J.,  {Croton} D.~J.,   {Poole} G.~B.,  2011, \mn@doi [\apj]
  {10.1088/0004-637X/736/2/84}, \href
  {https://ui.adsabs.harvard.edu/abs/2011ApJ...736...84M} {736, 84}

\bibitem[\protect\citeauthoryear{{Parsons} et~al.,}{{Parsons}
  et~al.}{2018}]{parsons2018}
{Parsons} H.,  et~al., 2018, \mn@doi [\apjs] {10.3847/1538-4365/aa989c}, \href
  {https://ui.adsabs.harvard.edu/abs/2018ApJS..234...22P} {234, 22}

\bibitem[\protect\citeauthoryear{{Pattle} et~al.,}{{Pattle}
  et~al.}{2015}]{pattle2015}
{Pattle} K.,  et~al., 2015, \mn@doi [\mnras] {10.1093/mnras/stv376}, \href
  {https://ui.adsabs.harvard.edu/abs/2015MNRAS.450.1094P} {450, 1094}

\bibitem[\protect\citeauthoryear{{Pattle}, {Gear}, {Redman}, {Smith}  \&
  {Greaves}}{{Pattle} et~al.}{2021}]{pattle2021}
{Pattle} K.,  {Gear} W.,  {Redman} M.,  {Smith} M. W.~L.,   {Greaves} J.,
  2021, \mn@doi [\mnras] {10.1093/mnras/stab1300}, \href
  {https://ui.adsabs.harvard.edu/abs/2021MNRAS.505..684P} {505, 684}

\bibitem[\protect\citeauthoryear{{Rieke}, {Alonso-Herrero}, {Weiner},
  {P{\'e}rez-Gonz{\'a}lez}, {Blaylock}, {Donley}  \& {Marcillac}}{{Rieke}
  et~al.}{2009}]{rieke2009}
{Rieke} G.~H.,  {Alonso-Herrero} A.,  {Weiner} B.~J.,  {P{\'e}rez-Gonz{\'a}lez}
  P.~G.,  {Blaylock} M.,  {Donley} J.~L.,   {Marcillac} D.,  2009, \mn@doi
  [\apj] {10.1088/0004-637X/692/1/556}, \href
  {https://ui.adsabs.harvard.edu/abs/2009ApJ...692..556R} {692, 556}

\bibitem[\protect\citeauthoryear{{Sadavoy} et~al.,}{{Sadavoy}
  et~al.}{2013}]{sadavoy2013}
{Sadavoy} S.~I.,  et~al., 2013, \mn@doi [\apj] {10.1088/0004-637X/767/2/126},
  \href {https://ui.adsabs.harvard.edu/abs/2013ApJ...767..126S} {767, 126}

\bibitem[\protect\citeauthoryear{{Schinnerer} et~al.,}{{Schinnerer}
  et~al.}{2017}]{schinnerer2017}
{Schinnerer} E.,  et~al., 2017, \mn@doi [\apj] {10.3847/1538-4357/836/1/62},
  \href {https://ui.adsabs.harvard.edu/abs/2017ApJ...836...62S} {836, 62}

\bibitem[\protect\citeauthoryear{{Schmidt}}{{Schmidt}}{1959}]{schmidt1959}
{Schmidt} M.,  1959, \mn@doi [\apj] {10.1086/146614}, \href
  {https://ui.adsabs.harvard.edu/abs/1959ApJ...129..243S} {129, 243}

\bibitem[\protect\citeauthoryear{{Scoville} et~al.,}{{Scoville}
  et~al.}{2000}]{scoville2000}
{Scoville} N.~Z.,  et~al., 2000, \mn@doi [\aj] {10.1086/301248}, \href
  {https://ui.adsabs.harvard.edu/abs/2000AJ....119..991S} {119, 991}

\bibitem[\protect\citeauthoryear{{Smith} et~al.,}{{Smith}
  et~al.}{2021}]{smith2021}
{Smith} M. W.~L.,  et~al., 2021, arXiv e-prints, \href
  {https://ui.adsabs.harvard.edu/abs/2021arXiv211000011S} {p. arXiv:2110.00011}

\bibitem[\protect\citeauthoryear{{Sorai} et~al.,}{{Sorai}
  et~al.}{2019}]{sorai2019}
{Sorai} K.,  et~al., 2019, \mn@doi [\pasj] {10.1093/pasj/psz115}, \href
  {https://ui.adsabs.harvard.edu/abs/2019PASJ...71S..14S} {71, S14}

\bibitem[\protect\citeauthoryear{{Tan} et~al.,}{{Tan} et~al.}{2013}]{tan2013}
{Tan} B.-K.,  et~al., 2013, \mn@doi [\mnras] {10.1093/mnras/stt1625}, \href
  {https://ui.adsabs.harvard.edu/abs/2013MNRAS.436..921T} {436, 921}

\bibitem[\protect\citeauthoryear{{Thilker} et~al.,}{{Thilker}
  et~al.}{2007}]{thilker2007}
{Thilker} D.~A.,  et~al., 2007, \mn@doi [\apjs] {10.1086/523853}, \href
  {https://ui.adsabs.harvard.edu/abs/2007ApJS..173..538T} {173, 538}

\bibitem[\protect\citeauthoryear{{Vlahakis}, {van der Werf}, {Israel}  \&
  {Tilanus}}{{Vlahakis} et~al.}{2013}]{vlahakis2013}
{Vlahakis} C.,  {van der Werf} P.,  {Israel} F.~P.,   {Tilanus} R.~P.~J.,
  2013, \mn@doi [\mnras] {10.1093/mnras/stt841}, \href
  {https://ui.adsabs.harvard.edu/abs/2013MNRAS.433.1837V} {433, 1837}

\bibitem[\protect\citeauthoryear{{Vollmer}, {Balkowski}, {Cayatte}, {van Driel}
   \& {Huchtmeier}}{{Vollmer} et~al.}{2004}]{vollmer2004}
{Vollmer} B.,  {Balkowski} C.,  {Cayatte} V.,  {van Driel} W.,   {Huchtmeier}
  W.,  2004, \mn@doi [\aap] {10.1051/0004-6361:20034552}, \href
  {https://ui.adsabs.harvard.edu/abs/2004A&A...419...35V} {419, 35}

\bibitem[\protect\citeauthoryear{{Waller} et~al.,}{{Waller}
  et~al.}{2001}]{waller2001}
{Waller} W.~H.,  et~al., 2001, \mn@doi [\aj] {10.1086/319384}, \href
  {https://ui.adsabs.harvard.edu/abs/2001AJ....121.1395W} {121, 1395}

\bibitem[\protect\citeauthoryear{{Walter}, {Brinks}, {de Blok}, {Bigiel},
  {Kennicutt}, {Thornley}  \& {Leroy}}{{Walter} et~al.}{2008}]{walter2008}
{Walter} F.,  {Brinks} E.,  {de Blok} W.~J.~G.,  {Bigiel} F.,  {Kennicutt}
  Robert~C. J.,  {Thornley} M.~D.,   {Leroy} A.,  2008, \mn@doi [\aj]
  {10.1088/0004-6256/136/6/2563}, \href
  {https://ui.adsabs.harvard.edu/abs/2008AJ....136.2563W} {136, 2563}

\bibitem[\protect\citeauthoryear{{Ward-Thompson} et~al.,}{{Ward-Thompson}
  et~al.}{2016}]{wardthompson2016}
{Ward-Thompson} D.,  et~al., 2016, \mn@doi [\mnras] {10.1093/mnras/stw1978},
  \href {https://ui.adsabs.harvard.edu/abs/2016MNRAS.463.1008W} {463, 1008}

\bibitem[\protect\citeauthoryear{{Williams}, {Gear}  \& {Smith}}{{Williams}
  et~al.}{2018}]{williams2018}
{Williams} T.~G.,  {Gear} W.~K.,   {Smith} M. W.~L.,  2018, \mn@doi [\mnras]
  {10.1093/mnras/sty1476}, \href
  {https://ui.adsabs.harvard.edu/abs/2018MNRAS.479..297W} {479, 297}

\bibitem[\protect\citeauthoryear{{Wilson} et~al.,}{{Wilson}
  et~al.}{2009}]{wilson2009}
{Wilson} C.~D.,  et~al., 2009, \mn@doi [\apj] {10.1088/0004-637X/693/2/1736},
  \href {https://ui.adsabs.harvard.edu/abs/2009ApJ...693.1736W} {693, 1736}

\bibitem[\protect\citeauthoryear{{Wilson} et~al.,}{{Wilson}
  et~al.}{2012}]{wilson2012}
{Wilson} C.~D.,  et~al., 2012, \mn@doi [\mnras]
  {10.1111/j.1365-2966.2012.21453.x}, \href
  {https://ui.adsabs.harvard.edu/abs/2012MNRAS.424.3050W} {424, 3050}

\bibitem[\protect\citeauthoryear{{Wong} \& {Blitz}}{{Wong} \&
  {Blitz}}{2000}]{wong2000}
{Wong} T.,  {Blitz} L.,  2000, \mn@doi [\apj] {10.1086/309368}, \href
  {https://ui.adsabs.harvard.edu/abs/2000ApJ...540..771W} {540, 771}

\bibitem[\protect\citeauthoryear{{Wong} \& {Blitz}}{{Wong} \&
  {Blitz}}{2002}]{wong2002}
{Wong} T.,  {Blitz} L.,  2002, \mn@doi [\apj] {10.1086/339287}, \href
  {https://ui.adsabs.harvard.edu/abs/2002ApJ...569..157W} {569, 157}

\bibitem[\protect\citeauthoryear{{Zabel} et~al.,}{{Zabel}
  et~al.}{2020}]{zabel2020}
{Zabel} N.,  et~al., 2020, \mn@doi [\mnras] {10.1093/mnras/staa1513}, \href
  {https://ui.adsabs.harvard.edu/abs/2020MNRAS.496.2155Z} {496, 2155}

\bibitem[\protect\citeauthoryear{{de Blok}, {Walter}, {Brinks}, {Trachternach},
  {Oh}  \& {Kennicutt}}{{de Blok} et~al.}{2008}]{deblok2008}
{de Blok} W.~J.~G.,  {Walter} F.,  {Brinks} E.,  {Trachternach} C.,  {Oh}
  S.~H.,   {Kennicutt} R.~C. J.,  2008, \mn@doi [\aj]
  {10.1088/0004-6256/136/6/2648}, \href
  {https://ui.adsabs.harvard.edu/abs/2008AJ....136.2648D} {136, 2648}

\makeatother
\end{thebibliography}

\clearpage

\onecolumn

\appendix

\section{Comparison of gas and dust emission}
\label{sec:appendix_g2d}

In this appendix we present comparisons of SCUBA-2 850$\mu$m surface brightness to atomic and molecular hydrogen column density, derived from VLA H\textsc{i} and Nobeyama 45m $^{12}$CO $J=1\to0$ data respectively, as described in Section~\ref{sec:g2d}.

\begin{figure*}
    \centering
    \includegraphics[width=\textwidth]{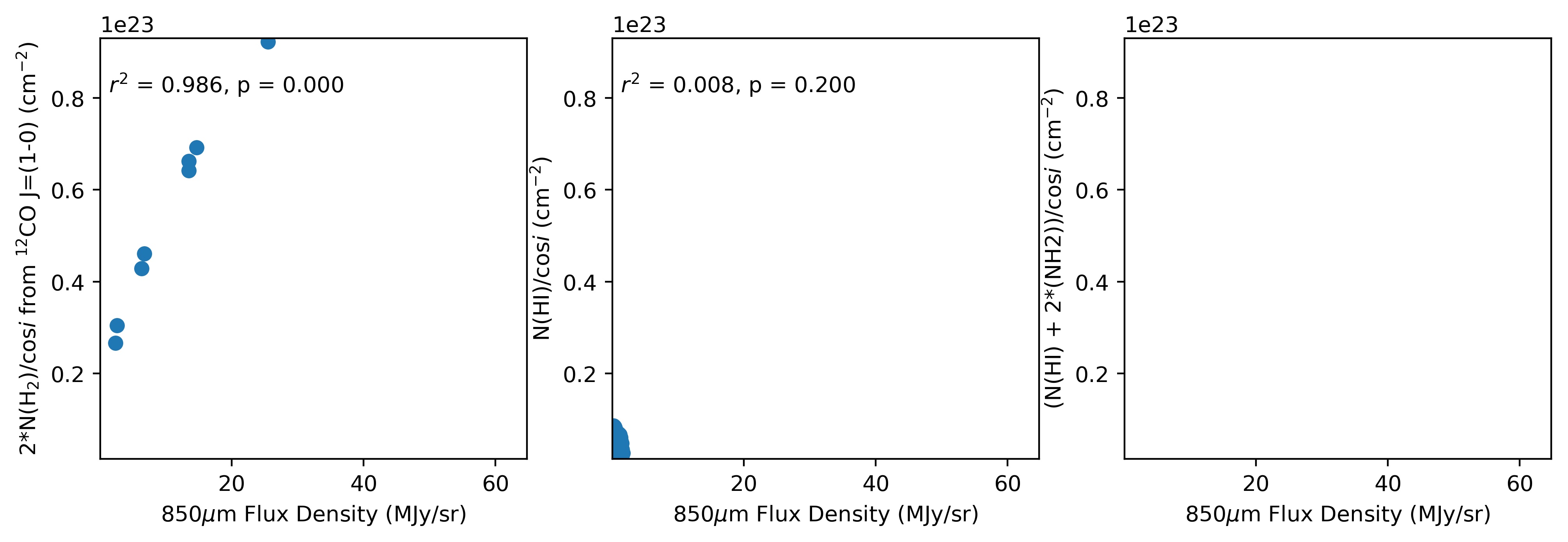}
    \caption{Comparison of dust and gas emission in NGC 3034. Left: SCUBA-2 850$\mu$m surface brightness compared to column density of molecular gas derived from Nobeyama 45m ${12}$CO $J=1\to 0$ measurements.  Centre: 850$\mu$m surface brightness compared to column density of {atomic} gas derived from VLA H\textsc{i} measurements. Right: SCUBA-2 850$\mu$m surface brightness compared to total gas column density.  Note that the VLA H\textsc{i} observations of NGC3034 are saturated towards the centre of the galaxy, resulting in there being no overlap between the two gas density tracers.}
    \label{fig:co_hi_ngc3034}
\end{figure*}

\begin{figure*}
    \centering
    \includegraphics[width=\textwidth]{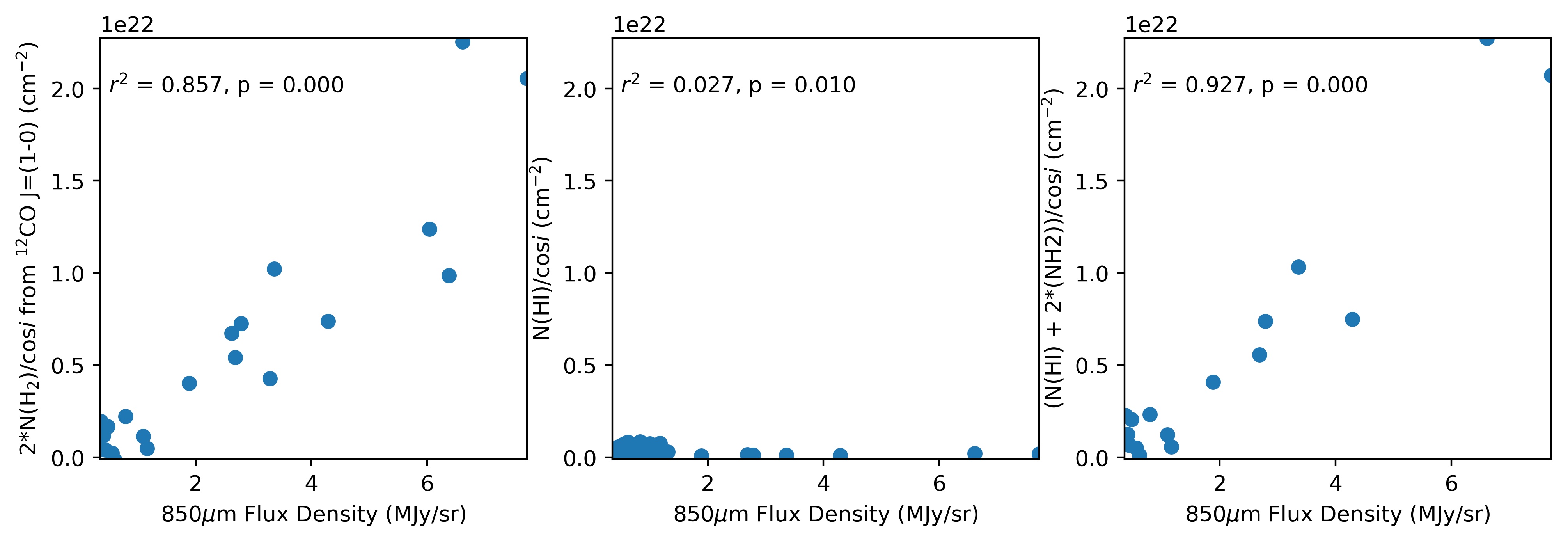}
    \caption{Comparison of dust and gas emission in NGC 3351.  Panels as in Figure~\ref{fig:co_hi_ngc3034}.}
    \label{fig:co_hi_ngc3351}
\end{figure*}

\begin{figure*}
    \centering
    \includegraphics[width=\textwidth]{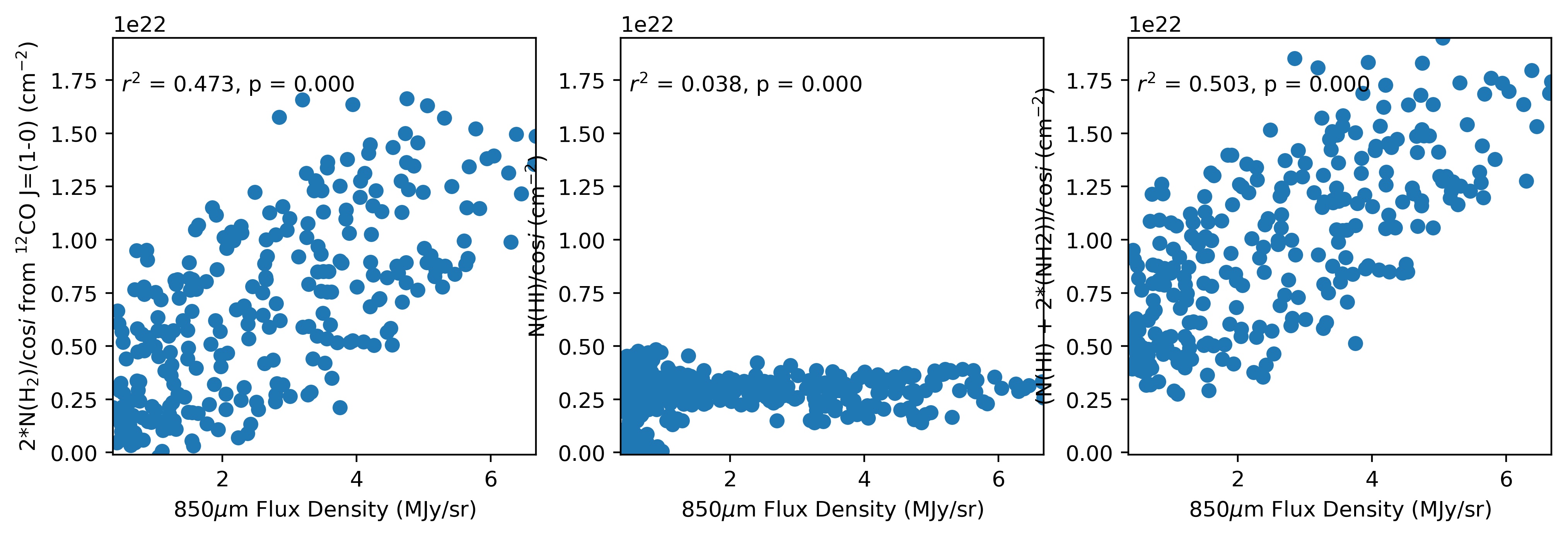}
    \caption{Comparison of dust and gas emission in NGC 3521.  Panels as in Figure~\ref{fig:co_hi_ngc3034}.}
    \label{fig:co_hi_ngc3521}
\end{figure*}

\begin{figure*}
    \centering
    \includegraphics[width=\textwidth]{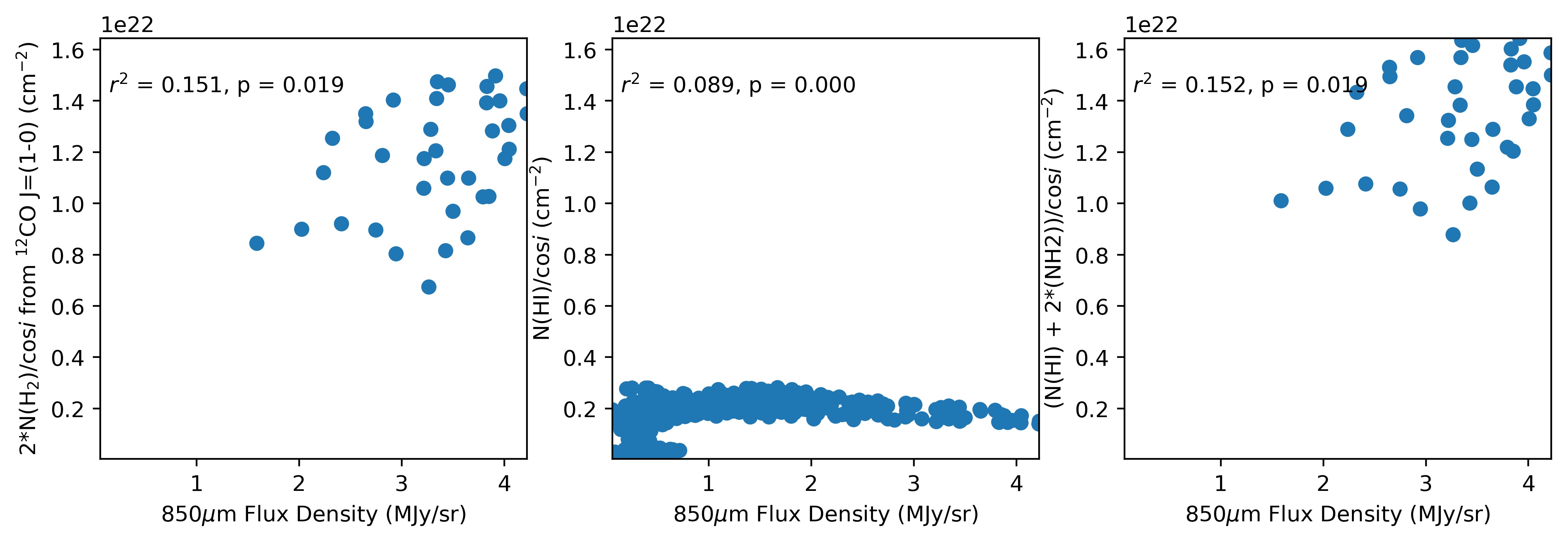}
    \caption{Comparison of dust and gas emission in NGC 4254.  Panels as in Figure~\ref{fig:co_hi_ngc3034}.}
    \label{fig:co_hi_ngc4254}
\end{figure*}

\begin{figure*}
    \centering
    \includegraphics[width=\textwidth]{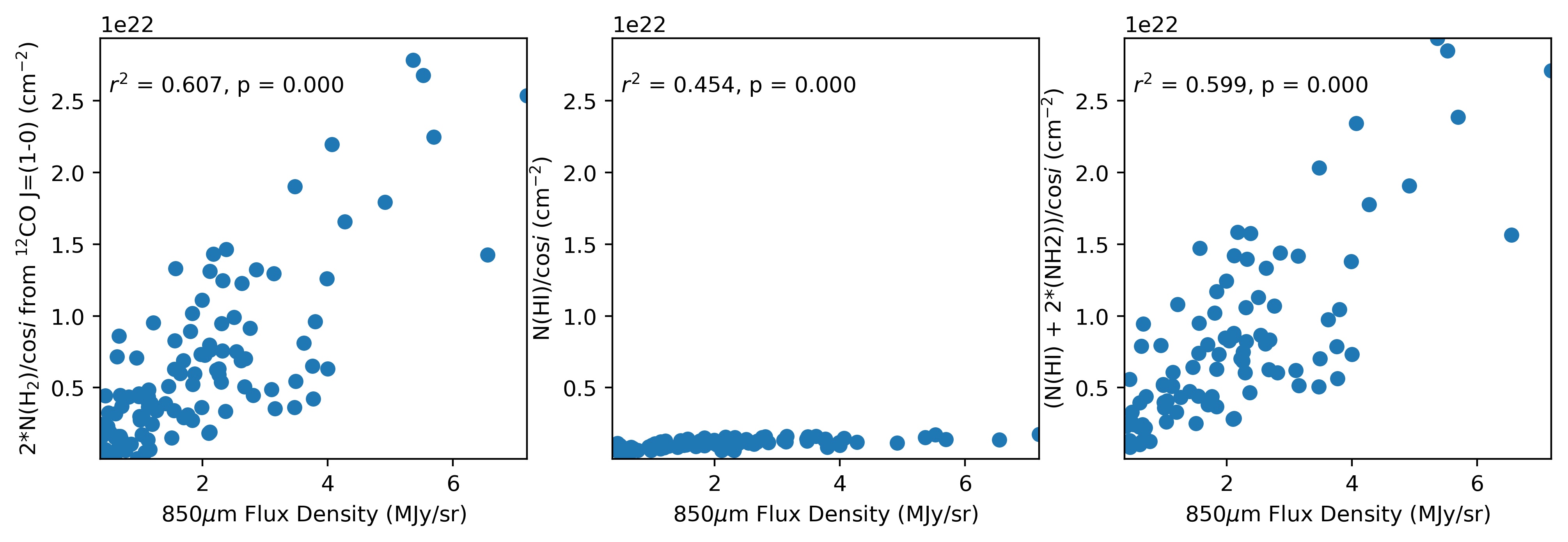}
    \caption{Comparison of dust and gas emission in NGC 4569.  Panels as in Figure~\ref{fig:co_hi_ngc3034}.}
    \label{fig:co_hi_ngc4569}
\end{figure*}

\begin{figure*}
    \centering
    \includegraphics[width=\textwidth]{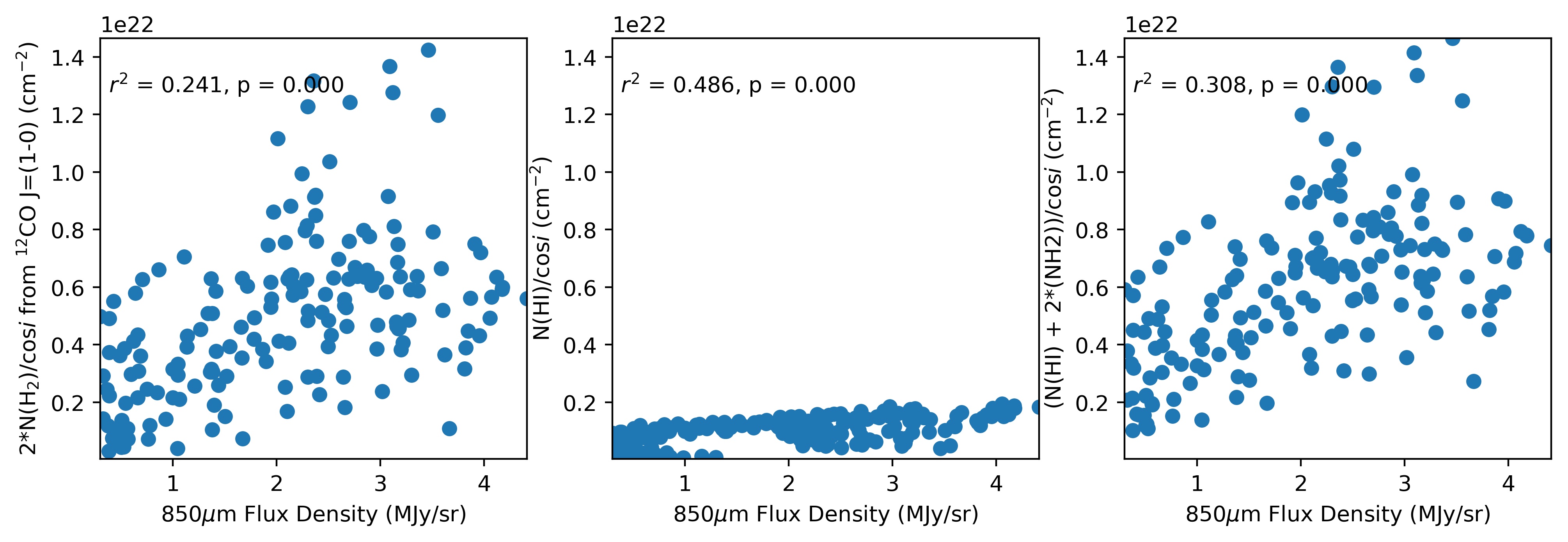}
    \caption{Comparison of dust and gas emission in NGC 4736.  Panels as in Figure~\ref{fig:co_hi_ngc3034}.}
    \label{fig:co_hi_ngc4736}
\end{figure*}

\begin{figure*}
    \centering
    \includegraphics[width=\textwidth]{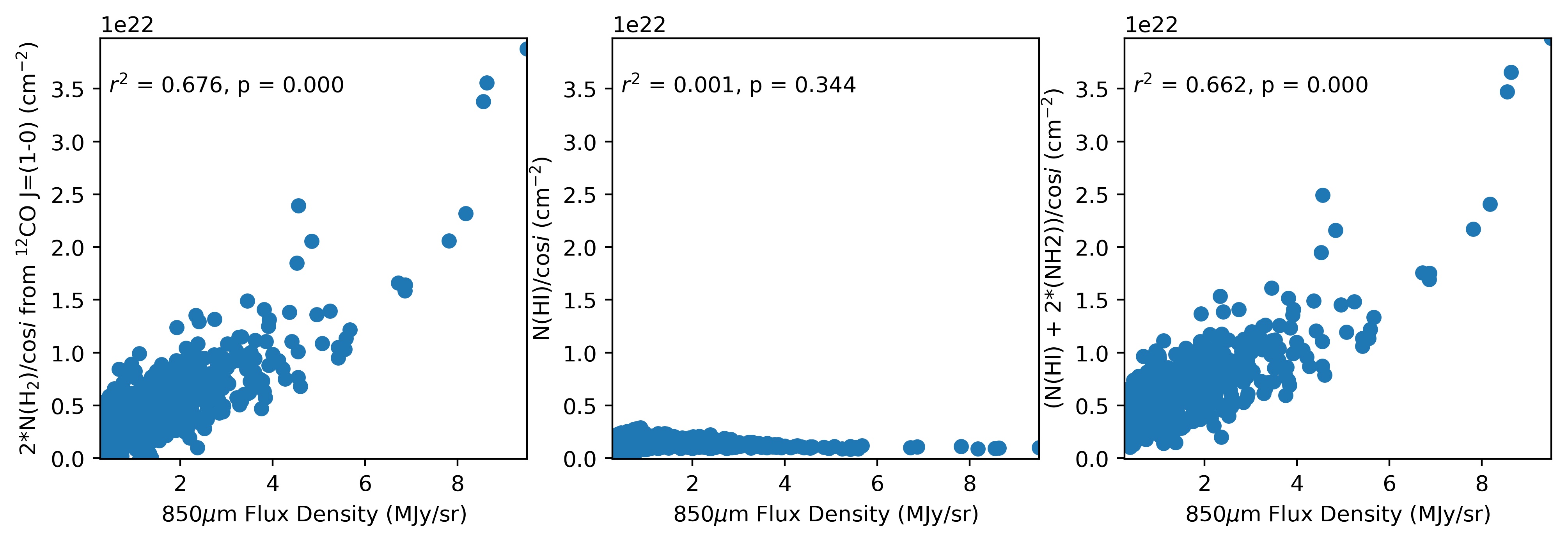}
    \caption{Comparison of dust and gas emission in NGC 5055.  Panels as in Figure~\ref{fig:co_hi_ngc3034}.}
    \label{fig:co_hi_ngc5055}
\end{figure*}

\begin{figure*}
    \centering
    \includegraphics[width=\textwidth]{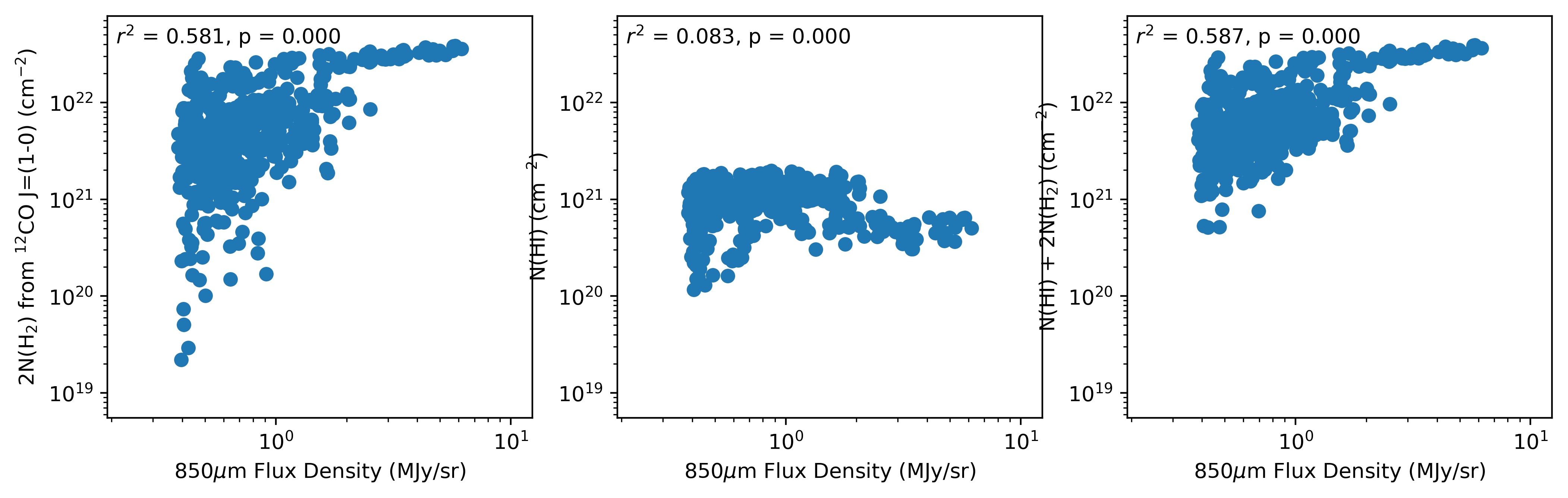}
    \caption{Comparison of dust and gas emission in NGC 5194.  Panels as in Figure~\ref{fig:co_hi_ngc3034}.}
    \label{fig:co_hi_ngc5194}
\end{figure*}

\clearpage

\section{SED fitting}
\label{sec:appendix_seds}

In this appendix we present maps of molecular hydrogen column density $N({\rm H}_2)$, dust temperature $T$ and dust opacity index $\beta$ resulting from the SED fitting described in Section~\ref{sec:seds}.  All maps are shown at a common resolution of 25.2$^{\prime\prime}$.

\begin{figure*}
    \centering
    \includegraphics[width=\textwidth]{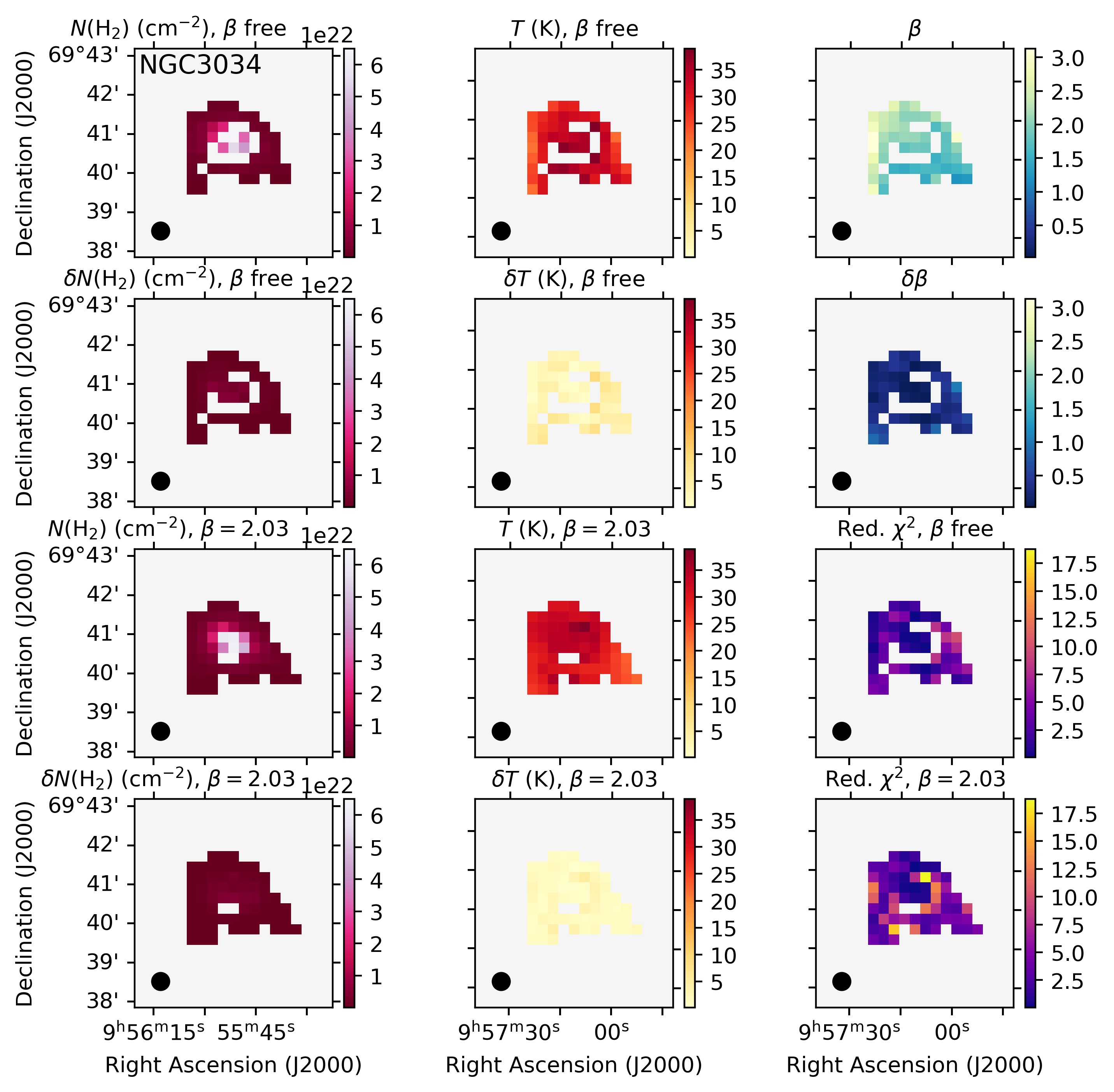}
    \caption{SED fitting results for NGC3034.  Left column: top, column density $N({\rm H}_2)$ in cm$^{-2}$, $\beta$-free case; upper centre, uncertainty on $N({\rm H}_2)$, $\beta$-free case; lower centre, $N({\rm H}_2)$, median-$\beta$ case; bottom, uncertainty on $N({\rm H}_2)$, median-$\beta$ case.  Centre column: top, dust temperature $T$ in K, $\beta$-free case; upper centre, uncertainty on $T$, $\beta$-free case; lower centre, $T$, median-$\beta$ case; bottom, uncertainty on $T$, median-$\beta$ case.  Right column: top, dust opacity index $\beta$, $\beta$-free case; upper centre, uncertainty on $\beta$, $\beta$-free case; lower centre, reduced $\chi^{2}$ values, $\beta$-free case; bottom, reduced $\chi^{2}$ values, median-$\beta$ case.}
    \label{fig:sed_ngc3034}
\end{figure*}

\begin{figure*}
    \centering
    \includegraphics[width=\textwidth]{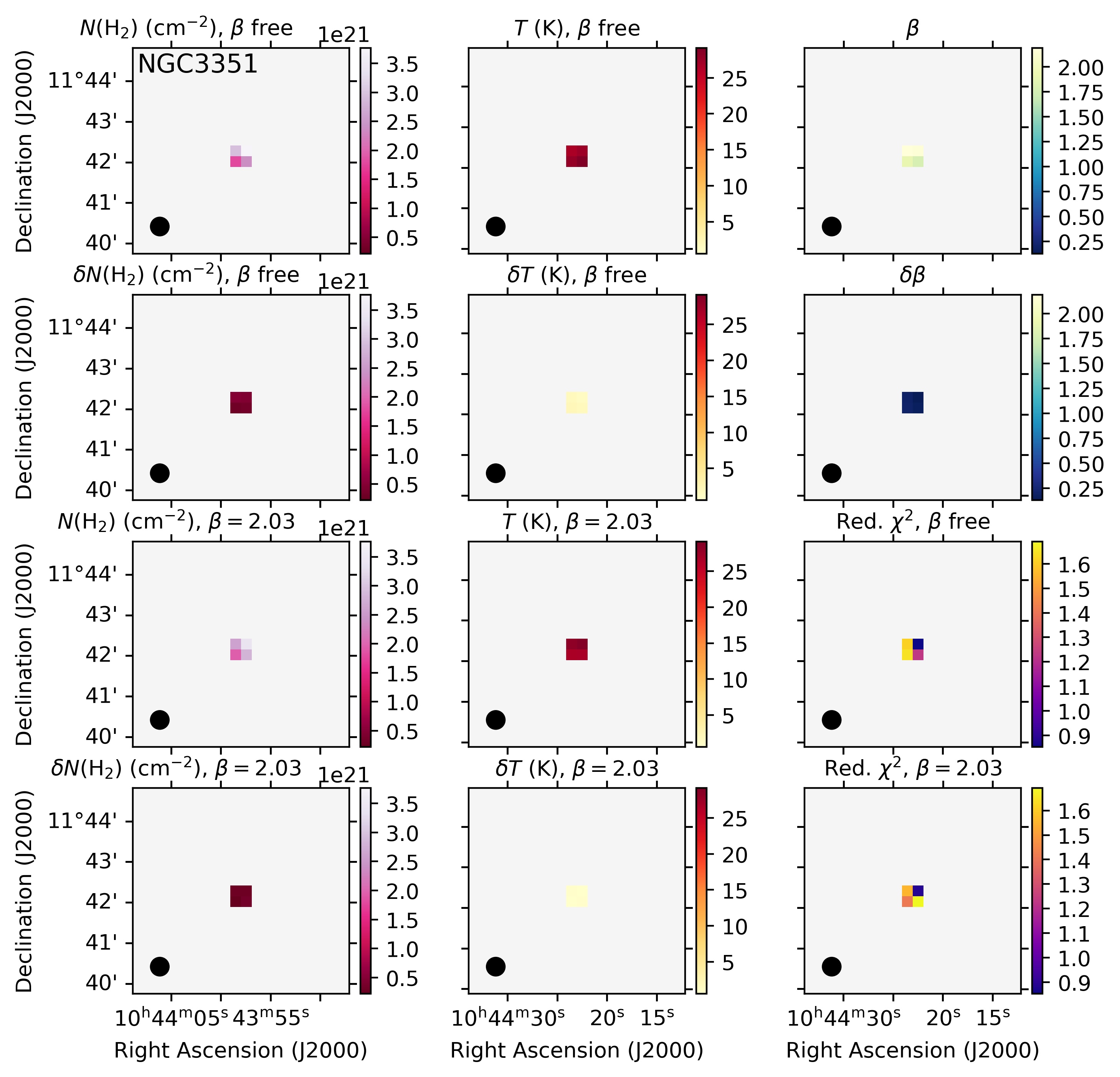}
    \caption{SED fitting results for NGC3351.  Panels as in Figure~\ref{fig:sed_ngc3034}.}
    \label{fig:sed_ngc3351}
\end{figure*}

\begin{figure*}
    \centering
    \includegraphics[width=\textwidth]{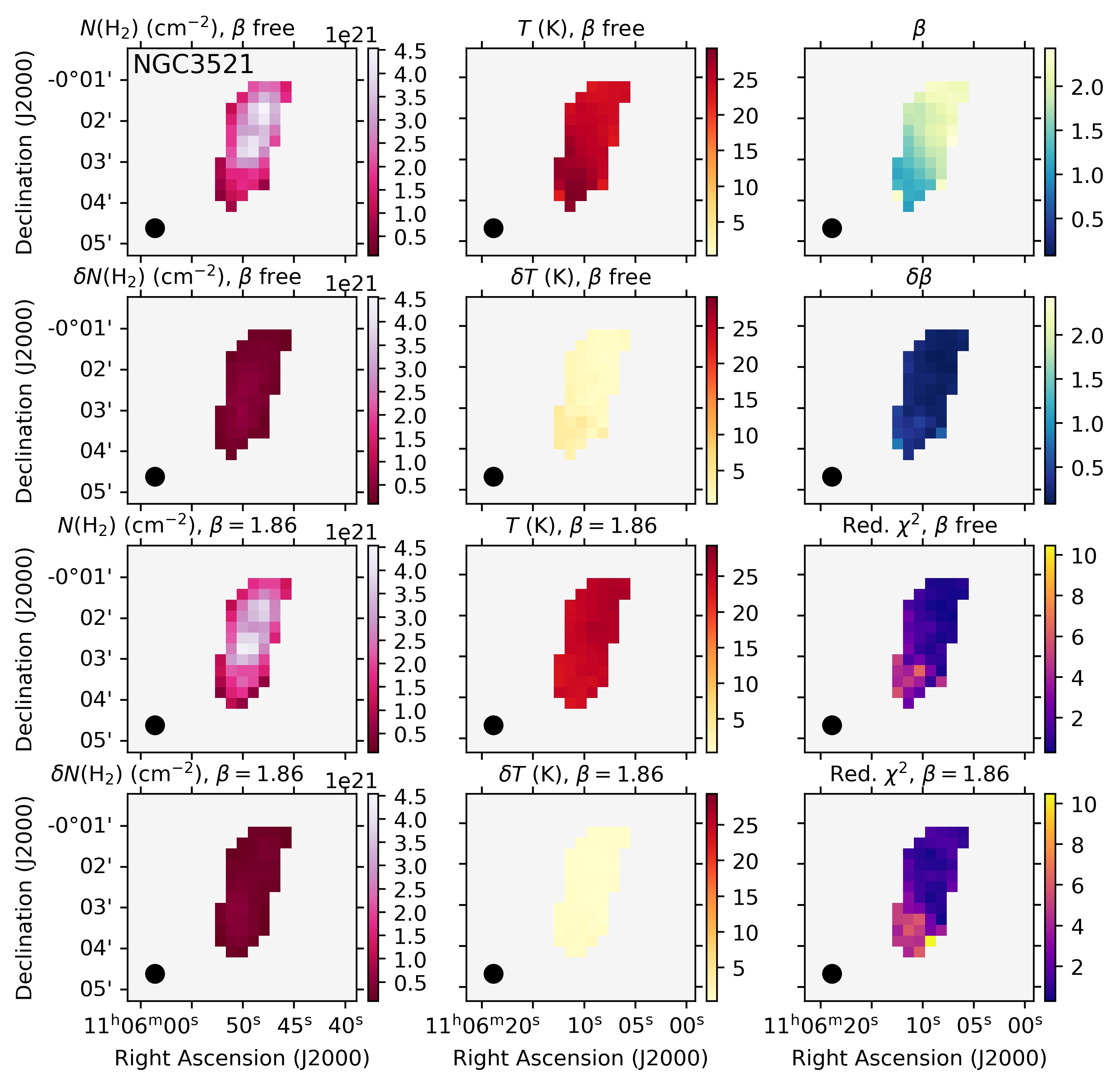}
    \caption{SED fitting results for NGC3521.  Panels as in Figure~\ref{fig:sed_ngc3034}.}
    \label{fig:sed_ngc3521}
\end{figure*}

\begin{figure*}
    \centering
    \includegraphics[width=\textwidth]{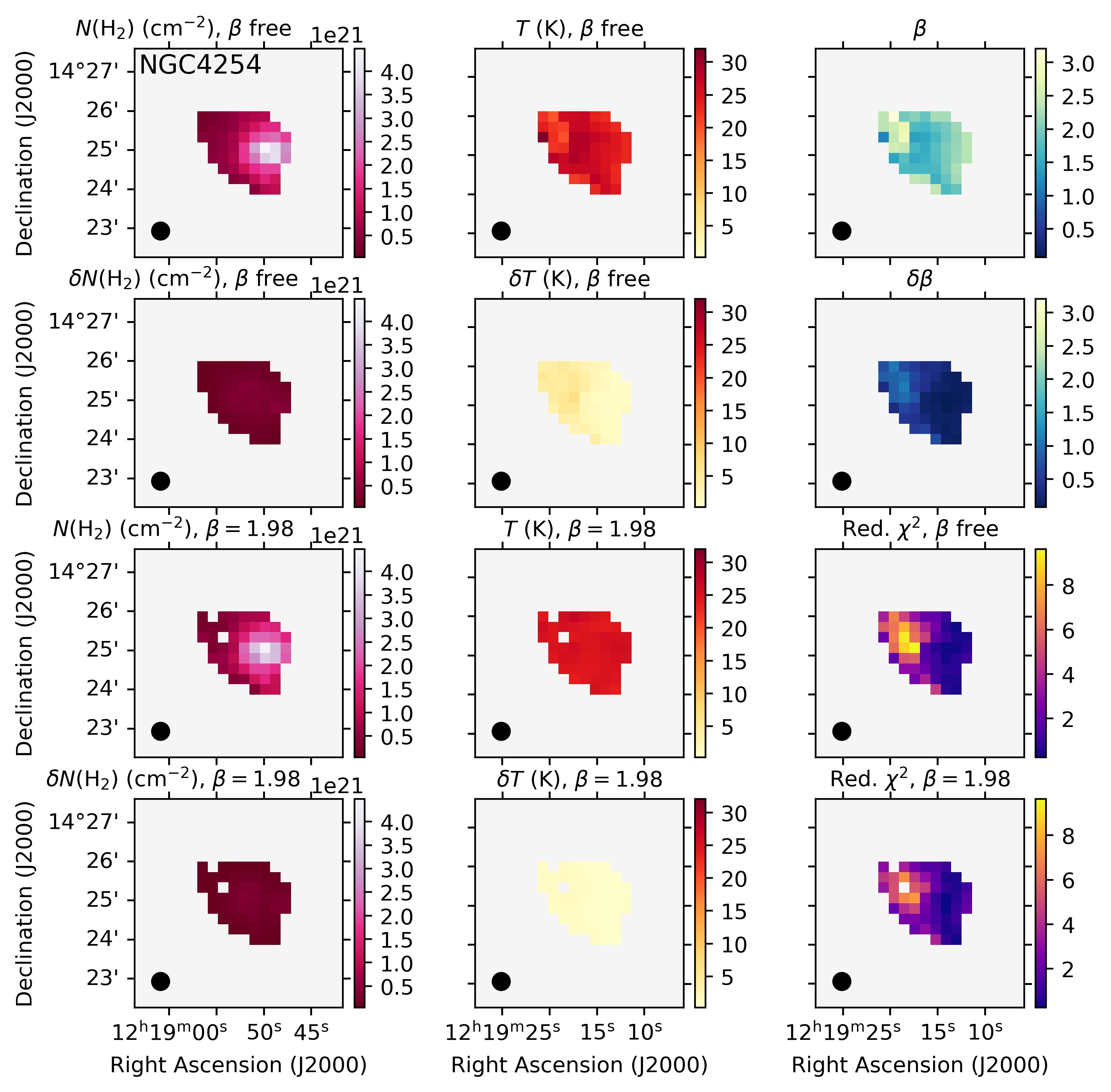}
    \caption{SED fitting results for NGC4254.  Panels as in Figure~\ref{fig:sed_ngc3034}.}
    \label{fig:sed_ngc4254}
\end{figure*}

\begin{figure*}
    \centering
    \includegraphics[width=\textwidth]{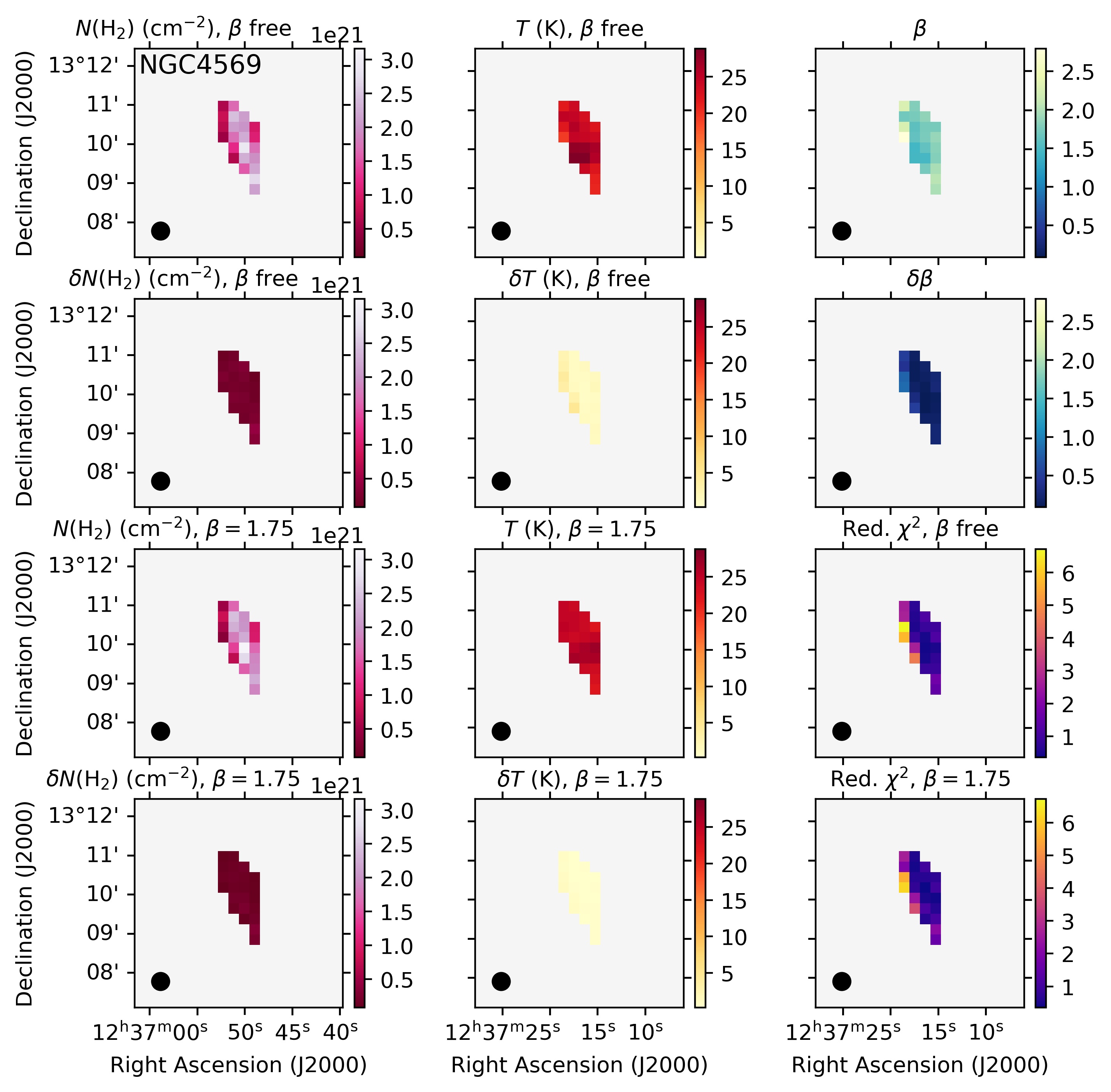}
    \caption{SED fitting results for NGC4569.  Panels as in Figure~\ref{fig:sed_ngc3034}.}
    \label{fig:sed_ngc4569}
\end{figure*}

\begin{figure*}
    \centering
    \includegraphics[width=\textwidth]{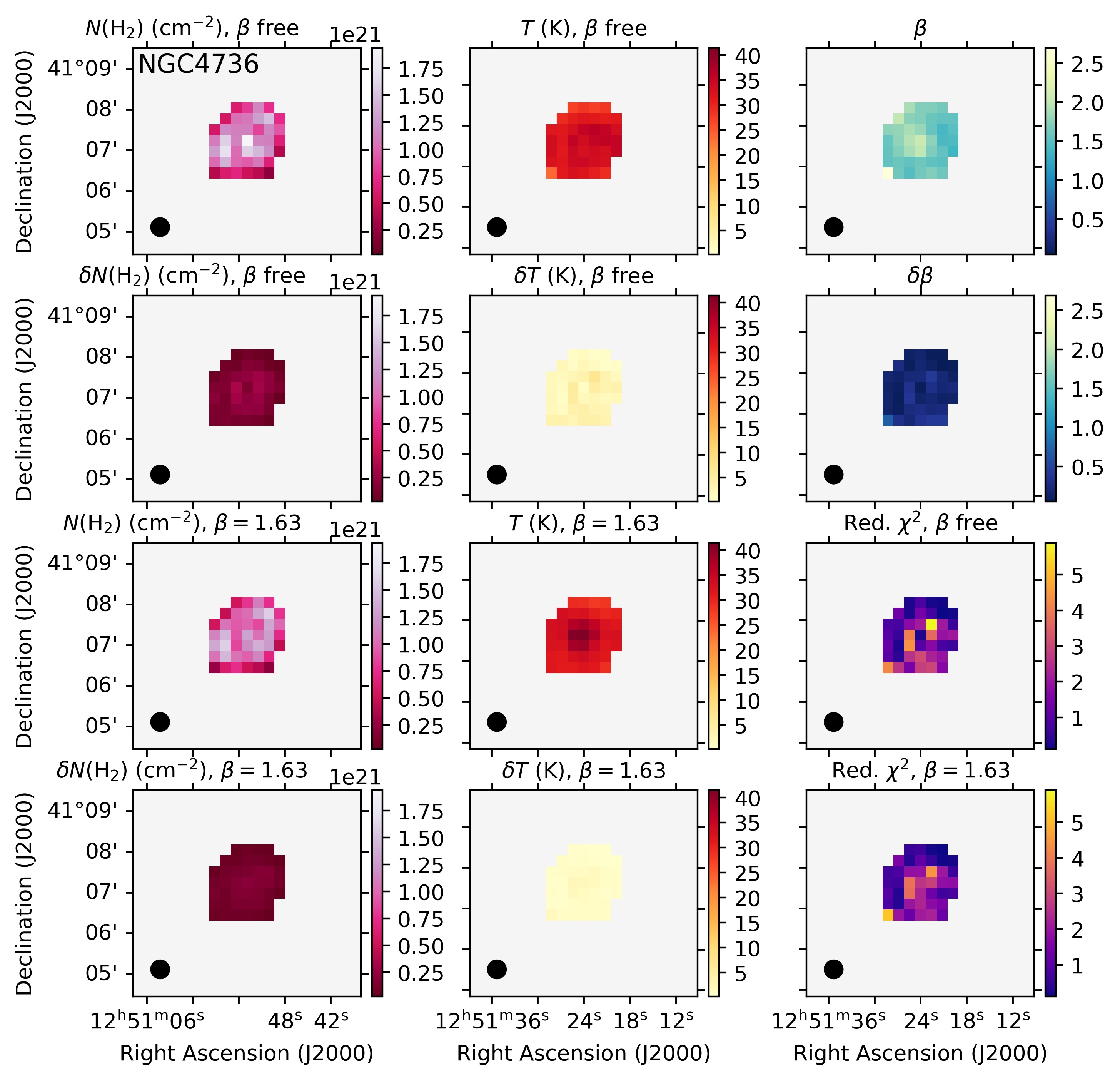}
    \caption{SED fitting results for NGC4736.  Panels as in Figure~\ref{fig:sed_ngc3034}.}
    \label{fig:sed_ngc4736}
\end{figure*}

\begin{figure*}
    \centering
    \includegraphics[width=\textwidth]{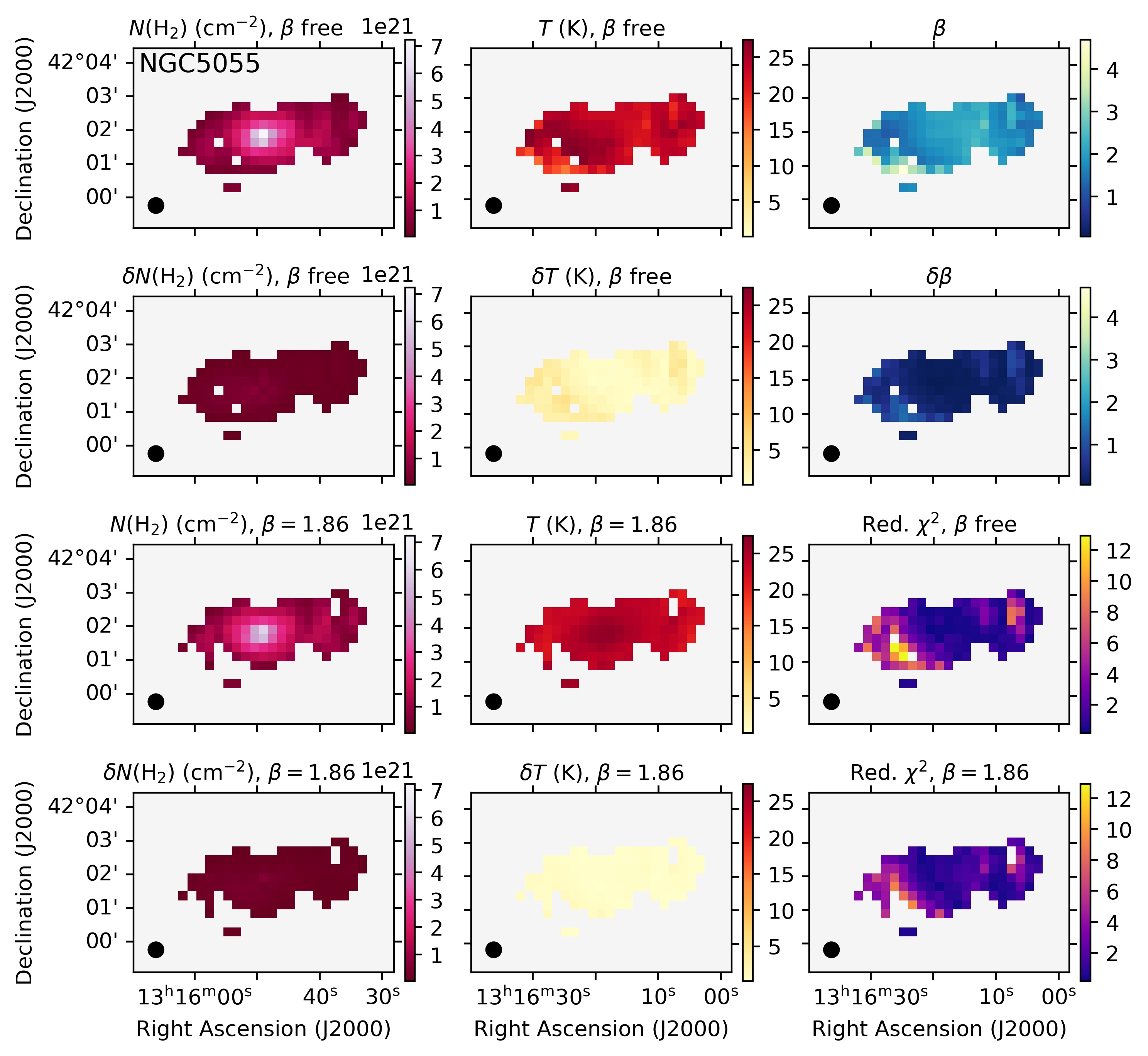}
    \caption{SED fitting results for NGC5055.  Panels as in Figure~\ref{fig:sed_ngc3034}.}
    \label{fig:sed_ngc5055}
\end{figure*}

\begin{figure*}
    \centering
    \includegraphics[width=\textwidth]{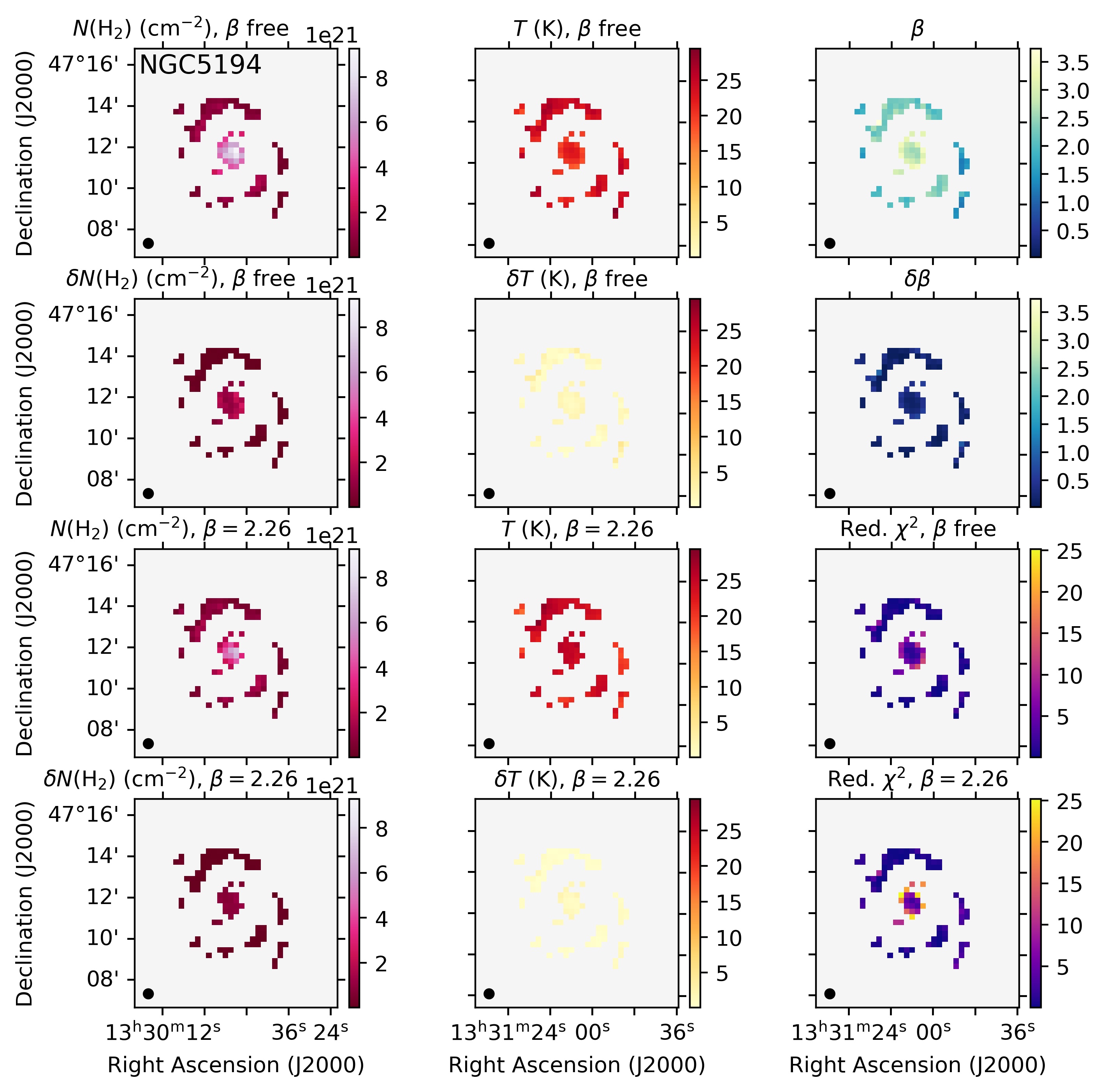}
    \caption{SED fitting results for NGC5194.  Panels as in Figure~\ref{fig:sed_ngc3034}.}
    \label{fig:sed_ngc5194}
\end{figure*}

\clearpage

\section{Multiwavelength imaging}
\label{sec:appendix_mwl}

In this appendix we present \textit{Spitzer} 24$\mu$m, \textit{Spitzer} 3.6$\mu$m, GALEX FUV and GALEX NUV imaging of each of the galaxies in our sample, taken from the Dustpedia database \citep{clark2018}.

\begin{figure*}
    \centering
    \includegraphics[width=\textwidth]{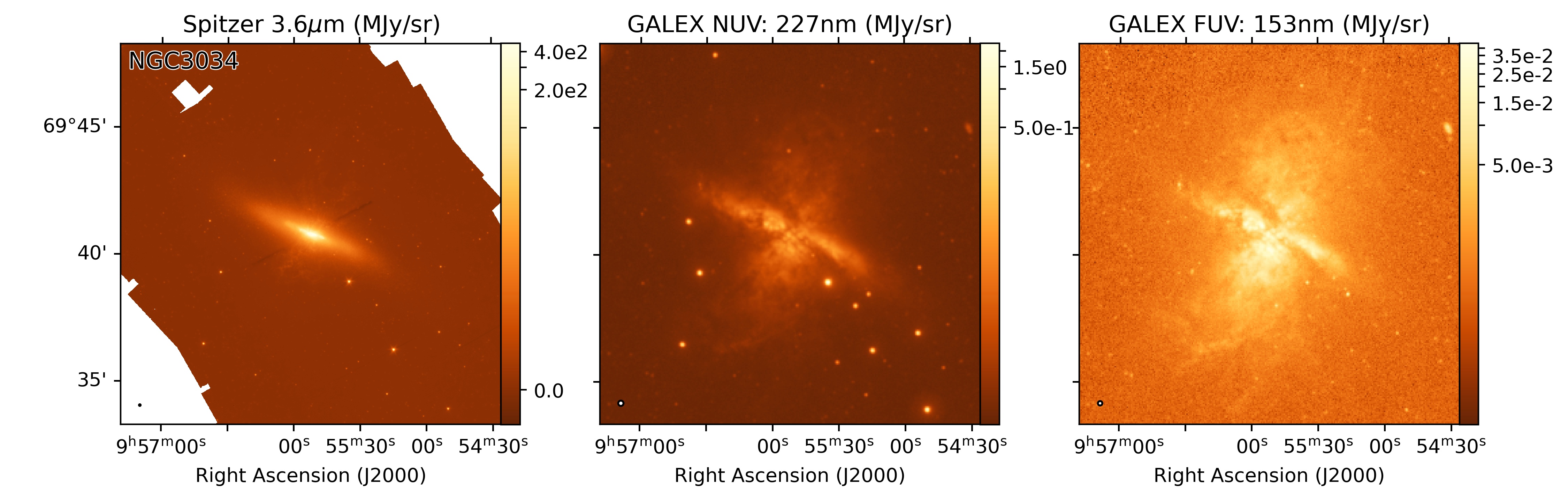}
    \caption{Multiwavelength observations of NGC3034, taken from the Dustpedia database \citep{clark2018}.  Left: \textit{Spitzer} 3.6\,$\mu$m emission.  Centre: GALEX NUV emission.  Right: GALEX FUV emission.  Beam sizes are shown in the lower left-hand corner of each image.}
    \label{fig:mwl_ngc3034}
\end{figure*}

\begin{figure*}
    \centering
    \includegraphics[width=\textwidth]{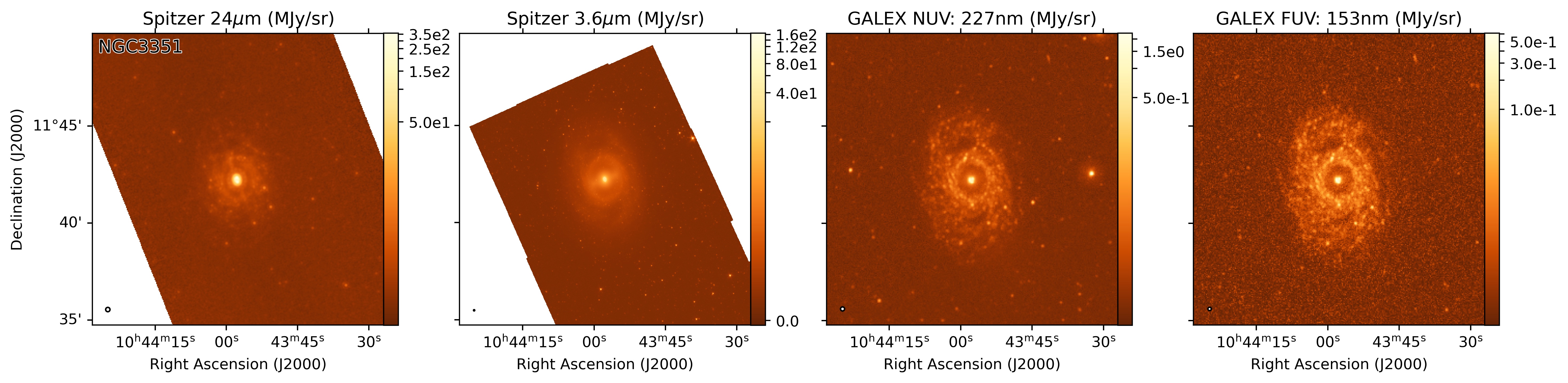}
    \caption{Multiwavelength observations of NGC3351, taken from the Dustpedia database \citep{clark2018}.  Far left: \textit{Spitzer} 24\,$\mu$m emission.  Centre left: \textit{Spitzer} 3.6\,$\mu$m emission.  Centre right: GALEX NUV emission.  Far right: GALEX FUV emission.  Beam sizes are shown in the lower left-hand corner of each image.}
    \label{fig:mwl_ngc3351}
\end{figure*}

\begin{figure*}
    \centering
    \includegraphics[width=\textwidth]{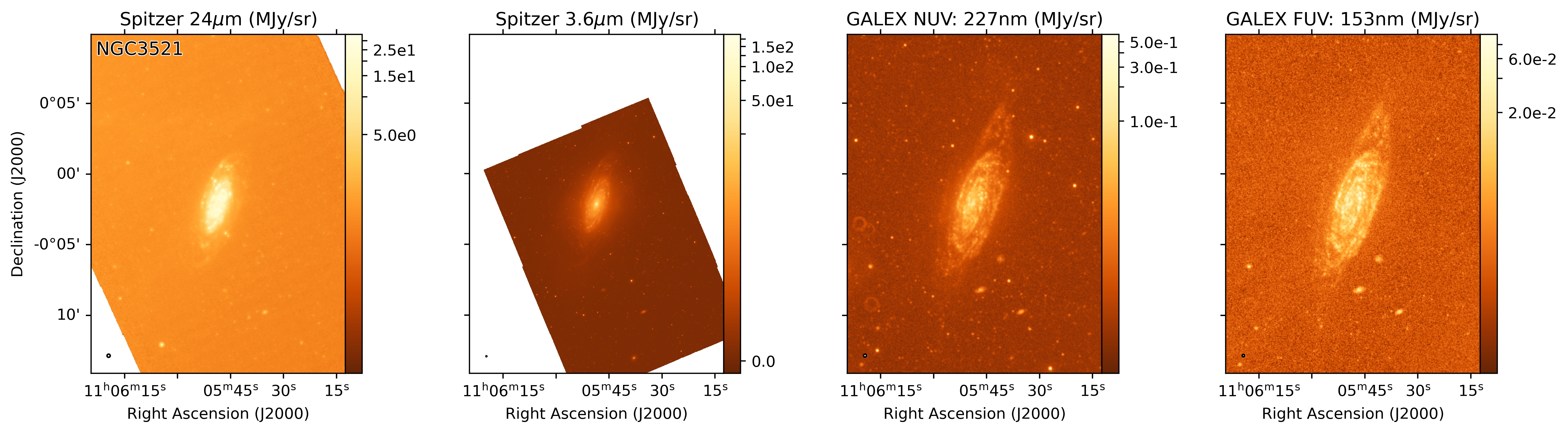}
    \caption{Multiwavelength observations of NGC3521, taken from the Dustpedia database \citep{clark2018}.  Panels as in Figure~\ref{fig:mwl_ngc3351}.}
    \label{fig:mwl_ngc3521}
\end{figure*}

\begin{figure*}
    \centering
    \includegraphics[width=\textwidth]{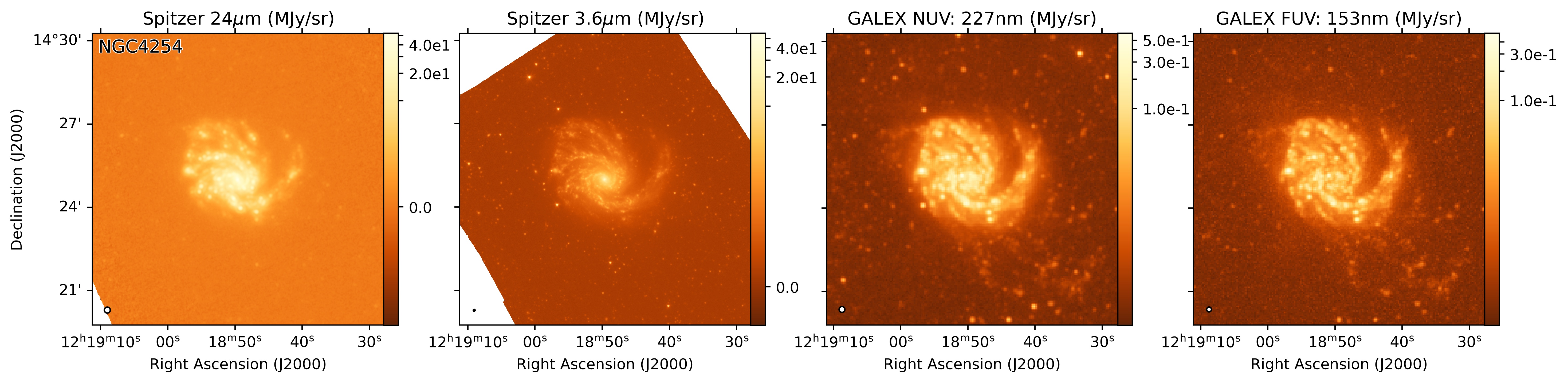}
    \caption{Multiwavelength observations of NGC4254, taken from the Dustpedia database \citep{clark2018}.  Panels as in Figure~\ref{fig:mwl_ngc3351}.}
    \label{fig:mwl_ngc4254}
\end{figure*}

\begin{figure*}
    \centering
    \includegraphics[width=\textwidth]{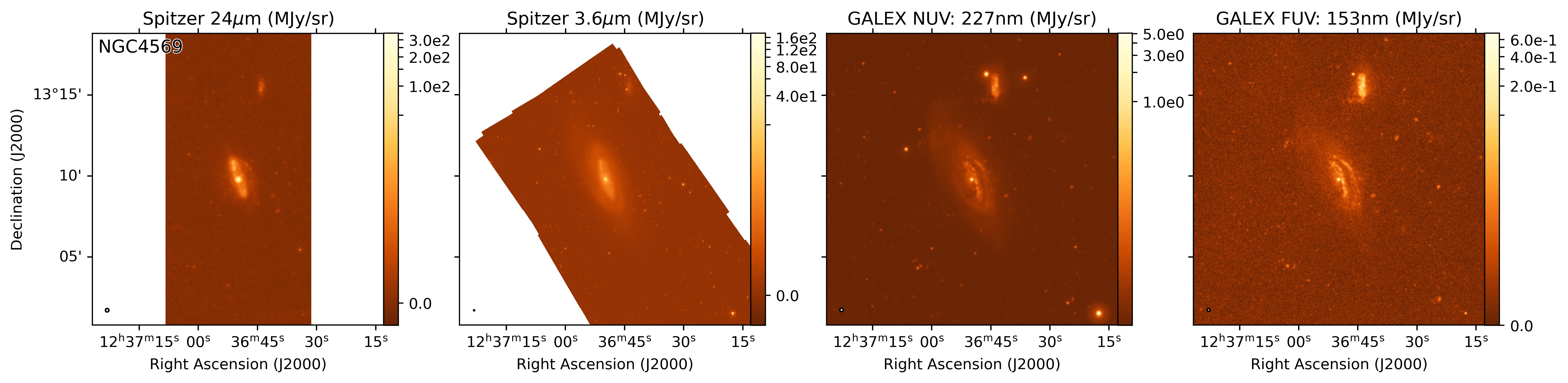}
    \caption{Multiwavelength observations of NGC4569, taken from the Dustpedia database \citep{clark2018}.  Panels as in Figure~\ref{fig:mwl_ngc3351}.}
    \label{fig:mwl_ngc4569}
\end{figure*}

\begin{figure*}
    \centering
    \includegraphics[width=\textwidth]{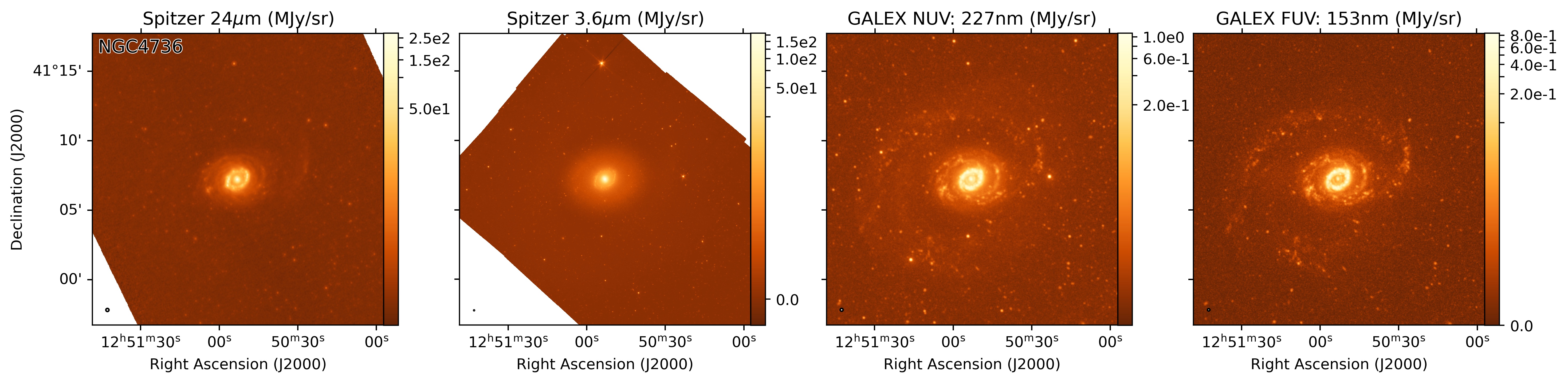}
    \caption{Multiwavelength observations of NGC4569, taken from the Dustpedia database \citep{clark2018}.  Panels as in Figure~\ref{fig:mwl_ngc3351}.}
    \label{fig:mwl_ngc4736}
\end{figure*}

\begin{figure*}
    \centering
    \includegraphics[width=\textwidth]{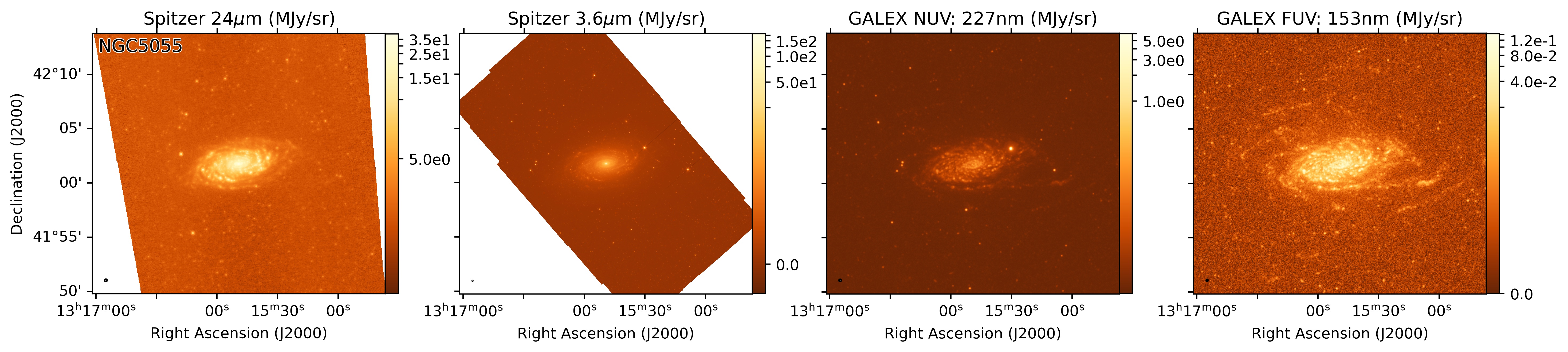}
    \caption{Multiwavelength observations of NGC5055, taken from the Dustpedia database \citep{clark2018}.  Panels as in Figure~\ref{fig:mwl_ngc3351}.}
    \label{fig:mwl_ngc5055}
\end{figure*}

\begin{figure*}
    \centering
    \includegraphics[width=\textwidth]{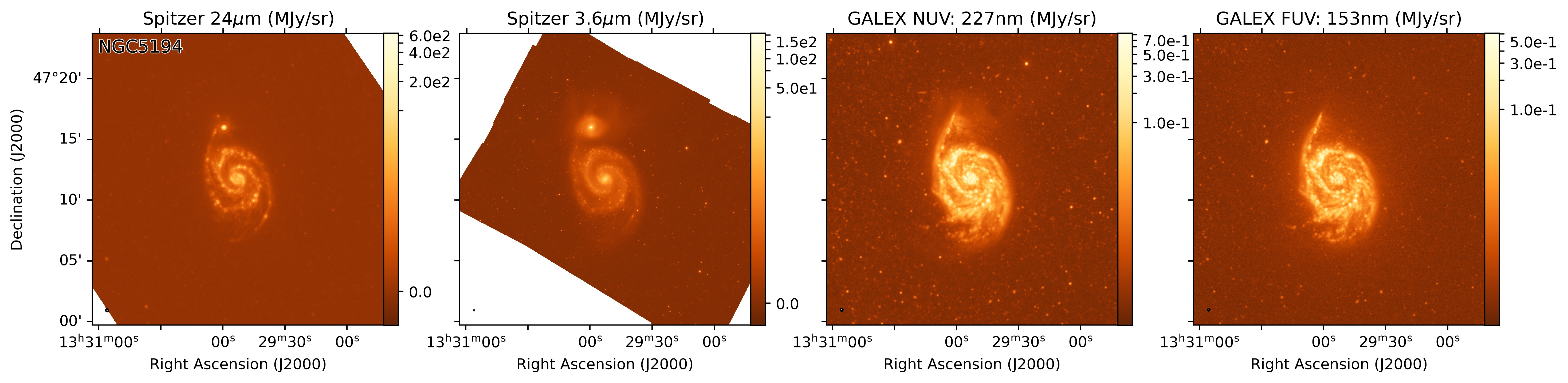}
    \caption{Multiwavelength observations of NGC5194, taken from the Dustpedia database \citep{clark2018}.  Panels as in Figure~\ref{fig:mwl_ngc3351}.}
    \label{fig:mwl_ngc5194}
\end{figure*}

\clearpage

\section{Star formation rates}
\label{sec:appendix_sfr}

In this appendix we present the GALEX FUV, Spitzer 24$\mu$m, Spitzer 3.6$\mu$m and star formation surface density maps for each galaxy in our sample, except NGC 3034, as described in Section~\ref{sec:ssfr}.  All maps are shown at a common resolution of 25.2$^{\prime\prime}$.

\begin{figure*}
    \centering
    \includegraphics[width=\textwidth]{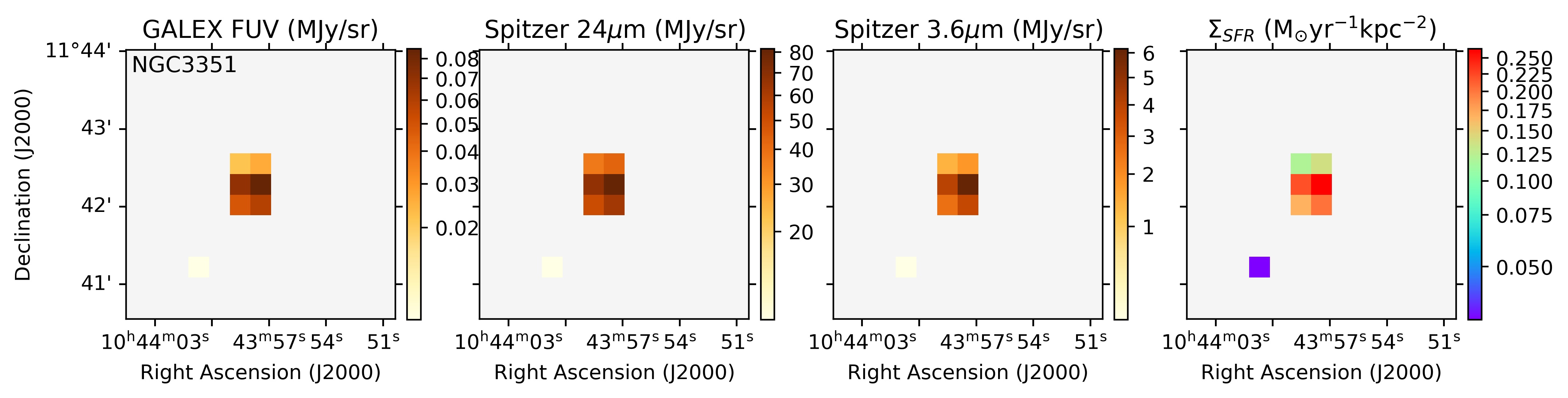}
    \caption{Surface density of star formation in NGC 3351.  Far left: GALEX FUV surface brightness; centre left: \textit{Spitzer} 24$\mu$m surface brightness; centre right: \textit{Spitzer} 3.6$\mu$m surface brightness.  All are taken from the Dustpedia database \citep{clark2018}, and smoothed to $25\farcs2$ resolution and gridded to 16$^{\prime\prime}$ pixels.  Far right: surface density of star formation.}
    \label{fig:ssfr_ngc3351}
\end{figure*}

\begin{figure*}
    \centering
    \includegraphics[width=\textwidth]{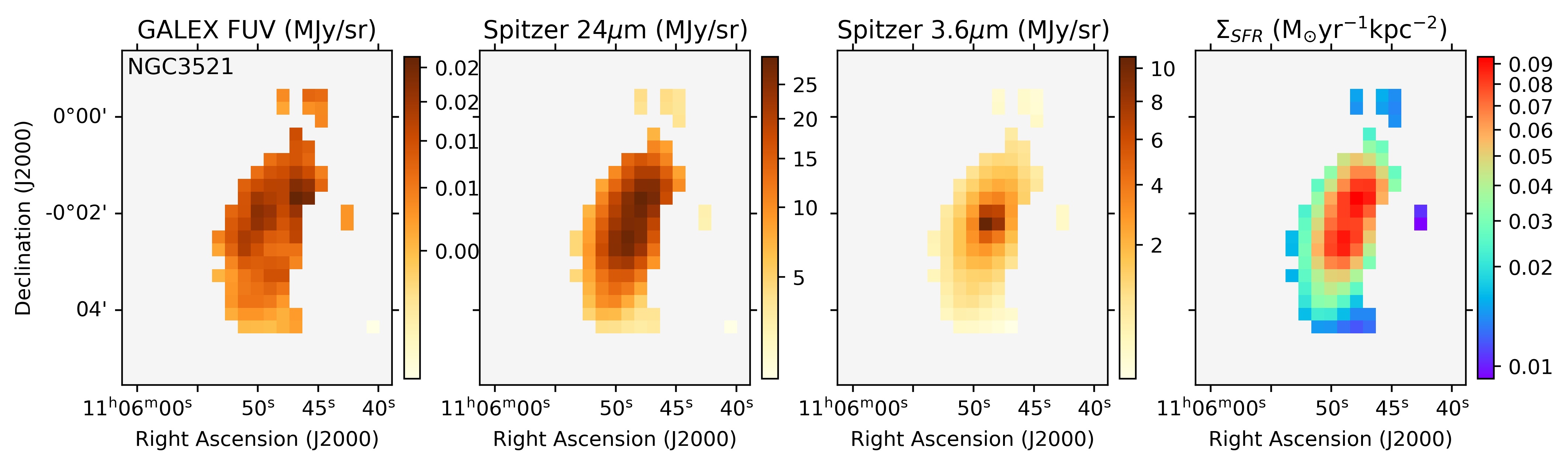}
    \caption{Surface density of star formation in NGC 3521.  Panels as in Figure~\ref{fig:ssfr_ngc3351}.}
    \label{fig:ssfr_ngc3521}
\end{figure*}

\begin{figure*}
    \centering
    \includegraphics[width=\textwidth]{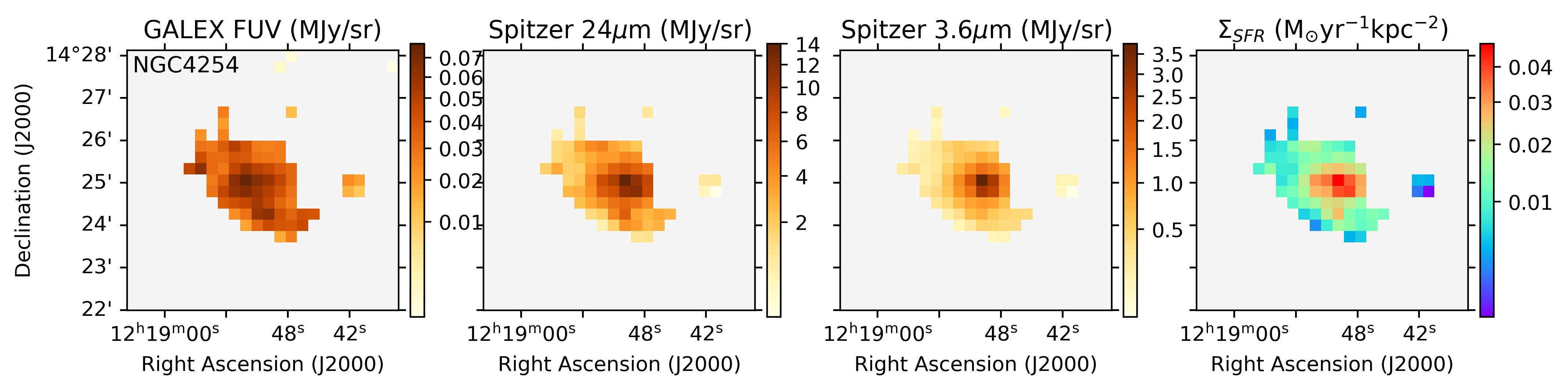}
    \caption{Surface density of star formaiton in NGC 4254.  Panels as in Figure~\ref{fig:ssfr_ngc3351}.}
    \label{fig:ssfr_ngc4254}
\end{figure*}

\begin{figure*}
    \centering
    \includegraphics[width=\textwidth]{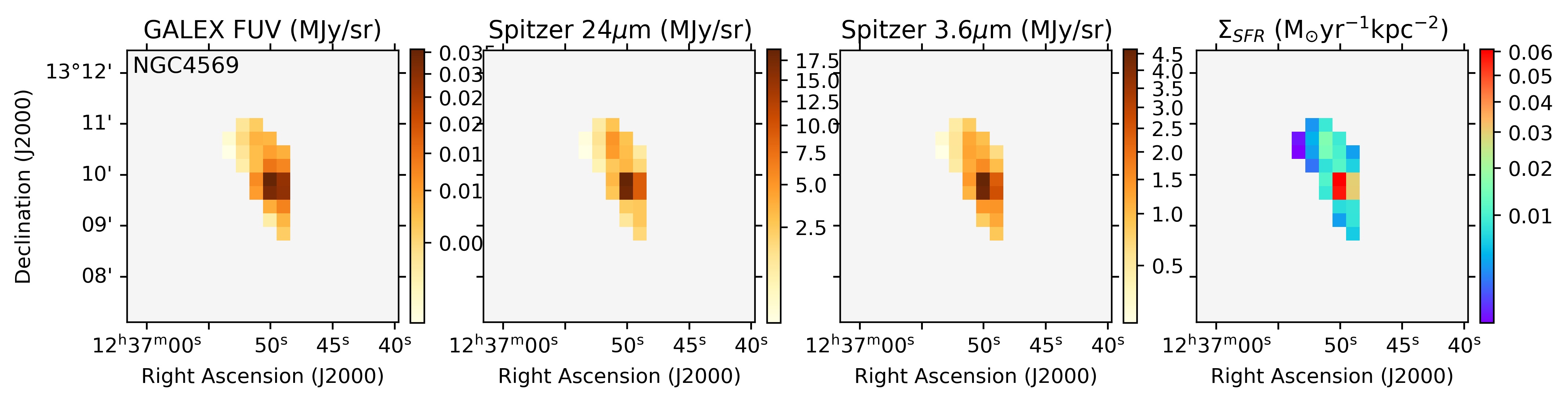}
    \caption{Surface density of star formation in NGC 4569.  Panels as in Figure~\ref{fig:ssfr_ngc3351}.}
    \label{fig:ssfr_ngc4569}
\end{figure*}

\begin{figure*}
    \centering
    \includegraphics[width=\textwidth]{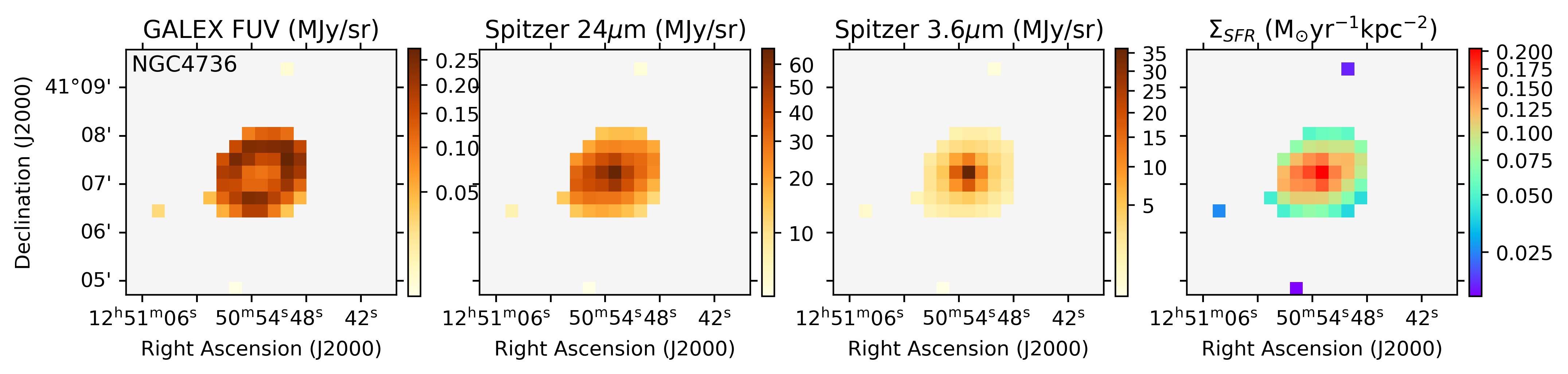}
    \caption{Surface density of star formation in NGC 4736.  Panels as in Figure~\ref{fig:ssfr_ngc3351}.}
    \label{fig:ssfr_ngc4736}
\end{figure*}

\begin{figure*}
    \centering
    \includegraphics[width=\textwidth]{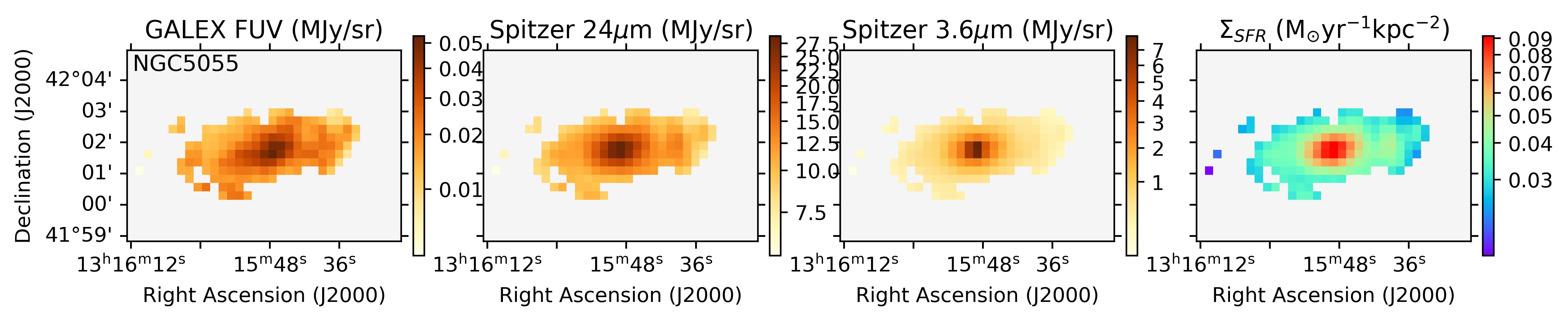}
    \caption{Surface density of star formation in NGC 5055.  Panels as in Figure~\ref{fig:ssfr_ngc3351}.}
    \label{fig:ssfr_ngc5055}
\end{figure*}

\begin{figure*}
    \centering
    \includegraphics[width=\textwidth]{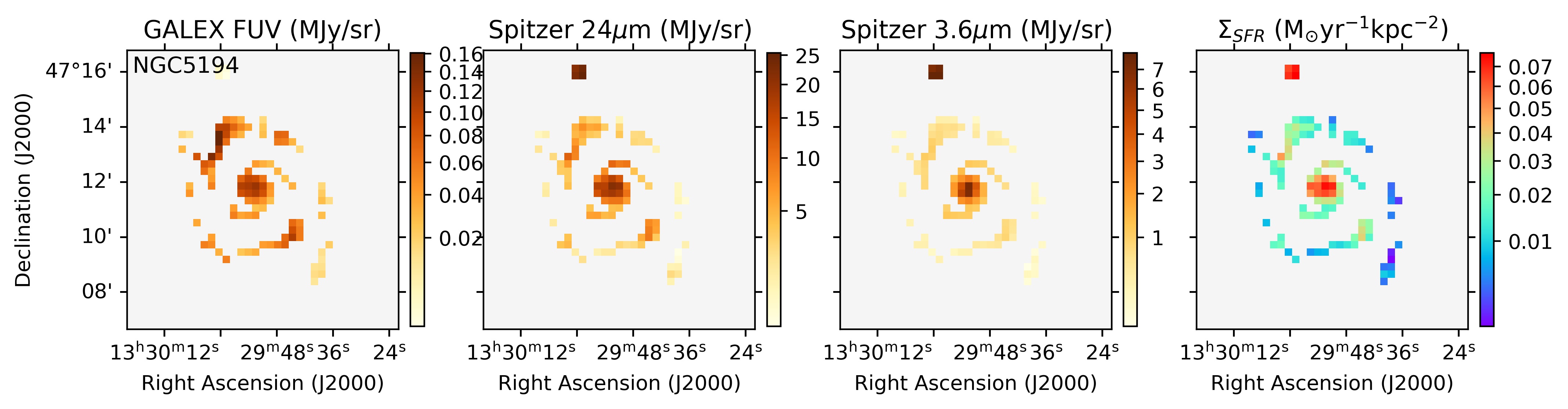}
    \caption{Surface density of star formation in NGC 5194.  Panels as in Figure~\ref{fig:ssfr_ngc3351}.}
    \label{fig:ssfr_ngc5194}
\end{figure*}

\bsp	
\label{lastpage}
\end{document}